\documentstyle[preprint, aps, epsf, floats]{revtex}
\renewcommand{\baselinestretch}{1.5}

\renewcommand{\theequation}{\arabic{section}.\arabic{equation}}
\newcommand{\mysection}[1]{\section{#1}\setcounter{equation}{0}}
\newcommand{\rf}[1]{(\ref{#1})}
\def\ca{\mbox{${\cal A}$}}

\def\cf{\mbox{${\cal F}$}}

\def\cm{\mbox{${\cal M}$}}
\def\cn{\mbox{${\cal N}$}}

\def\cs{\mbox{${\cal S}$}}
\def\cz{\mbox{${\cal Z}$}}
\def\mn{{\mu\nu}}

\newcommand{\gb}{\bar{g}}
\newcommand{\hb}{\bar{h}}
\newcommand{\lb}{\bar{\lambda}}
\newcommand{\be}{\begin{equation}}
\newcommand{\ee}{\end{equation}}
\newcommand{\bea}{\begin{eqnarray}}
\newcommand{\eea}{\end{eqnarray}}
\newcommand{\wh}{\widehat{h}}

\newcommand{\Fbeta}{\mbox{\boldmath $\beta$}}
\newcommand{\kh}{\breve{k}}
\newcommand{\lh}{\breve{\lambda}}

\newcommand{\p}[4]{\Phi^{#1}_{#2}({#3})^{\mbox{\rm \scriptsize {#4}}}}

\def\w{\bar{w}} 
\def\wb{\bar{w}}   
\def\wh{\breve{w}}
\def\v{\bar{u}}  
\def\vh{\breve{u}}
\def\mp{m_{\rm Pl}}
\def\yh{\hat{y}}

\def\Vb{\bar{V}}
\def\hk{\hat{k}}

\def\ub{\bar{\upsilon}}
\def\uh{\breve{\upsilon}}

\def\ob{\sigma}

\def\Tr{\mbox{\rm Tr}}
\def\tr{\mbox{\rm tr}}

\def\ln{\mbox{\rm ln}}
\def\y{y}

\def\rh{\breve{r}}
\def\rhm{\breve{r}_{\rm min}}

\begin{document}
\begin{titlepage}
\renewcommand{\baselinestretch}{1}
\small\normalsize
\begin{flushright}
hep-th/0206145\\
MZ-TH/02-09 \\
\end{flushright}

\vspace{0.2cm}

\begin{center}
{\LARGE \sc A class of nonlocal truncations in\\[1mm]
Quantum Einstein Gravity and its\\[4mm]
renormalization group behavior}

\vspace{0.8cm}
{\large M. Reuter}\\

\vspace{0.0cm}
\noindent
{\it Institute of Physics, University of Mainz\\
Staudingerweg 7, D-55099 Mainz, Germany}\\

\vspace{0.8cm}
{\large F. Saueressig}\\

\vspace{0.0cm}
{\it Institute of Theoretical Physics, University of Jena \\
Max-Wien-Platz 1, D-07743 Jena, Germany}\\
\end{center}

\vspace*{0.0cm}
\begin{abstract}
Motivated by the conjecture that the cosmological constant problem could be solved by strong quantum effects in the infrared we use the exact flow equation of Quantum Einstein Gravity to determine the renormalization group behavior of a class of nonlocal effective actions. They consist of the Einstein-Hilbert term and a general nonlinear function $\cf_k(V)$ of the Euclidean space-time volume $V.$ A partial differential equation governing its dependence on the scale $k$ is derived and its fixed point is analyzed. For the more restrictive truncation of theory space where $\cf_k(V)$ is of the form $V$+$V \ln V$, $V$+$V^2$, and $V$+$\sqrt{V}$, respectively, the renormalization group equations for the running couplings are solved numerically. The results are used in order to determine the $k$-dependent curvature of the $S^4$-type Euclidean space-times which are solutions to the effective Einstein equations, i.e. stationary points of the scale dependent effective action. For the $V$+$V \ln V$-invariant (discussed earlier by Taylor and Veneziano) we find that the renormalization group running enormously suppresses the value of the renormalized curvature which results from Planck-size bare parameters specified at the Planck scale. Hence one can obtain very large, almost flat universes without fine-tuning the cosmological constant.
\end{abstract}
\end{titlepage}
\mysection{Introduction}
Exact renormalization group (RG) equations \cite{bag} provide a powerful tool for the nonperturbative investigation of both fundamental (renormalizable) and effective quantum field theories. In particular the RG equation of the effective average action \cite{ber} has been applied to a variety of matter field theories as well as to Quantum Einstein Gravity \cite{ERGE,ERGE3a}.

The main ingredient in this approach is the effective average action $\Gamma_k$, a Wilsonian coarse grained free energy functional which has a built-in infrared (IR) cutoff at a variable mass scale $k$. The $k$-dependence of $\Gamma_k$ is governed by an exact functional RG equation. In any realistic theory it is impossible to solve this equation exactly. But by appropriately truncating the space of action functionals (``theory space'') one can obtain nonperturbative approximate solutions which do not rely upon small expansion parameters. The truncation is carried out by making an ansatz for $\Gamma_k$ which contains a finite or infinite set of $k$-dependent parameters (``coupling constants'') $\mbox{g}_i(k)$. Upon inserting this ansatz into the functional RG equation and projecting the RG flow onto the truncation subspace one obtains either a partial differential equation or a coupled system of ordinary differential equations for the running couplings.

In the case of Euclidean quantum gravity the effective average action and its RG equation have been constructed in ref. \cite{ERGE}. This first construction used a cutoff of ``TYPE A'' which is formulated in terms of the complete metric fluctuation $h_\mn$. The resulting RG equation has been used to derive the flow equation for the running Newton constant $G_k$ and the cosmological constant $\lb_k$ on the theory space spanned by the ``Einstein-Hilbert truncation'', i.e. by the invariants $\int d^dx \sqrt{g} R$ and $\int d^dx \sqrt{g}$. In \cite{myself} the resulting coupled differential equations have been solved numerically, leading to the complete classification of the RG flow of the Einstein-Hilbert truncation. In order to extend the truncated theory space to invariants containing higher powers of the curvature such as $\int d^dx \sqrt{g} R^2$, for instance, refs. \cite{oliver,oliver2,oliver2a,oliver3} introduce a new cutoff of ``TYPE B'' which is natural to use if one employs the transverse-traceless decomposition of $h_{\mn}$ \cite{tt}.

One of the remarkable results found in \cite{ERGE,myself,oliver,oliver2,oliver3,souma} is that the high energy behavior of 4 dimensional Quantum Einstein Gravity seems to be governed by a non-Gaussian  fixed point (NGFP) which is ultraviolet (UV)-attractive for both the dimensionless Newton constant $g(k) \equiv k^{d-2} G_k$ and cosmological constant $\lambda(k) \equiv \lb_k / k^2$. If this result also holds true for the exact theory, it is likely that Quantum Einstein Gravity is renormalizable at the nonperturbative level \cite{weinbergcc}. In this case it would provide us with a fundamental rather than merely effective theory of quantum gravity which is mathematically consistent and predictive at arbitrarily small distances. In this theory the Newton constant is asymptotically free: near the NGFP, $G_k \approx g_* / k^{d-2}$ vanishes for $k \rightarrow \infty$ (if $d>2$).

\begin{figure}[t]
\renewcommand{\baselinestretch}{1}
\epsfxsize=0.99 \textwidth
\centerline{\epsfbox{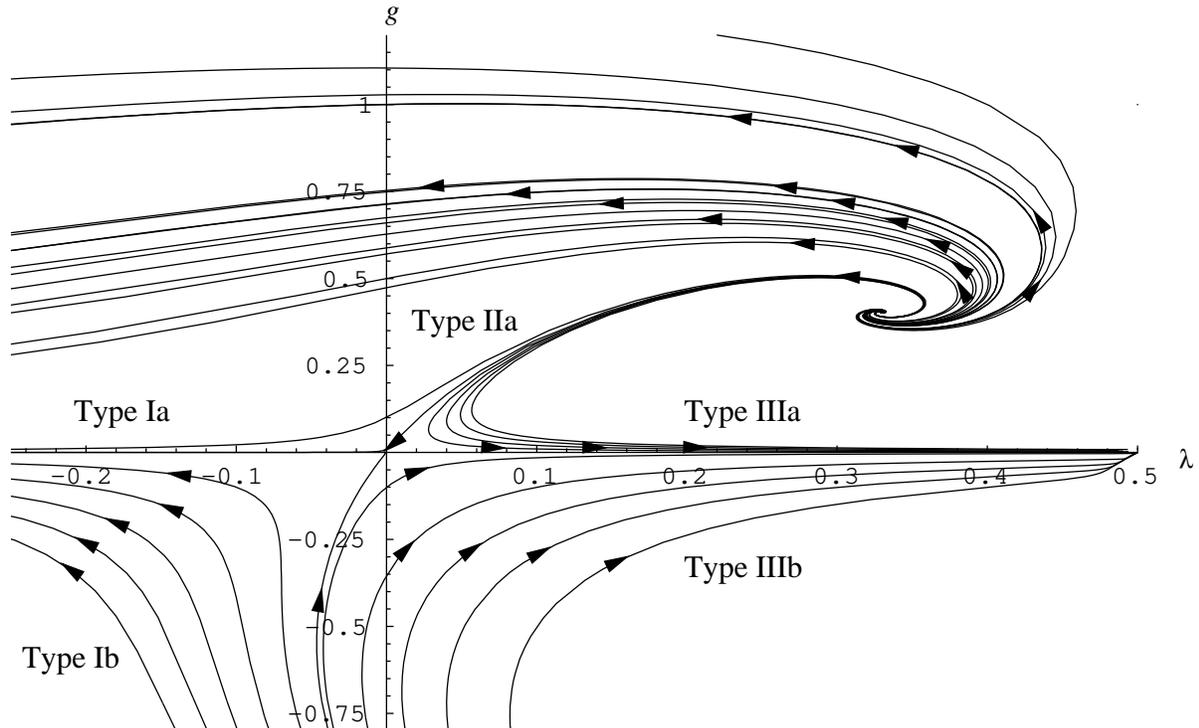}}
\parbox[c]{\textwidth}{\caption{\label{1.eins}{Part of coupling constant space of the Einstein-Hilbert truncation with its RG flow. The arrows point in the direction of decreasing values of $k$. The flow pattern is dominated by a non-Gaussian fixed point in the first quadrant and a trivial one at the origin.}}}
\end{figure}

In ref. \cite{myself} the flow equations of the Einstein-Hilbert truncation were simplified by introducing a technically convenient sharp cutoff and then used in order to continue the trajectories emanating from the NGFP towards the IR (decreasing $k$). The resulting RG flow in 4 dimensions is shown in Fig. \ref{1.eins}.
We see that the IR behavior of these trajectories is governed by the Gaussian fixed point (GFP) located at the origin of the $\lambda$-$g$--plane. The cross-over between the two scaling regions governed by the Gaussian and non-Gaussian fixed point, respectively, takes place at a scale $k_{\rm asym}$ which is of the order of the Planck mass $m_{\rm Pl} \equiv G^{-1/2}_0$. 

Depending on whether the trajectories run to the left, right, or on the stability axis of the GFP we find trajectories of the Types Ia, IIIa, and IIa, respectively. Only the trajectories Ia and IIa lead to well-defined $k \rightarrow 0$-limits of $G_k$ and $\lb_k$, yielding $\lb_0 < 0$ and $\lb_0 = 0$, respectively. The trajectories IIIa terminate at a finite value of $k$ at the boundary of the $\lambda$-$g$--space at $\lambda = 1/2$. Beyond this boundary the $\Fbeta$-functions of $g$ and $\lambda$ are undefined. Therefore the RG trajectories found in the Einstein-Hilbert truncation can only lead to negative or vanishing values for the renormalized cosmological constant $\lb_0 \equiv \lim_{k \rightarrow 0} \lb_k$. While the ``separatrix'' connecting the NGFP to the GFP (trajectory IIa) leads to $\lb_0 = 0$, the trajectories of Type Ia generically give rise to (negative) renormalized cosmological constants of the order of $m_{\rm Pl}^2$. Smaller values can be obtained only by an extreme fine-tuning.

Even though recent observations of high redshift supernovae \cite{sn1,sn2,sn3} and measurements of the power spectrum of the cosmic microwave radiation \cite{boom,boom2} seem to indicate that our universe is characterized by a small (positive) cosmological constant, it is clear that a renormalized value $\lb_0 = {\rm O}(m_{\rm Pl}^2)$ is phenomenological unacceptable. In the present universe the vacuum energy density due to $\lb$ can be of the order of the matter energy density at most. Hence $\lb$ must be smaller than its ``natural'' value $m_{\rm Pl}^2$ by about 120 orders of magnitude. This is the famous cosmological constant problem \cite{wb1a}.

A notorious difficulty which any attempt at solving this naturalness problem will face is that its solution cannot come from any remote mechanism which is operative near the Planck scale only. In fact, even if for some reason there is no bare cosmological constant of the order $m_{\rm Pl}^2$, the vacuum condensates associated with the electroweak or QCD phase transition induce contributions to $\lb_0$ which are still many orders of magnitude larger than the experimental bound. As a consequence, if one tries to solve the fine-tuning problem by invoking some dynamical mechanism giving rise to a sufficiently small $\lb_0$ automatically \cite{rwcos} then this mechanism must be effective at very low energy scales, the familiar scales of standard particle physics and below.

It is a very attractive speculation that strong renormalization effects of quantum gravity in the infrared are responsible for the tiny value of $\lb_0$ \cite{tsamis,mottola}. In fact, since perturbation theory with $\lb \not= 0$ is much more IR divergent than with $\lb = 0$, one could argue that if quantum gravity is capable of curing its IR diseases dynamically by nonperturbative effects, then these effects should drive $\lb_k$ to zero for $k \rightarrow 0$. 

Strong IR quantum effects of this kind would be comparable to phenomena like confinement or the formation of bound states in QCD. It is clear that such strong coupling phenomena are much harder to understand than the weakly coupled asymptotic scaling region where $k \gg \Lambda_{\rm QCD}$ in the case of QCD or $k \gg m_{\rm Pl}$ for gravity \cite{myself}. In QCD fairly simple, local truncations are sufficient for a reliable description of the asymptotic scaling region, while in the IR much more complicated, nonlocal truncations are needed \cite{gluco}. Likewise it seems that in gravity the Einstein-Hilbert truncation or simple local extensions of it are appropriate close to the NGFP \cite{oliver,oliver2}, but in the IR, i.e. for scales $k \lesssim m_{\rm Pl}$, we expect that much more complicated, presumably nonlocal truncations are needed for a proper description. The termination of the Type IIIa trajectories at a nonzero $k_{\rm term} = {\rm O}(m_{\rm Pl})$ is a typical symptom showing that the Einstein-Hilbert truncation becomes insufficient below a certain critical scale \cite{myself}.

Which invariants could be important in the IR of quantum gravity? As we are trying to understand its large distance behavior (on the scale set by the Planck length $\ell_{\rm Pl} \equiv 1/m_{\rm Pl} \equiv \sqrt{G_0}$) it is clear that terms with higher powers of the curvature ($\int d^4x \, \sqrt{g} \, R_{\mn\rho\sigma}R^{\mn\rho\sigma}$, $\int d^4x \, \sqrt{g} \, R^3$, etc.) are of no help here. For a large universe, after the last cosmological phase transition, say, the contributions of these terms to the effective Einstein equations are negligible compared to those from $\int d^4x \, \sqrt{g} R$. They are suppressed by inverse powers of $m_{\rm Pl}$. What one needs are invariants which grow and become important when the universe gets large. Typical invariants which meet this requirement are nonlocal functionals of the metric. As an example, the term $\int d^4x \, \sqrt{g} \, R (-D^2)^{-1} R$ was added to $\int d^4x \, \sqrt{g} R$ in ref. \cite{cw}, and it was shown that the resulting modification of general relativity is phenomenological acceptable for a wide range of parameters. In ref. \cite{mottola} the IR physics resulting from the 4D ``induced gravity'' action $\int d^4x \, \sqrt{g} \, R \, \ln(-D^2) \, R$ was analyzed.

In the present paper we shall perform a first investigation of the RG behavior of nonlocal actions in the effective average action framework. As we are mainly interested in getting a first idea about the qualitatively new effects which can occur, and in order to avoid the extreme algebraic complexity of calculations involving terms like $\int d^{4}x \sqrt{g} R (-D^2)^{-1} R$, we shall focus on a simpler class of nonlocal terms. We are going to study invariants of the type $\cf_k(V)$ where $\cf_k$ is a nonlinear function of the Euclidean space-time volume $V \equiv \int d^dx \sqrt{g}$. In particular we will investigate the RG flow of effective actions containing the invariants $V \ln(V/V_0)$ and $V^2$ which, in the context of wormhole-physics, have already been discussed in \cite{TVII} and \cite{coleman,TVI,adler}, respectively. (Here $V_0$ denotes an arbitrary reference volume.) The corresponding approximations of $\Gamma_k$ will be called the ``$V$+$V \ln V$--'' and ``$V$+$V^2$--truncation'' and are given by
\be\label{1.1}
\Gamma_k[g] = \frac{1}{16 \pi G} \, \int d^dx \sqrt{g} \left(-R + 2 \lb_k \right) \, + \, \frac{1}{16 \pi G} \, \v_k \, V  \, \ln(V / V_0) 
\ee
and
\be\label{1.1a}
\nonumber \Gamma_k\left[g \right] = \frac{1}{16 \pi G} \, \int d^dx \sqrt{g} \left( -R + 2 \bar{\lambda}_k \right) +\frac{1}{16 \pi G} \, \w_k V^2, 
\ee
respectively. Inspired by the fixed point properties of the RG flow we shall also discuss the ``$V$+$\sqrt{V}$--truncation''.

As we are mostly interested in the IR regime we neglect the running of Newton's constant in the present investigation: $G_k \approx G_0 \equiv G$. At least according to the results from the Einstein-Hilbert truncation this is a sensible first approximation if $k \lesssim m_{\rm Pl}$.

We shall impose initial conditions for $\lb_k$ and $\v_k$ or $\w_k$ at some scale $k = \hk$, usually at $\hk = m_{\rm Pl}$, and then use the RG equation in order to evolve the parameters towards smaller values of $k$. It will turn out that if the new couplings $\v_k$ and $\w_k$ are put to zero at the starting point, they will continue to vanish at all lower scales $k < \hk$. The RG flow resulting from the truncations \rf{1.1} or \rf{1.1a} cannot generate the new invariants if no initial ``seed'' is present. However, we expect that with a more general truncation nonlocal terms, perhaps of a more complicated structure, are generated dynamically along the RG trajectories emanating from the NGFP \cite{olivercond}. Imposing nonzero initial values $\v_{\hk}$ or $\w_{\hk}$ in our calculation mimics this more complicated dynamics to some extent.

Another way of looking at our results is to interpret them in the spirit of effective field theories. For instance, one could imagine that string theory prepares the initial conditions for a low energy field theory description of gravity, and that these initial conditions include $\v_{\hk} \not= 0$ or $\w_{\hk} \not= 0$ at the Planck scale.

Including the new couplings in the effective action of gravity gives rise to various theoretical possibilities that might lead to a solution of the cosmological constant problem. First, as speculated in \cite{astro}, the extension of the theory space could generate a new fixed point which is IR-attractive for the dimensionless cosmological constant: $\lambda(k) \rightarrow \lambda^*_{\rm IR}$ as $k \rightarrow 0$. In this case the renormalized coupling constant $\lb_0 \equiv \lim_{k \rightarrow 0} \lambda(k) k^2$ would vanish with $\lb_k$ proportional to $k^2$ for all trajectories reaching the basin of attraction of the fixed point. In this manner the cosmological evolution could be understood as a ``cross-over'' from the NGFP in the UV to the new fixed point in the IR\footnote{It is known \cite{liou} that 2D Liouville quantum gravity describes a similar cross-over between two conformal field theories.}. (In ref. \cite{cosmo1} it has been shown by means of a straightforward RG improvement\footnote{For a similar RG improvement in black hole physics see ref. \cite{bh}.} that the NGFP in the UV leads to a very interesting cosmology of the Planck era which might provide a solution of the flatness and the horizon problem.)

A different scenario using nonlocal effective actions in order to solve the cosmological constant problem has been proposed by Taylor and Veneziano \cite{TVII,TVI}. We will now briefly review this mechanism using the example of the $V$+$V \ln V$--truncation.

Varying \rf{1.1} with respect to the metric $g_\mn$ leads to the following modified equation of motion\footnote{We ignore the $k$-dependence of $\lb$ and $\v$ for the time being.}:
\be\label{1.2}
R^{\mn} \, - \, \frac{1}{2} \, R \, g^{\mn} = - \left[\lb + \frac{\v}{2} + \frac{\v}{2} \, \ln(V / V_0) \right] \, g^{\mn} 
\ee
The structure of this equation suggests defining an effective cosmological constant, $\lambda_{\rm eff}(V)$, as
\be\label{1.3}
\lambda_{\rm eff}(V) \equiv \lb + \frac{\v}{2} + \frac{\v}{2} \, \ln(V / V_0),
\ee
which depends on the volume of the space-time. Thus eq. \rf{1.2} takes on the usual form of the Einstein equation without matter:
\be\label{1.4}
R^{\mn} \, - \, \frac{1}{2} \, R \, g^{\mn} = - \lambda_{\rm eff}(V) \, g^{\mn} 
\ee
Contracting \rf{1.4} with $g_\mn$ and substituting the resulting equation $R = 4 \lambda_{\rm eff}$ back into \rf{1.4} leads to
\be\label{1.5}
R_{\mn} = \lambda_{\rm eff} \, g_{\mn}
\ee
Obviously all solutions to the modified Einstein equation are Einstein spaces. In the following we specialize for maximally symmetric solutions. In particular the 4-sphere $S^4$ of radius $r$ has the following properties:
\be\label{1.6}
R^{\mn}= \frac{3}{r^2} \, g^{\mn}, \quad R = \frac{12}{r^2}, \quad V = \ob_4 \, r^4, \quad \ob_4 \equiv \frac{8 \pi^2}{3}
\ee
Substituting the 4-sphere into \rf{1.4} we see that it is a solution to the equations of motion provided
\be\label{1.7}
\frac{3}{r^2} = \lambda_{\rm eff}(\ob_4 \, r^4)
\ee
The resulting equation for the radius $r$ reads\footnote{As a consequence of the ``principle of symmetric criticality'' \cite{palais} inserting the $S^4$ ansatz \rf{1.4} into the action \rf{1.1} and extremizing the resulting function of $r$ leads to the same equation for the ``on-shell'' value of the radius $r$.}:
\be\label{1.8}
\v \, r^2 + \v \, r^2 \, \ln(\ob_4 \, r^4 / V_0) + 2 \, \lb \, r^2 - 6 = 0
\ee
Reexpressing $r^2$ by the effective cosmological constant and considering $\lambda_{\rm eff}$ the independent variable we find the condition:
\be\label{1.9}
\lambda_{\rm eff} = \lb + \frac{\v}{2} \left[ \ln\left( \frac{9 \ob_4}{\lambda_{\rm eff}^2 V_0} \right) + 1 \right]
\ee

This equation has very interesting properties. Assuming $\v > 0$ (which, as we will see, is a reasonable assumption) one finds the following relations between the cosmological constant proper, $\lb$, and the effective cosmological constant \cite{TVII}:
\bea\label{1.10}
\nonumber \lambda_{\rm eff} \approx \lb &\quad& \mbox{if} \quad  \lb > 0 \\
\lambda_{\rm eff} \approx \left( \frac{9 \ob_4}{V_0} \right)^{1/2} \exp\left[ \frac{\lb}{\v} + \frac{1}{2} \right] &\quad& \mbox{if} \quad \lb < 0 
\eea
For $\lb > 0$ the cosmological constant and the effective cosmological constant are of the same order of magnitude while for a negative sign of $\lb$ the effective cosmological constant is exponentially suppressed (``quenched''). Therefore this mechanism can provide a satisfactorily small effective cosmological constant, i.e. small space-time curvature, without the need of the cosmological constant itself being small. 

In the work of Taylor and Veneziano \cite{TVII,TVI}, $\lb$ and $\v$ are classical parameters and no evolution effects are taken into account. In the present paper we investigate the ``RG improvement'' of the mechanism reviewed above. We shall replace $\lb$ and $\v$ by their running counterparts $\lb_k$ and $\v_k$. In this manner the radius $r$ of the $S^4$, too, becomes a function of $k$. We may expect that for a Euclidean universe of radius $r$ the relevant effective action is $\Gamma_k$ at $k \approx 1/r$. Thus, when looking for large universes, we should use the renormalized couplings $\lb_0$, $\v_0$ rather than the bare ones, $\lb_{\hk}$ and $\v_{\hk}$.

For a given trajectory $k \mapsto (\lb_k, \v_k)$ we shall solve eq. \rf{1.8} with the running couplings inserted and obtain the radius $r \equiv r(k)$ of the $S^4$ which is the stationary point of $\Gamma_k$. We can then compare the ``bare'' radius $r(\hk)$ to the ``renormalized'' one, $r(k = 0)$. As we shall see, the inclusion of the RG running leads to a tremendous ``inflation'' of the universe: $r(k=0) \gg r(\hk)$. The renormalization effects greatly facilitate obtaining large, essentially flat universes from generic Planck-size initial values $(\lb_{\hk}, \v_{\hk})$ specified at $\hk = m_{\rm Pl}$. In this way one can effectively solve the cosmological constant problem even though $\lb_0$ is not small. 

The remaining sections of this paper are organized as follows. In Section II we use the exact RG equation with the ``TYPE A'' cutoff \cite{ERGE} to derive a partial differential equation which governs the $k$-dependence of an arbitrary nonlocal invariant of the form $\cf_k(V)$. The flow equations of the $V$+$V \ln V$-- and $V$+$V^2$--truncations are then derived by specializing $\cf_k$ to these truncations. In Section III we investigate the RG flow of the coupling constants in the $V$+$V \ln V$--truncation by numerically solving the flow equations with the sharp cutoff introduced in \cite{myself}. In Section IV we investigate the impact of the running coupling constants in the $V$+$V \ln V$--truncation on the radius of the ``classical'' $S^4$ solution of the modified Einstein equations. In Section V we discuss the properties of the modified GFP which follow from the partial differential equation for $\cf_k(V)$. Motivated by the results of this analysis we investigate the RG flow of the $V$+$\sqrt{V}$--truncation in Section VI. In Appendix A we briefly summarize the results for the RG flow and classical solutions in the $V$+$V^2$--truncation.

\mysection{Flow equations with nonlocal invariants}
In order to derive the nonperturbative partial differential equation describing the RG flow of an effective action of quantum gravity which includes arbitrary nonlocal invariants $\cf_k(V)$ we use the effective average action approach to Quantum Einstein Gravity \cite{ERGE}.

The main ingredient of this method is the exact evolution equation for the effective average action $\Gamma_k[g_{\mn}]$ for gravity which, in its original formulation, has been constructed in \cite{ERGE}. The derivation of this evolution equation parallels the approach already successfully tested for Yang-Mills theories \cite{ym,ym2}. In principle it is straightforward to include the additional renormalization effects coming from matter fields \cite{percacci,odintsov}, but these are not included in the present derivation.

In the construction of $\Gamma_k[g]$ one starts out with the usual path integral of $d$-dimensional Euclidean gravity. It is gauge fixed by  using the background field method \cite{abb,adl} and employing a background gauge fixing condition. A priori the effective average action $\Gamma_k[g; \gb]$ depends on both the ``dynamical'' metric $g$ and the background metric $\gb$. The conventional effective action $\Gamma[g]$ is regained as the $k\rightarrow 0$--limit of $\Gamma_k[g] \equiv \Gamma_k[g;\gb=g]$ where the two metrics have been identified. By this construction $\Gamma_k[g]$ becomes invariant  under general coordinate transformations. 

The crucial new component in the construction of $\Gamma_k[g, \gb]$ is the $k$-dependent IR-cutoff term $\Delta_k S$ added to the action under the path integral. This term discriminates between the high $(p^2>k^2)$ and low-momentum modes $(p^2<k^2)$. It suppresses the contribution of the low-momentum modes to the path integral by adding a momentum dependent mass term
\be\label{2.1}
\Delta_k S[h,C,\bar{C};\gb]
= \frac{1}{2}\kappa^2\int d^dx \, \sqrt{\gb}\, h_\mn R^{\rm grav}_k[\gb]^{\mn\rho\sigma}h_{\rho\sigma} +\sqrt{2}\int d^dx\, \sqrt{\gb}\, \bar{C}_\mu R^{\rm gh}_k[\gb]C^\mu
\ee
Here $\kappa \equiv (32 \pi G)^{-1/2}$, and the first and second term on the RHS provide the cutoff for the fluctuations of the metric, $h_{\mn} = g_{\mn} - \gb_{\mn}$, and the ghost fields $\bar{C}_{\mu}, C^{\mu}$, respectively. In this paper we choose the following form of the cutoff 
operators $R^{\rm grav}_k$ and $R^{\rm gh}_k$ \cite{ERGE}: 
\be\label{2.2}
 R^{\rm grav}_k[\gb] = \cz^{\rm grav}_k k^2 R^{(0)}(-\bar{D}/k^2), \qquad R^{\rm gh}_k[\gb] = k^2 R^{(0)}(-\bar{D}/k^2)
\ee
Here $(\cz^{\rm grav}_k)^{\mn\rho\sigma} = \left[ (I-P_{\phi})^{\mn\rho\sigma} - (d-2)/2 P_{\phi}^{\mn\rho\sigma}  \right] Z_{Nk}$ is a matrix acting on $h_{\mn}$. In this expression $(P_\phi)^{\mn\rho\sigma} \equiv d^{-1} \, \gb^{\mn} \, \gb^{\rho\sigma}$ projects $h_{\mn}$ onto its trace part $\phi$. In the terminology of ref. \cite{oliver}, this form of $\Delta_k S$ defines the ``cutoff of TYPE A''. The so-called ``shape function'' $R^{(0)}$ is essentially arbitrary except that it has to satisfy the conditions
\be\label{2.3}
R^{(0)}(0) = 1, \qquad R^{(0)}(z \rightarrow \infty) = 0
\ee

Neglecting the evolution of the ghost sector which corresponds to a first truncation of the general structure of $\Gamma_k$, one finds that $\Gamma_k[g, \gb]$ satisfies the following flow equation \cite{ERGE}:
\bea\label{2.4}
\nonumber \partial_t \Gamma_k\left[g, \gb \right] &=& \frac{1}{2} \Tr \left[ \left(\kappa^{-2} \Gamma^{(2)}_k +  R^{\rm grav}_k\left[ \bar{g} \right] \right)^{-1} \, \partial_t R^{\rm grav}_k\left[ \gb \right] \right] \\
&& - \Tr\left[ \left( -\cm \left[ g, \gb \right] + R^{\rm gh}_k\left[ \gb \right] \right)^{-1} \partial_t R^{\rm gh}_k \left[ \gb \right] \right]
\eea
Here $\Gamma_k^{(2)}[g, \gb]$ denotes the Hessian of $\Gamma_k\left[g, \gb \right]$ with respect to $g_{\mn}$ at fixed background field $\gb_{\mn}$, and $t \equiv \ln(k/\hat{k})$ is the  ``renormalization group time'' with respect to the reference scale $\hat{k}$. Furthermore, $\cm$ represents the Faddeev-Popov ghost operator. 

This equation is our starting point to derive the partial differential equation describing  the RG flow of an effective action including an arbitrary function $\cf_k(V k^d)$ of the volume $V \equiv \int d^dx \sqrt{g}$. In this course we approximate $\Gamma_k[g; \gb]$ by the following truncation ansatz: 
\be\label{2.5}
\Gamma_k[g; \gb] = 2 \, \kappa^2 \, \left\{ \int d^dx \sqrt{g} \, (-R) + 2 \, \cf_k(V k^d) \, \right\} + \mbox{classical gauge fixing term}
\ee
The function $\cf_k$ includes a running cosmological constant, but we neglect the running of Newton's constant in the present investigation. The parameter $\kappa$ is treated as a constant in the following.

Later on it will be easy to derive the ordinary differential equations governing the scale-dependence of the coupling constants in the $V$+$V \ln V$--truncation \rf{1.1} and the $V$+$V^2$--truncation \rf{1.1a}. They correspond to specializing
\be\label{2.6}
\cf_k(V k^d) = \lb_k \, V + \frac{1}{2} \, \v_k \, V \, \ln(V/V_0) 
\ee
and
\be\label{2.7}
\cf_k(V k^d) = \lb_k \, V + \frac{1}{2} \, \wb_k \, V^2
\ee
respectively.

Let us now derive the RG equation for $\cf_k$. Substituting our ansatz for $\Gamma_k$, eq. \rf{2.5}, into \rf{2.4} we find the following expression for the LHS of the flow equation:
\be\label{2.10}
 \frac{d}{dt} \, \Gamma_k[g; \gb] = 4 \, \kappa^2 \, \frac{d}{dt} \, \cf_k(V k^d) \equiv \cs_L[g]
\ee
The derivative with respect to $t$ acts on both the implicit and the explicit $k$-dependence of $\cf_k$.

For the evaluation of the RHS, $\cs_R$, of eq. \rf{2.4} we first calculate the second functional derivative of $\Gamma_k[g;\gb]$ at fixed background metric $\gb$. We therefore decompose $g = \gb + \hb$ into the background metric and an arbitrary fluctuation $\hb$ and expand $\Gamma_k[\hb;\gb]$ in powers of $\hb$, $\Gamma_k[\gb+\bar{h};\gb]= \Gamma_k[\gb;\gb] + O(\bar{h})+\Gamma_k^{ \rm quad}[\bar{h};\gb]+O(\bar{h}^3)$. Splitting $\hb_{\mn}$ into its traceless part $\widehat{h}_{\mn}$ and its trace part using
\be\label{2.12}
\hb_{\mu\nu}=\widehat{h}_{\mu\nu} + d^{-1} \, \gb_{\mu\nu} \, \phi, \quad \gb^{\mu\nu} \, \widehat{h}_{\mu\nu} = 0 \qquad \mbox{and} \qquad \phi \equiv \gb^{\mn} \bar{h}_{\mn}
\ee
we find the following quadratic term $\Gamma^{\rm quad}_k[\hb;\gb]$:
\bea\label{2.13}
\nonumber \Gamma^{\rm quad}_k[\hb; \gb] &=& \kappa^2 \, \int d^dx \sqrt{\gb} \, \bigg\{\frac{1}{2} \, \widehat{h}_{\mn} \, \left[- \bar{D}^2 - 2 \, \cf_k^{\prime}(\Vb k^d) \,  k^d + \bar{R} \right] \, \widehat{h}^{\mn} \\ 
\nonumber && \qquad - \, \frac{d-2}{4d} \, \phi \, \left[ -\bar{D}^2 - 2 \, \cf_k^{\prime}(\Vb k^d) \,  k^d + \frac{d-4}{d} \, \bar{R} \right] \, \phi \\
\nonumber && \qquad - \bar{R}_{\mn} \, \widehat{h}^{\nu\rho} \, \widehat{h}^{\mu}_{~\rho} + \bar{R}_{\alpha\beta\nu\mu} \, \widehat{h}^{\beta\nu} \, \widehat{h}^{\alpha\mu} + \frac{d-4}{d} \, \phi \, \bar{R}_{\mn} \, \widehat{h}^{\mn} \, \bigg\} \\
&& + \frac{1}{2} \, \kappa^2 \, \int d^dx \sqrt{\gb} \, \int d^dy \sqrt{\gb} \, \phi(x) \, \cf_k^{\prime \prime} (\Vb k^d) \, k^{2d} \, \phi(y) 
\eea
Here all the bared quantities are constructed from the background metric, and the prime denotes the derivative of $\cf_k$ with respect to its argument. From the quadratic form \rf{2.13} we can read off the Hessian $\Gamma^{(2)}_k$ to be used under the trace on the RHS of the flow equation. 

This trace is a complicated functional of both $g$ and $\bar{g}$. Our next task is to project this functional onto the truncation subspace parameterized by the ansatz \rf{2.5}. Since neither the term $\int d^dx \, \sqrt{g} \, R$ nor the classical gauge fixing term contain information about $\cf_k$ we do not have to project out those terms. This means in particular that after having performed the second variation we may set $g = \gb$, in which case the gauge fixing term vanishes \cite{ERGE}: 
\be\label{2.14}
\Gamma_k[\gb;\gb] = 2 \kappa^2 \left\{ \int d^dx \, \sqrt{\gb} (- \bar{R}) + 2 \, \cf_k(\bar{V} k^d) \right\}
\ee
For $g = \gb$, the traces $\Tr[\ldots]$ are functionals of $\gb$ alone. In order to determine the running of $\cf_k$ we have to expand these functionals and retain only the terms without any curvature quantity but with an arbitrary dependence on $\bar{V}$.

 This is most easily achieved by the following technical trick. We choose $\gb$ to be a one-parameter family of flat metrics on the torus ${\rm T}^d$, the free parameter being its volume $V[\gb] = \int \, d^dx \, \sqrt{\gb} = \bar{V}$. From now on we consider $\bar{V}$ a pure number rather than a functional of the metric. Thus our problem boils down to computing the dependence of the traces $\Tr[\ldots]$ on the parameter $\bar{V}$. Upon equating the result to $\cs_L[\gb] = 4 \kappa^2 \frac{d}{dt} \cf_k(\bar{V} k^d)$ we obtain the desired equation for $\cf_k$. Since for the flat torus $\bar{R}^{\rm torus}_{\mn\rho\sigma}=0, \bar{R}^{\rm torus}_{\mn} = 0, \bar{R}^{\rm torus} = 0$ it is clear that this method projects out precisely the right terms from $\Gamma_k$.

Substituting the ${\rm T}^d$-metric, eq. \rf{2.13} simplifies to
\bea\label{2.15}
\nonumber \Gamma^{\rm quad}_k[\hb; \gb] &=& \kappa^2 \, \int d^dx \sqrt{\gb} \, \bigg\{\frac{1}{2} \, \widehat{h}_{\mn} \, \left[- \bar{D}^2 - 2 \, \cf_k^{\prime}(\Vb k^d) \,  k^d \right] \, \widehat{h}^{\mn} \\ 
\nonumber && \qquad \qquad \qquad - \, \frac{d-2}{4d} \, \phi \, \left[ -\bar{D}^2 - 2 \, \cf_k^{\prime}(\Vb k^d) \,  k^d \right] \, \phi \bigg\} \\
 && + \frac{1}{2} \, \kappa^2 \, \int d^dx \sqrt{\gb} \, \int d^dy \sqrt{\gb} \, \phi(x) \, \cf_k^{\prime \prime} (\Vb k^d) \, k^{2d} \, \phi(y) 
\eea

In order to diagonalize this quadratic form in field space we expand the trace $\phi$ in terms of  harmonic functions $T_{lm}$ on the torus. The functions $T_{lm}$ are eigenfunctions of the operator $-\bar{D}^2$ constructed from the background metric, with eigenvalues $\mu_l$ and degeneracies $\delta_l$: 
\be\label{2.16}
-\bar{D}^2 \, {\rm T}_{lm} = \mu_l \, {\rm T}_{lm} \qquad \mbox{with } m=1, \cdots ,\delta_l \mbox{ and} \quad \left\{ \begin{array}{rcl}  \mu_0 = 0 & \mbox{for} & l=0 \\ \mu_l > 0 & \mbox{for} & l>0 \end{array} \right.
\ee
Here $T_{0m} \equiv \frac{1}{\sqrt{V}}$ denotes the nondegenerate zero-mode of $-\bar{D}^2$. Using the orthonormality relation for the $T_{lm}$, $\int d^dx \, \sqrt{\gb(x)} \, {\rm T}_{lm} \, {\rm T}_{l^{\prime}m^{\prime}} = \delta_{ll^{\prime}}\, \delta_{mm^{\prime}}
$, we can decompose the scalar field $\phi(x)$ into its zero-mode $\varphi$ and the higher modes $\widehat{\phi}(x)$:
\be\label{2.17}
\phi(x) = c_{00} {\rm T}_{00} + \; \sum^{\infty}_{l = 1} \, \sum^{\delta_l}_{m=1} c_{lm}{\rm T}_{lm}(x)  \equiv \varphi +  \widehat{\phi}(x)
\ee

Applying this decomposition to eq. \rf{2.15}, we find that the result is now diagonal in the fields $\widehat{h}_{\mn}, \widehat{\phi}$ and $\varphi$:
\bea\label{2.18}
\Gamma^{\rm quad}_k[\hb; \gb] &=& \kappa^2 \, \int d^dx \sqrt{\gb} \, \bigg\{\frac{1}{2} \, \widehat{h}_{\mn} \, \left[- \bar{D}^2 - 2 \, \cf_k^{\prime}(\Vb k^d) \,  k^d \right] \, \widehat{h}^{\mn} \\ 
\nonumber && \qquad \qquad \qquad - \, \frac{d-2}{4d} \, \widehat{\phi} \, \left[ -\bar{D}^2 - 2 \, \cf_k^{\prime}(\Vb k^d) \,  k^d \right] \, \widehat{\phi} \\
\nonumber && \hspace{-5mm} - \, \frac{d-2}{4d} \, \varphi \left[ -\bar{D}^2 - 2 \, \cf_k^{\prime}(\Vb k^d) \,  k^d -  \frac{2d}{d-2} \, \cf_k^{\prime \prime}(\Vb k^d) \, k^{2d} \, \Vb \right] \varphi \bigg\}
\eea
From this quadratic form we can read off the operators which appear in the trace over the metric degrees of freedom in eq. \rf{2.4}:
\bea\label{2.19}
\left( \kappa^{-2} \Gamma^{(2)}_k [g; g] + R^{\rm grav}_k \right)_{\widehat{h}\widehat{h}} &=& 
\left[ -D^2 + k^2 \, R^{(0)}\left(-D^2/k^2 \right) - 2 \cf_k^{\prime}(V k^d) k^d \right] \\[1.5ex]
\nonumber \left( \kappa^{-2} \Gamma^{(2)}_k [g; g] + R^{\rm grav}_k \right)_{\widehat{\phi}\widehat{\phi}} &=& 
- \, \frac{d-2}{2d} \, \left[ -D^2 + k^2 \, R^{(0)}\left(-D^2/k^2 \right) - 2 \cf_k^{\prime}(V k^d) k^d \right] \\[1.5ex]
\nonumber \left( \kappa^{-2} \Gamma^{(2)}_k [g; g] + R^{\rm grav}_k \right)_{\varphi\varphi} &=& 
- \frac{d-2}{2d} \left[k^2 - 2 \, \cf_k^{\prime}(V k^d) \, k^d - \frac{2d}{d-2} \, \cf_k^{\prime \prime}(V k^d) \, k^{2d} \, V \right]  
\eea
To derive the last line, we have used the fact that $-\bar{D}^2 \varphi = 0$ and $R^{(0)}(0)=1$. Since we anyhow identified $g \equiv \gb$ after carrying out the variation with respect to $\hb$, we dropped the bars on the quantities constructed from the background metric.

After adding the ghost contribution already found in \cite{ERGE} the RHS of eq. \rf{2.4} takes the following form: 
\bea\label{2.20}
\nonumber \cs_{\rm R} &=& \Tr_T \left[ \frac{\cn_0}{\ca} \right] + \Tr_S \left[ \frac{\cn_0}{\ca } \right] - 2 \Tr_V \left[ \frac{\cn_0}{\ca_0 } \right] - \, \frac{k^2}{k^2 - 2 \, \cf_k^{\prime}(V k^d) \, k^d} \\ 
&& + \frac{k^2}{k^2 - 2 \, \cf_k^{\prime}(V k^d) \, k^d - \frac{2d}{d-2} \, \cf_k^{\prime \prime}(V k^d) \, k^{2d} \, V}
\eea
Here the subscripts $T, S$, and $V$ on the traces indicate sums over traceless tensor, scalar and vector harmonics, respectively. Furthermore, we have introduced the following notations:
\bea\label{2.21}
\nonumber \cn_0 &=& \frac{1}{2} \, \partial_t (k^2 \, R^{(0)}(-D^2/k^2)) \\[1.5ex]
\nonumber \ca &=& -D^2 + k^2 \, R^{(0)}(-D^2/k^2) - 2 \, \cf_k^{\prime}(V k^d) \, k^d \\[1.5ex]
\ca_0 &=& -D^2 + k^2 \, R^{(0)}(-D^2/k^2) 
\eea
In writing down eq. \rf{2.20} we have completed the trace over the $\widehat{\phi}$-modes by adding and subtracting the corresponding zero-mode contribution, so that the scalar trace now runs over the complete set of scalar harmonics, including $l=0$.

The traces in \rf{2.20} can be evaluated using the first term of the standard heat-kernel expansion,
\be\label{2.22}
\Tr\left[e^{-isD^2}\right] = \left(\frac{i}{4\pi s}\right)^{d/2} \tr(I) \int d^dx\, \sqrt{g},
\ee
which is exact for a flat metric. Here $I$ denotes the unit matrix of the space of fields on which $-D^2$ acts. Hence we have
\be\label{2.23}
\tr_S[I] = 1, \qquad \tr_V[I] = d, \qquad \tr_T[I] = \frac{1}{2} (d-1)(d+2)
\ee
For an arbitrary function $W$ with Fourier transform $\widetilde{W}$ the trace
\be\label{2.24}
\nonumber \Tr\left[ W(-D^2) \right] = \int^{\infty}_{-\infty} \!\! ds \, \widetilde{W}(s) \, \Tr \left[ e^{-isD^2} \right] 
\ee
yields, in flat space,
\be\label{2.25}
\Tr[W(-D^2)] = (4\pi)^{-d/2} \, \tr(I) \,  Q_{d/2}[W] \, \int d^dx \, \sqrt{g}
\ee
with
\be\label{2.26}
Q_n[W] \equiv \int_{-\infty}^{\infty} ds\, (-is)^n \, \widetilde W(s)
\ee
Reexpressing $Q_n$ in terms of $W$ leads to a Mellin transform $(n \ge 0)$:
\be\label{2.27}
Q_0[W] = W(0), \qquad Q_n[W] = \frac{1}{\Gamma(n)} \int_0^\infty dz\, z^{n-1} W(z)
\ee

In order to write down the evolution equation for $\cf_k$ it is convenient to introduce the following dimensionless standard threshold functions ($p = 1,2, \cdots, n>0$):
\be\label{2.28}
\Phi^p_n(w) \equiv \frac{1}{\Gamma(n)} \int^{\infty}_{0} \!\! dz z^{n-1} \frac{ R^{(0)}(z) \, - \, z\, R^{(0)\prime}(z)}{\left[ z +  R^{(0)}(z) + w\right]^p} 
\ee
By substituting the definitions of $\cn, \cn_0$ and $\ca_0$ from \rf{2.21} into \rf{2.27} we find the following relation between the Mellin transforms and the threshold functions:
\bea\label{2.29}
\nonumber Q_{d/2} \left[ \frac{\cn_0}{\ca^p} \right] &=& k^{d-2p+2} \, \p{p}{d/2}{-2 \, \cf_k^{\prime}(V k^d) \, k^{d-2}}{} \\[1.5ex]
Q_{d/2} \left[ \frac{\cn_0}{\ca^p_0} \right] &=& k^{d-2p+2} \, \p{p}{d/2}{0}{} 
\eea
Evaluating the traces in \rf{2.20} using \rf{2.25} and then reexpressing the Mellin transforms in terms of the threshold functions \rf{2.28}, the RHS of eq. \rf{2.4} yields:
\bea\label{2.30}
\nonumber \cs_{\rm R} &=& (4 \pi)^{-d/2} \, k^d \, V \, \left\{ \frac{d\,(d+1)}{2} \p{1}{d/2}{-2 \, \cf_k^{\prime}(V k^d) \, k^{d-2}}{} - 2 \, d \, \p{1}{d/2}{0}{} \right\} \hspace{12mm} \\
&& \hspace{-2mm} - \, \frac{k^2}{k^2 - 2 \, \cf_k^{\prime}(V k^d) \, k^d} + \frac{k^2}{k^2 - 2 \, \cf_k^{\prime}(V k^d) \, k^d - \frac{2d}{d-2} \, \cf_k^{\prime \prime}(V k^d) \, k^{2d} \, V}
\eea
Since both the LHS and RHS of eq. \rf{2.4} are already projected onto the subspace of action functionals under consideration, the evolution equation of $\cf_k$ is simply found by equating \rf{2.30} to $\cs_L$ of \rf{2.10}:
\bea\label{2.31}
\nonumber \frac{1}{8 \pi G} \, \frac{d}{dt} \, \cf_k(V k^d) &=& \\
\nonumber && \hspace{-35mm} = (4 \pi)^{-d/2} \, k^d \, V \, \left\{ \frac{d \, (d+1)}{2} \, \p{1}{d/2}{-2 \, \cf_k^{\prime}(V k^d) \, k^{d-2}}{} - 2 \, d \, \p{1}{d/2}{0}{} \right\} \hspace{5mm} \\
&& \hspace{-35mm} \; \; - \, \frac{k^2}{k^2 - 2 \, \cf_k^{\prime}(V k^d) \, k^d} + \frac{k^2}{k^2 - 2 \, \cf_k^{\prime}(V k^d) \, k^d - \frac{2d}{d-2} \, \cf_k^{\prime \prime}(V k^d) \, k^{2d} \, V}
\eea

Introducing the dimensionless function
\be\label{2.31a}
f_k(\vartheta) \equiv k^{d-2} \cf_k(V k^d)
\ee
depending on the dimensionless argument $\vartheta \equiv V k^d$, eq. \rf{2.31} takes on the following final form:
\bea\label{2.31b}
\nonumber (\partial_t \, f_k)(\vartheta) &=& (d-2) \, f_k(\vartheta) - d \, \vartheta \, f_k^{\prime}(\vartheta) \\
\nonumber && + \frac{2}{(4 \pi)^{d/2-1}} \, G \, \vartheta \, k^{d-2} \, \left[ \frac{d \,(d+1)}{2} \, \p{1}{d/2}{-2 \, f_k^{\prime}(\vartheta)}{} - 2 \, d \, \p{1}{d/2}{0}{} \right] \\
&& + 8 \pi \, G \, k^{d-2} \, \left[ - \, \frac{1}{1 - 2 \, f_k^{\prime}(\vartheta)} + \frac{1}{1 - 2 \, f_k^{\prime}(\vartheta) - \frac{2d}{d-2} \, f_k^{\prime \prime}(\vartheta) \, \vartheta } \right]
\eea
This is the partial differential equation we wanted to derive. 

We observe that the flow equation \rf{2.31b} does not generate any nonlinear terms in $f_k$ unless we start the evolution with a $f_k$ which is nonlinear at the initial point already. In fact, if $f_{\hk}(\vartheta) = \hk^{-2} \, \lambda(\hk) \, \vartheta$ without any nonlocality, then the RHS of \rf{2.31b} is linear in $\vartheta$, too, so that $f_k(\vartheta) \propto \vartheta$ for all $k < \hk$.

In its general form eq. \rf{2.31b} will be analyzed in Section V. Here we use it in order to derive the flow equations for the coupling constants of the $V$+$V \ln V$-- and $V$+$V^2$--truncations, respectively.

In the case of the $V$+$V \ln V$--truncation, $\cf_k$ has been given in eq. \rf{2.6}. By substituting this ansatz into \rf{2.31} and projecting the resulting RHS onto the invariants $V$ and $V \ln(V/V_0)$ we obtain the following flow equation for the cosmological constant $\lb_k$ and the coefficient of the $V \ln(V/V_0)$-term, $\v_k$:
\bea\label{2.32}
\nonumber \partial_t \lb_k &=& (4\pi)^{1-d/2} \, G \, k^d \, \left\{ d\,(d+1) \p{1}{d/2}{-2 \lb_k/k^2 - \v_k /k^2 }{} \, - \, 4d \, \p{1}{d/2}{0}{} \right\} \\[1.5ex]
\partial_t \v_k &=& (4\pi)^{1-d/2} \, 2d(d+1) \,G \, \v_k \, k^{d-2} \, \p{2}{d/2}{-2 \lb_k/k^2 - \v_k /k^2 }{} 
\eea
In the expansion of the RHS we made use of the fact that $\ln(V/V_0)$ has no power series expansion about  $V=0$, so that we can expand the RHS treating $V$ and $\ln(V/V_0)$ as independent variables.

The flow equation for the coupling constants in the $V$+$V^2$--truncation are found in an analogous way. Substituting the $\cf_k$ given in eq. \rf{2.7} into \rf{2.31} and projecting the RHS onto the subspace spanned by the invariants $V$ and $V^2$ we find:
\bea\label{2.33}
\nonumber \partial_t \, \lb_k &=& (4 \pi)^{1-d/2} \, G \, k^d \left\{ d \, (d+1) \, \p{1}{d/2}{-2 \lb_k / k^2}{} - 4 \, d \, \p{1}{d/2}{0}{} \right\} \\
\nonumber && + 16 \pi \, G \, \frac{d}{d-2} \, \frac{\w_k}{k^2} \, \frac{1}{(1 - 2 \lb_k/k^2)^2} \\[1.5ex]
\nonumber \partial_t \w_k &=& (4 \pi)^{1-d/2} \, 4d(d+1) \,  G \,  \w_k  \, k^{d-2} \,\p{2}{d/2}{-2 \lb_k / k^2}{} \\
&& + 128 \pi \, G \, \frac{d\,(3d-4)}{(d-2)^2} \, \frac{\w_k^{~2}}{k^4} \, \frac{1}{(1 - 2 \lb_k / k^2)^3}
\eea

The properties of the RG flow \rf{2.32} are discussed in the main part of this paper. The analogous discussion of the flow \rf{2.33} is summarized in the Appendix.
\mysection{Renormalization Group Flow in the $V$+$V \ln V$--Truncation}
We now analyze the $V$+$V \ln V$--truncation governed by eq. \rf{2.32}. We first investigate its fixed point structure before we proceed to discussing the properties of the numerical solutions of the flow equation.

Since we set $G_k \equiv G = \mbox{const}$, the flow equation \rf{2.32} is expected to be valid only on scales $k \lesssim m_{\rm Pl}$ where, at least according to the Einstein-Hilbert truncation, the running of $G_k$ is negligible \cite{oliver}. We do not expect that the solutions of \rf{2.32} can be continued up to arbitrarily high values of $k \gg m_{\rm Pl}$, since for a proper description of the RG flow in this region the running of $G_k$ is an essential effect. In the following we therefore only investigate the RG flow in the region $k \lesssim m_{\rm Pl}$.

\begin{subsection}{The Gaussian Fixed Point} 
In order to investigate the fixed point structure of \rf{2.32} we introduce the following ``$k$-scaled'' dimensionless coupling constants
\be\label{3.1}
g(k) \equiv G \, k^{d-2}, \qquad \lambda(k) \equiv \lb_k \, k^{-2}, \qquad u(k) \equiv \v_k \, k^{-2} 
\ee
They allow us to write the flow equation in a $k$-independent, autonomous way: 
\be\label{3.2}
\partial_t \, g(k) = \Fbeta_g(\lambda, g, u), \qquad \partial_t \, \lambda(k) = \Fbeta_\lambda(\lambda, g, u), \qquad \partial_t \, u(k) = \Fbeta_u(\lambda, g, u)
\ee
The $\Fbeta$-functions resulting from substituting \rf{3.1} into \rf{2.32} are
\bea\label{3.3}
\nonumber \Fbeta_g(\lambda, g, u) &=& (d-2) \, g \\[1.5ex]
\nonumber \Fbeta_\lambda(\lambda, g, u) &=& -2 \lambda + 
(4 \pi)^{1-d/2}\,g\, \left\{ d \, (d+1) \, \p{1}{d/2}{-2 \lambda - u}{} - 4 \, d \, \p{1}{d/2}{0}{} \right\} \\[1.5ex]
\Fbeta_u(\lambda, g, u) &=& -2 u + (4 \pi)^{1-d/2} \, 2 \, d \, (d+1) \, g \, u \, \p{2}{d/2}{-2 \lambda - u}{} 
\eea
The fixed point equation
\be\label{3.4}
\Fbeta_\lambda(\lambda^*, g^*, u^*) = 0, \quad \Fbeta_g(\lambda^*, g^*, u^*) = 0, \quad \Fbeta_u(\lambda^*, g^*, u^*) = 0
\ee
has only the trivial solution
\be\label{3.5}
g^* = 0, \qquad \lambda^* = 0, \qquad u^* = 0
\ee
which corresponds to the Gaussian fixed point. The corresponding stability matrix ${\bf B}_{ij} = \partial_j \Fbeta_i|_{g=0, \lambda = 0, u = 0}$ is given by: 
\be\label{3.6}
{\bf B}_{\rm GFP} = \left[ 
\begin{array}{ccc}  \hspace{5mm} -2 \hspace{5mm} & (4 \pi)^{1-d/2} \, d(d-3) \,  \p{1}{d/2}{0}{} & \hspace{5mm} 0 \hspace{5mm} \\ 
0 & (d-2) & 0 \\ 
0 & 0 & -2 \end{array} \right]
\ee
Here $i,j \in \{\lambda, g, u \}$. By diagonalizing \rf{3.6} we find the following stability coefficients $\theta_I$ and associated right eigenvectors $V^I$ satisfying $ {\bf B} V^I = - \theta_I V^I$: 
\bea\label{3.7}
\nonumber \theta_1 = +2\qquad &\mbox{with}& \qquad V^1 = \left( 1, \quad 0, \quad 0 \right)^{\sf T} \\[1.5ex]
\nonumber \theta_2 = -(d-2) \qquad &\mbox{with}& \qquad V^2 = \left( (4 \pi)^{1-d/2} \, (d-3) \, \p{1}{d/2}{0}{} , \quad 1 , \quad 0 \right)^{\sf T} \\[1.5ex]
\theta_3 = +2 \qquad &\mbox{with}& \qquad V^3 = \left( 0 , \quad 0 , \quad 1  \right)^{\sf T} 
\eea

We can now use these results to write down the linearized RG flow ${\rm g}_j(t) = {\rm g}^*_j + \sum_{I = 1}^3 \alpha_I \exp \left[ - \theta_I t \right] V_j^I$ for the couplings ${\rm g}_1 \equiv \lambda, {\rm g}_2 \equiv g, {\rm g}_3 \equiv u$ in the vicinity of the GFP:
\bea\label{3.8}
\nonumber \lambda(t) &=& \alpha_1 \, e^{-2 t} + \alpha_2 \, e^{(d-2) t} \, (4 \pi)^{1-d/2} \, (d-3) \, \p{1}{d/2}{0}{} \\
\nonumber g(t) &=& \alpha_2 \, e^{(d-2) t} \\
u(t) &=& \alpha_3 \, e^{-2 t}
\eea
Here the $\alpha_I$'s are constants which allow for adjusting the general solution of the linearized flow equation to given initial conditions. Replacing the RG time $t \equiv \ln(k / m_{\rm Pl})$ by the scale $k$ and rewriting \rf{3.8} in terms of the dimensionful couplings $G_k, \lb_k$ and $\v_k$ we find:
\bea\label{3.9}
\nonumber G_k &=& G \\
\nonumber \lb_k &=& \lb_0 + (4 \pi)^{1-d/2} \, (d-3) \, G \, k^d \, \p{1}{d/2}{0}{}\\
\v_k &=& \v_0 
\eea
Here we expressed the constants $\alpha_I$ by the $k=0$-values of the corresponding coupling constants: $\alpha_1 \equiv \lb_0, \alpha_2 \equiv G$ and $\alpha_3 \equiv \v_0$.

Equation \rf{3.9} shows that including the new invariant $V \ln(V/V_0)$ in the truncation does not change the linearized RG flow of the cosmological constant in the vicinity of the Gaussian fixed point. In the limit $k \rightarrow 0$ all coupling constants run towards constant, but in general nonzero values $G, \lb_0$ and $\v_0$. Hence the modified GFP certainly does not provide a solution to the cosmological constant problem.
\end{subsection}
\begin{subsection}{Numerical solution of the flow equation}
We now investigate the numerical solutions of the flow equation \rf{2.32} in $d=4$ dimensions using the sharp cutoff introduced in ref. \cite{myself}. For this special choice of the shape function $R^{(0)}$ the integrals appearing in the threshold functions \rf{2.28} can be evaluated analytically \cite{opt}. They read \cite{myself}:
\bea\label{3.9a}
\nonumber \p{1}{d/2}{w}{sc} &=& - \, \frac{1}{\Gamma(d/2)} \, \ln(1+w) + \varphi_{d/2} \\
\p{p}{d/2}{w}{sc} &=& \frac{1}{\Gamma(d/2)} \, \frac{1}{p-1} \, \frac{1}{(1+w)^{p-1}} \quad {\rm for} \quad p>1 
\eea
Here the $\varphi_{d/2}$'s are a priori arbitrary positive constants which reflect the residual cutoff scheme dependence which is still present after having opted for a sharp cutoff. In the following we will make the ``canonical'' choice $ \varphi_{2} \equiv 2 \, \zeta(3)$ where $\zeta$ is the Riemann $\zeta$-function. (See \cite{myself} for a detailed discussion.)

Introducing the ``$\mp$--scaled'' dimensionless coupling constants
\be\label{3.9b}
\lh(\kh) = \frac{\lb_k}{m_{\rm Pl}^2}, \quad \vh(\kh) = \frac{\v_k}{m_{\rm Pl}^2}
\ee
and the dimensionless scale variable
\be\label{3.9c}
 y \equiv \kh^2 \equiv \frac{k^2}{m_{\rm Pl}^2} 
\ee
the flow equation \rf{2.32} with the sharp cutoff becomes in $d=4$ : 
\bea\label{3.10}
\nonumber \frac{d \, \lh(y)}{d \, y} &=& \, \frac{y}{2 \pi} \, \left\{ - 5 \, \ln\left[1 - 2 \lh(y)/y - \vh(y) / y \right] + \varphi_2 \right\} \\[1.5ex]
\frac{d \, \vh(y)}{d \, y} &=& \, \frac{5}{\pi} \, y \, \frac{\vh(y)}{y-2\lh(y)-\vh(y)}
\eea

The interesting property of this equation is that the $V \ln(V/V_0)$-term in the truncation leads to a modification of the boundary $\lh = y/2$ (corresponding to $\lambda = 1/2$) which was found in the Einstein-Hilbert truncation. The new boundary is located at
\be\label{3.11}
y = 2 \lh + \vh
\ee
We find that for coupling constants in the region $y \le 2 \lh + \vh$ the RHS of eq.  \rf{3.10} is not defined, so that there are no RG trajectories in this region.

Compared to the Einstein-Hilbert truncation, the most important new property of eq. \rf{3.10} is that it is now possible to obtain positive IR values $\lh(0) > 0$. They can be compensated by negative values $\vh(0) < 0$ which prevent the trajectory from running into the boundary \rf{3.11}.

Moreover, \rf{3.10} shows that the $\Fbeta$-function of $\vh$ vanishes at $\vh = 0$ for any $\lh$, i.e. a trajectory starting at some initial point $y = \yh$ with $\vh(\yh) = 0$ does not dynamically generate a nonzero coupling $\vh$ by the RG flow:
\be\label{3.12}
\vh(\yh) = 0 \quad \Rightarrow \quad \vh(y) = 0 \quad \forall \, y < \yh
\ee 
Because of the vanishing $\Fbeta$-function, the trajectories cannot cross the $\vh = 0$--line. This line separates trajectories with $\vh(y) > 0$ and $\vh(y) < 0$.

Unless stated otherwise, we shall specify initial conditions for $\vh$ and $\lh$ at the Planck scale: $\hk = m_{\rm Pl}, \kh = 1, \yh \equiv \hk^2/m_{\rm Pl}^2 = 1$. 

In order to disentangle the various effects which contribute to the running of $\vh$ and $\lh$ we first investigate the decoupled flow equation for $\vh$ with $\lh$ set to a constant value. In the second step we then drop this approximation. 
\begin{subsubsection}{The decoupled flow equation}
Approximating
\be\label{3.13}
\lh(y) \approx \mbox{const} \equiv \lh 
\ee
in the flow equation \rf{3.10} leads to the following decoupled differential equation for $\vh(y)$:
\be\label{3.14}
\frac{d \, \vh(y)}{d \, y} = \, \frac{5}{\pi} \, y \, \frac{\vh(y)}{y-2\lh-\vh(y)}
\ee
The characteristic properties of this equation are shown in Fig. \ref{3.eins} where we have solved \rf{3.14} for the parameters $\lh = -0.1$ and $\lh = + 0.1$ and various initial values $\vh(\yh)$.
\begin{figure}[t]
\renewcommand{\baselinestretch}{1}
\epsfxsize=0.49\textwidth
\begin{center}
\leavevmode
\epsffile{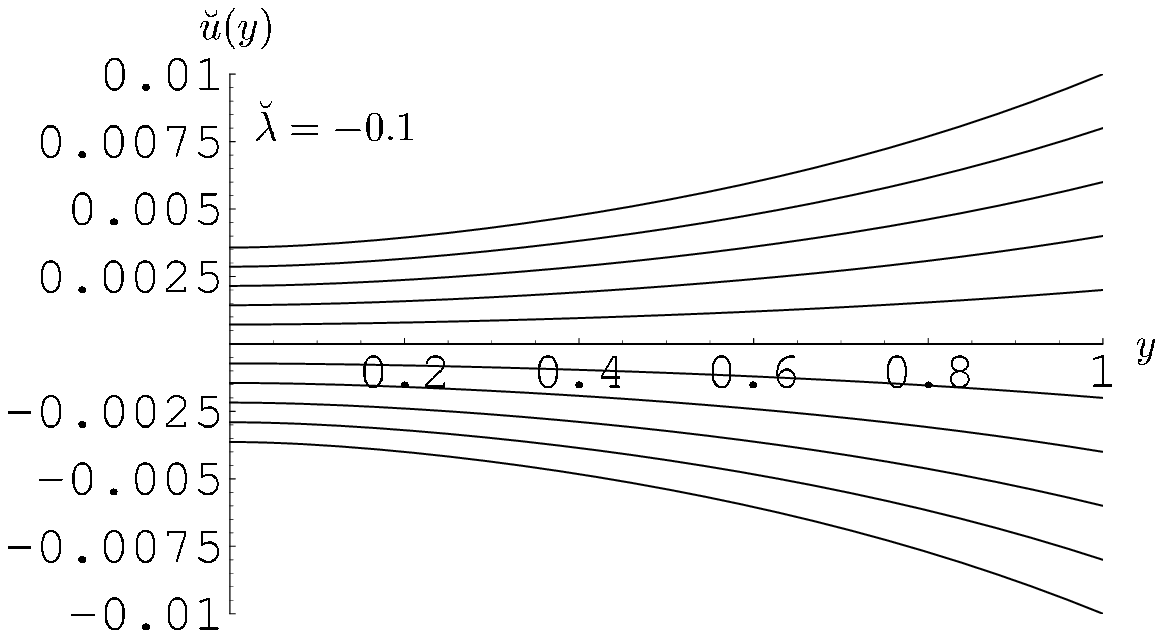}
\epsfxsize=0.48\textwidth
\epsffile{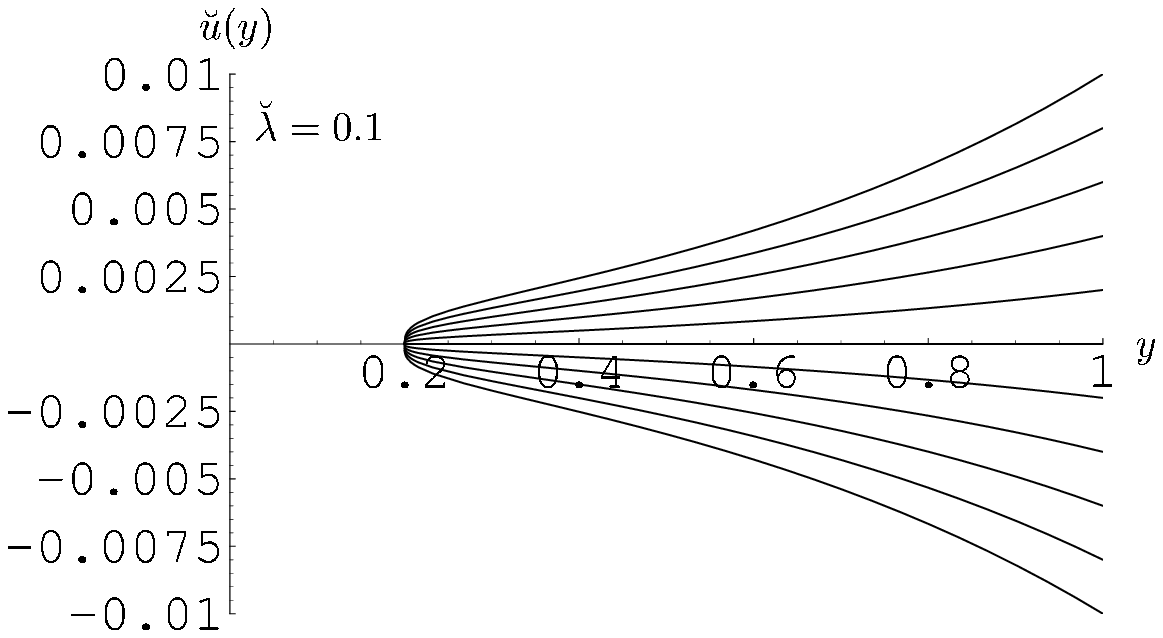}
\end{center}
\parbox[c]{\textwidth}{\caption{\label{3.eins}{\footnotesize Solutions of the flow equation \rf{3.14} for various values $\vh(\yh)$ and $\lh = -0.1$, $\lh = +0.1$ in the left and right diagram, respectively. The solutions with $\lh = -0.1$ all lead to a finite IR value $|\vh(0)| > 0$. For $\lh = +0.1$ all trajectories terminate at $\yh_{\rm term} > 0$ with a vanishing  $\vh(\yh_{\rm term})$.}}}
\end{figure}

In the case of $\lh = -0.1$ we find that all trajectories can be continued to $y = 0$ and lead to nonvanishing IR values $|\vh(0)|>0$. The values of $\vh(y)$ along the trajectory are bounded by the initial values:
\be\label{3.15}
|\vh(y)| < |\vh(\yh)|, \qquad \forall \, y < \yh
\ee

In the case of $\lh = +0.1$ all trajectories terminate at a finite value $y_{\rm term}>0$ because the trajectories hit the boundary \rf{3.11}. In this course the coupling $\vh(y)$ vanishes identically: $\vh(\y_{\rm term}) = 0$. Substituting $\vh(y_{\rm term}) = 0$ into eq. \rf{3.11} we see that $y_{\rm term} = 2 \lh$.

In order to understand the vanishing of $\vh(y_{\rm term})$ analytically we observe that the trajectories in the right diagram of Fig. \ref{3.eins} terminate due to the denominator of \rf{3.14} becoming zero, i.e. the corresponding trajectory runs into the boundary \rf{3.11}. Close to $y_{\rm term}$ we can approximate in the denominator of eq. \rf{3.14}
\be\label{3.16}
2 \lh + \vh(y \approx y_{\rm term}) \approx y_{\rm term} = \mbox{const}
\ee  
Using this approximation, the flow equation \rf{3.14} simplifies to
\be\label{3.17}
\frac{d \, \vh(y)}{d \, y} \approx \, \frac{5}{\pi} \, \frac{y \, \vh(y)}{y - y_{\rm term} },
\ee
It is easily integrated:
\be\label{3.18}
\vh(y) = \vh(y_0) \, e^{5 y/\pi} \, (y - y_{\rm term} )^{5 y_{\rm term}/\pi} \quad \mbox{with} \quad y \gtrsim y_{\rm term}, \quad y_{\rm term} > 0 
\ee
Taking the limit $y \rightarrow y_{\rm term}$ shows that $\vh(y_{\rm term})$ vanishes, independently of the sign of $\vh(y_0)$. This exactly matches the behavior found in Fig. \ref{3.eins}. 
\end{subsubsection}
\begin{subsubsection}{The full flow equation}
We now drop the approximation \rf{3.13} and investigate the coupled system of equations for $\vh(y)$ and $\lh(y)$. We solve the flow equation \rf{3.10} for the initial conditions $\lh(\yh) = 0.1$ and various positive and negative values of $\vh(\yh)$. The resulting trajectories are shown in Fig. \ref{3.zwei}.
\begin{figure}[t]
\renewcommand{\baselinestretch}{1}
\epsfxsize=0.49\textwidth
\begin{center}
\leavevmode
\epsffile{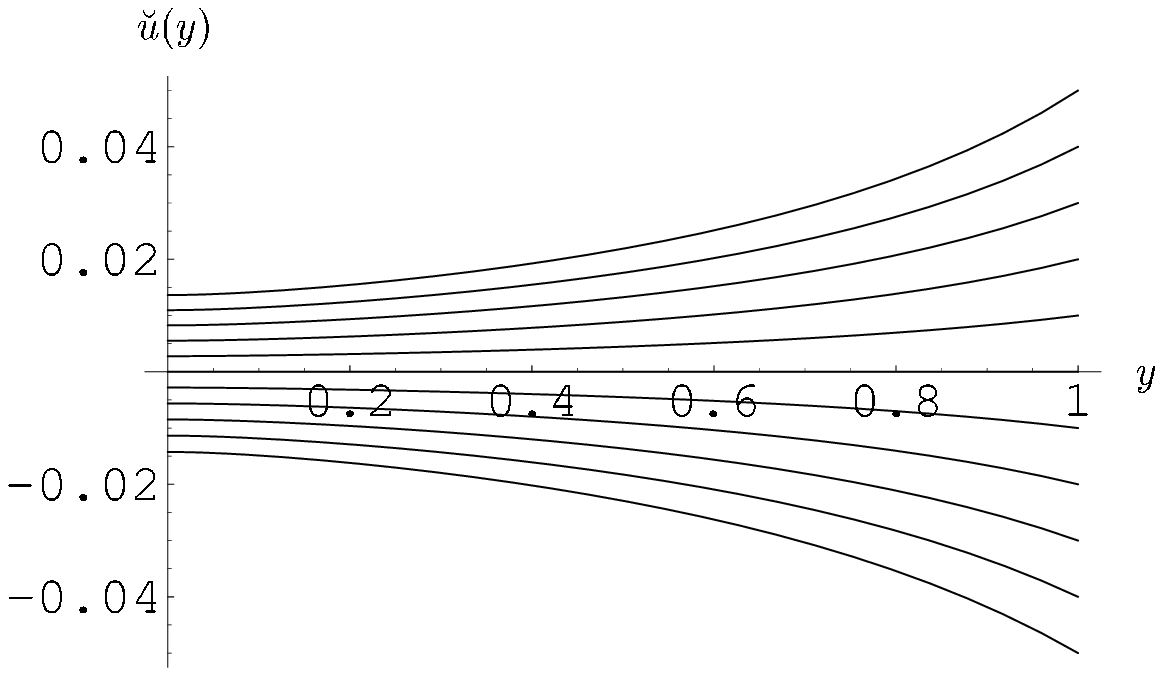}
\epsfxsize=0.48\textwidth
\epsffile{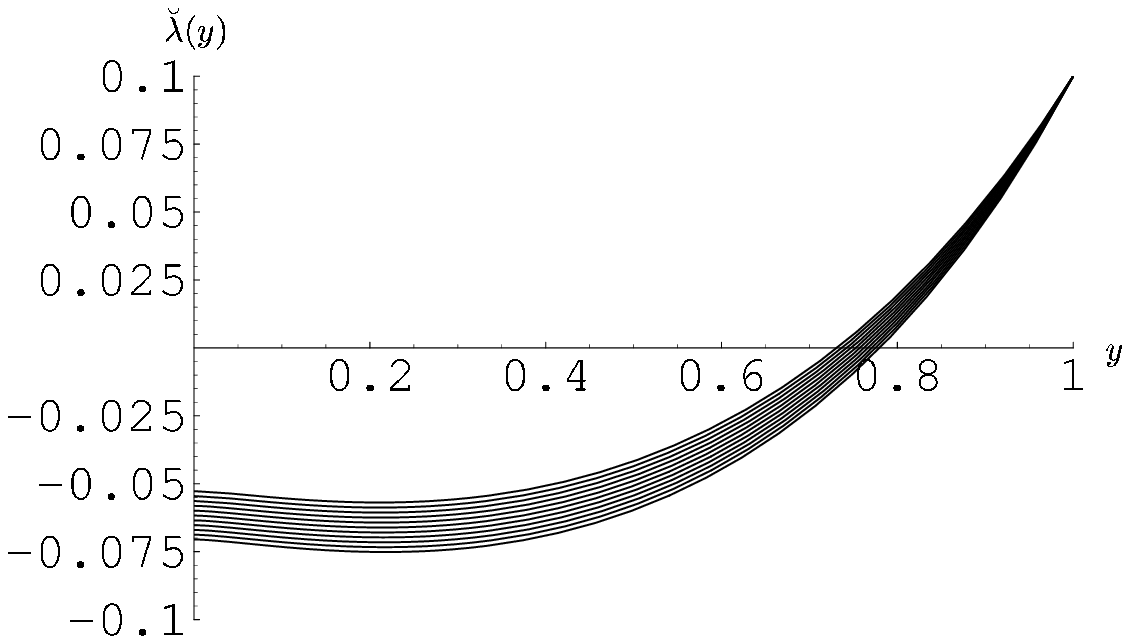}
\end{center}
\parbox[c]{\textwidth}{\caption{\label{3.zwei}{\footnotesize Numerical solutions to the full flow equation \rf{3.10} with initial conditions $\lh(\yh) = 0.1$ and selected positive and negative values $\vh(\yh)$. The initial conditions all lead to trajectories of Type Ia, yielding negative values for $\lh(0)$. There is no qualitative difference between the trajectories starting with $\vh(\yh) > 0$ and $\vh(\yh) < 0$. Compared to the trajectory starting with $\vh(\yh)=0$ the solutions with $\vh(\yh) > 0$ and $\vh(\yh) < 0$ lead to decreased and increased values of $\lh(0)$, respectively.}}}
\end{figure}

Here we see that there is no qualitative difference between the trajectories starting with positive and negative values of $\vh(\yh)$. All trajectories can be continued to $y=0$ and yield a negative value for the renormalized cosmological constant, $\lh(0) < 0$. This is characteristic of trajectories of Type Ia.

Regarding the dependence of $\lh(0)$ on the initial value $\vh(\yh)$ we find that, compared to the trajectory with $\vh(\yh) = 0$, negative values $\vh(\yh) < 0$ lead to less negative cosmological constants, $\lh(0)_{\vh(\yh)<0} > \lh(0)_{\vh(\yh)=0}$, while positive values $\vh(\yh) > 0$ drive $\lh(0)$ further away from zero: $\lh(0)_{\vh(\yh)>0} < \lh(0)_{\vh(\yh)=0}$.

A consequence of the general trend that negative values $\vh(\yh) < 0$ shift $\lh(0)$ upwards is the existence of trajectories with {\it positive} renormalized cosmological constant: $\lh(0) > 0$. Trajectories of this type did not exist in the Einstein-Hilbert truncation. This new mechanism is illustrated in Fig. \ref{3.drei}.
\begin{figure}[t]
\renewcommand{\baselinestretch}{1}
\epsfxsize=0.49\textwidth
\begin{center}
\leavevmode
\epsffile{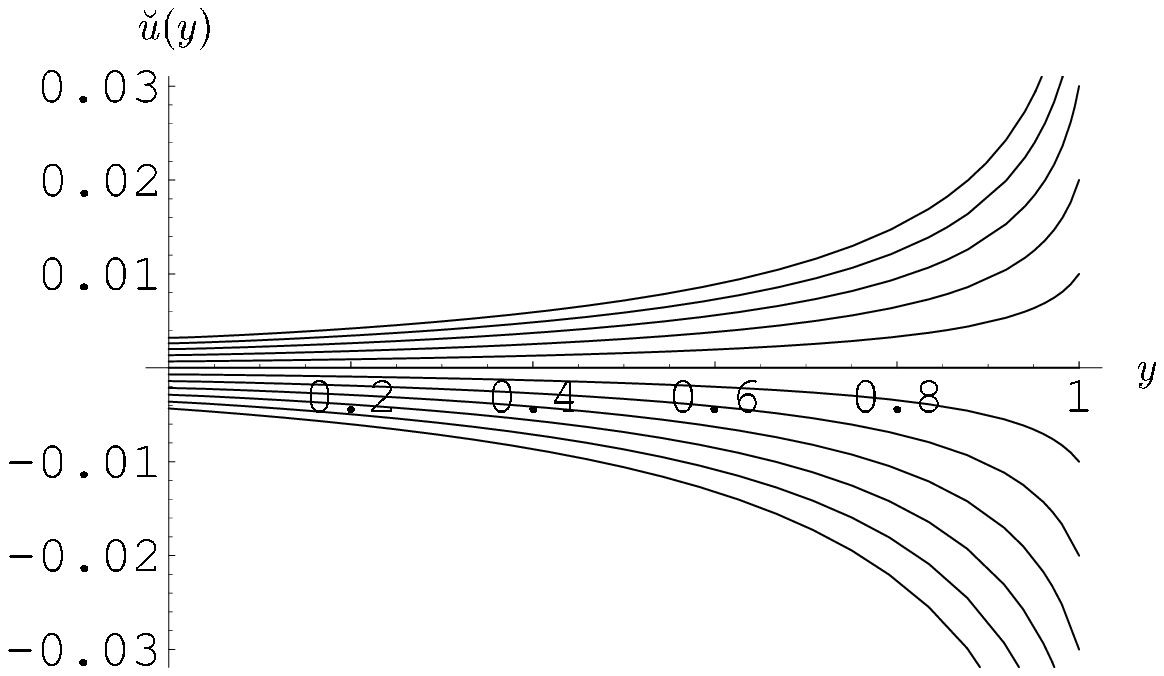}
\epsfxsize=0.48\textwidth
\epsffile{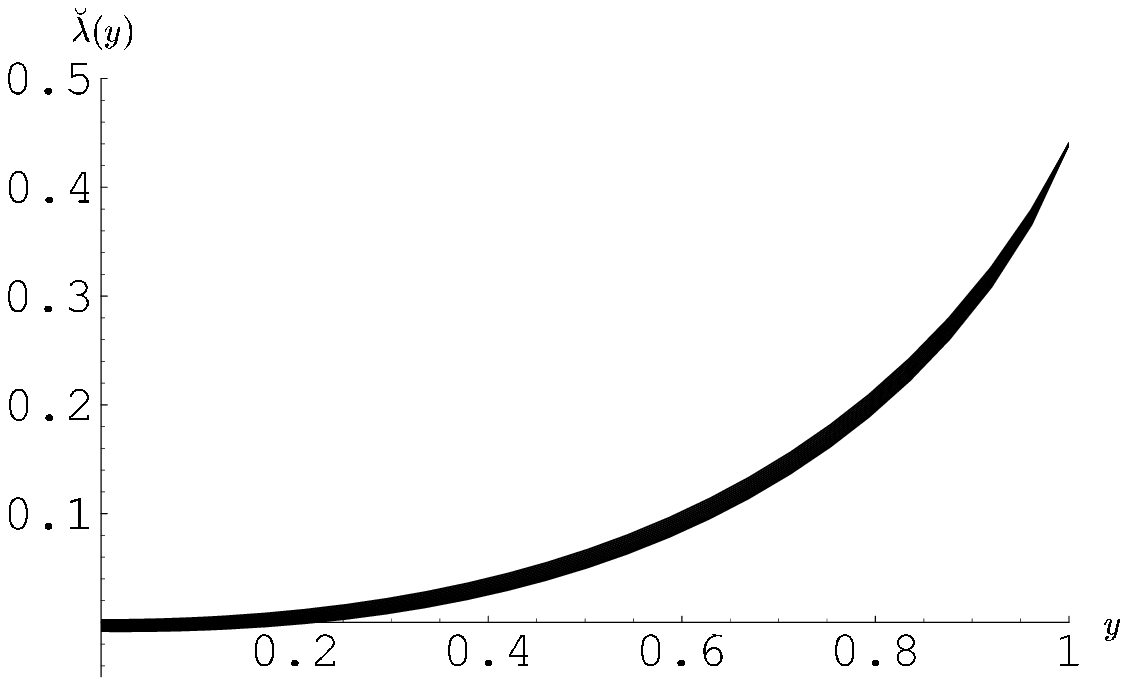}
\end{center}
\vspace{2mm}
\epsfxsize=0.49\textwidth
\begin{center}
\leavevmode
\epsffile{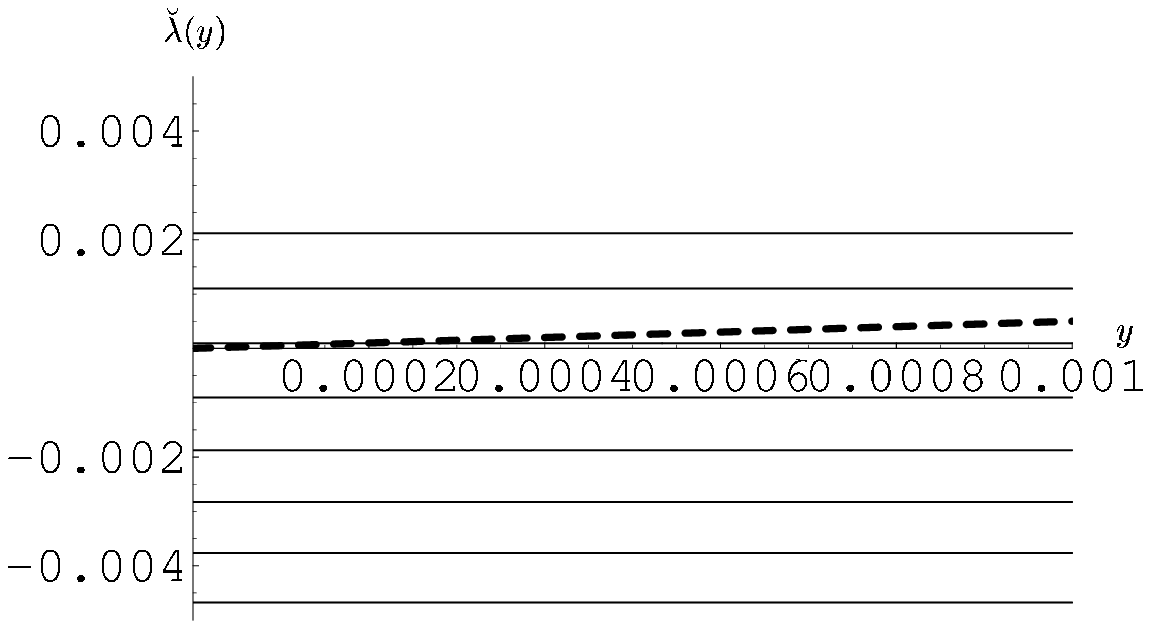}
\end{center}
\parbox[c]{\textwidth}{\caption{\label{3.drei}{\footnotesize Numerical solutions to the full flow equation with initial conditions $\lh(\yh) = 0.44$ and $\vh(\yh) = 0.05$. Decreasing $\vh(\yh)$ by steps of $\Delta \vh(\yh) = 0.01$ leads to an increase of $\lh(0)$. Fine-tuning $\vh(\yh) = \vh(\yh)_{\rm crit}$ results in a trajectory with $\lh(0) = 0$. Initial conditions $\vh(\yh) < \vh(\yh)_{\rm crit}$ yield trajectories of Type VIa with $\lh(0) > 0$. The dashed line indicates the former boundary of $\lh$--space at $\lh = y/2$.}}}
\end{figure}

Here we first consider the trajectory starting with $\lh(\yh) = 0.44, \vh(\yh) = 0.05$ which results in a trajectory with a small negative value $\lh(0) \lesssim 0$. We then lower $\vh(\yh)$ in steps of $\Delta \vh(\yh) = 0.01$ which leads to an increase of $\lh(0)$. The result is shown in the third diagram of Fig. \ref{3.drei}.

We find that by fine-tuning the initial value to $\vh(\yh) = \vh(\yh)_{\rm crit}$ a trajectory that was originally Type Ia with $\lh(0) < 0$ can be turned into a trajectory of Type IIa with vanishing $\lh(0)$. By a further decrease of $\vh(\yh)$ to $\vh(\yh) < \vh(\yh)_{\rm crit}$ we find a new type of RG trajectories with a positive value $\lh(0) > 0$. Trajectories of this type will be referred to as solutions of Type VIa. These trajectories provide a significant new feature of the RG flow of the cosmological constant, since without the $V \ln(V/V_0)$-invariant $\lh(0)$ could only assume negative values.

Another important issue is the impact of a nonzero coupling $\vh(\yh)$ on the trajectories of Type IIIa which, without the new coupling, terminate in the boundary $\lh = y /2$. Since there are no initial conditions given at $\yh = 1$ that result in trajectories of this type, we choose $\yh = 0.5, \hk \approx 0.7 m_{\rm Pl}$. The effect of the nonzero $\vh(\yh)$ is illustrated in Fig. \ref{3.vier} for trajectories starting with $\lh(\yh) = 0.2$ and various positive and negative values $\vh(\yh)$.
\begin{figure}[t]
\renewcommand{\baselinestretch}{1}
\epsfxsize=0.49\textwidth
\begin{center}
\leavevmode
\epsffile{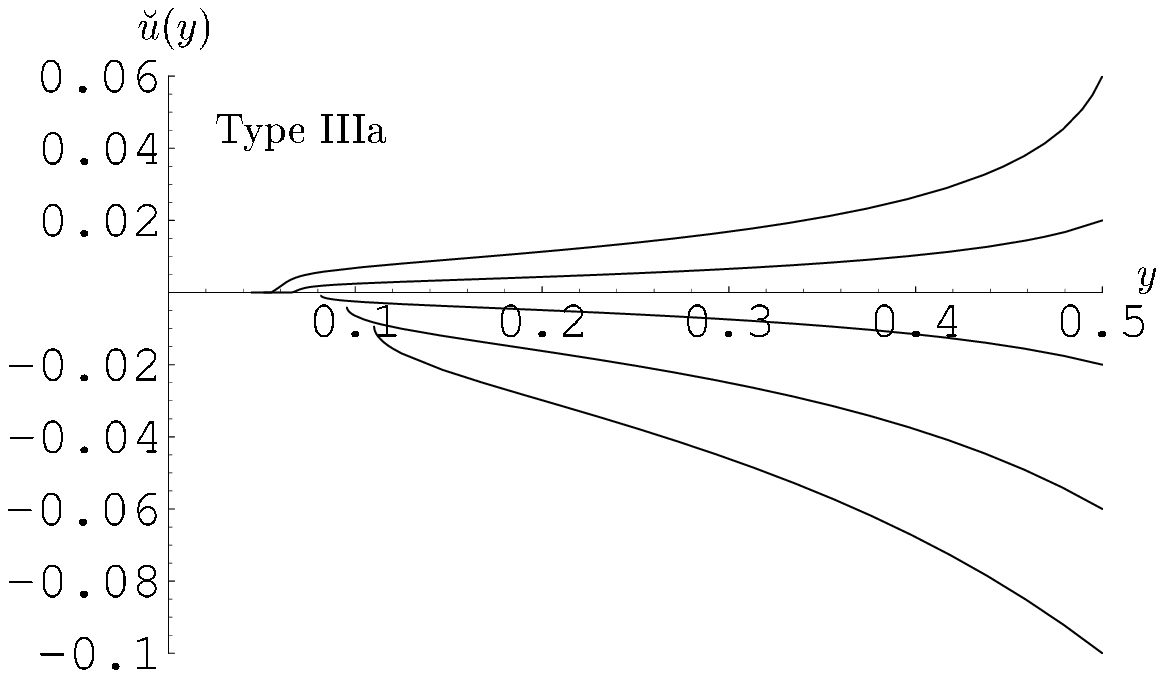}
\epsfxsize=0.48\textwidth
\epsffile{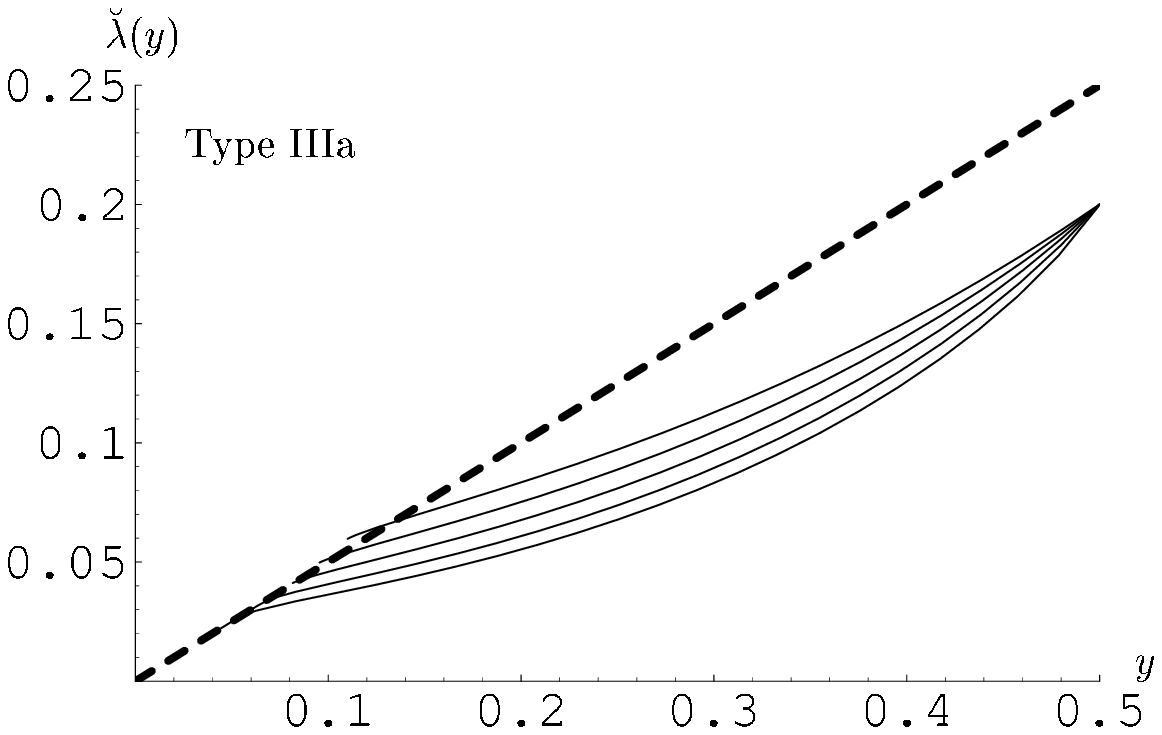}
\end{center}
\begin{center}
\leavevmode
\epsfxsize=0.60\textwidth
\epsffile{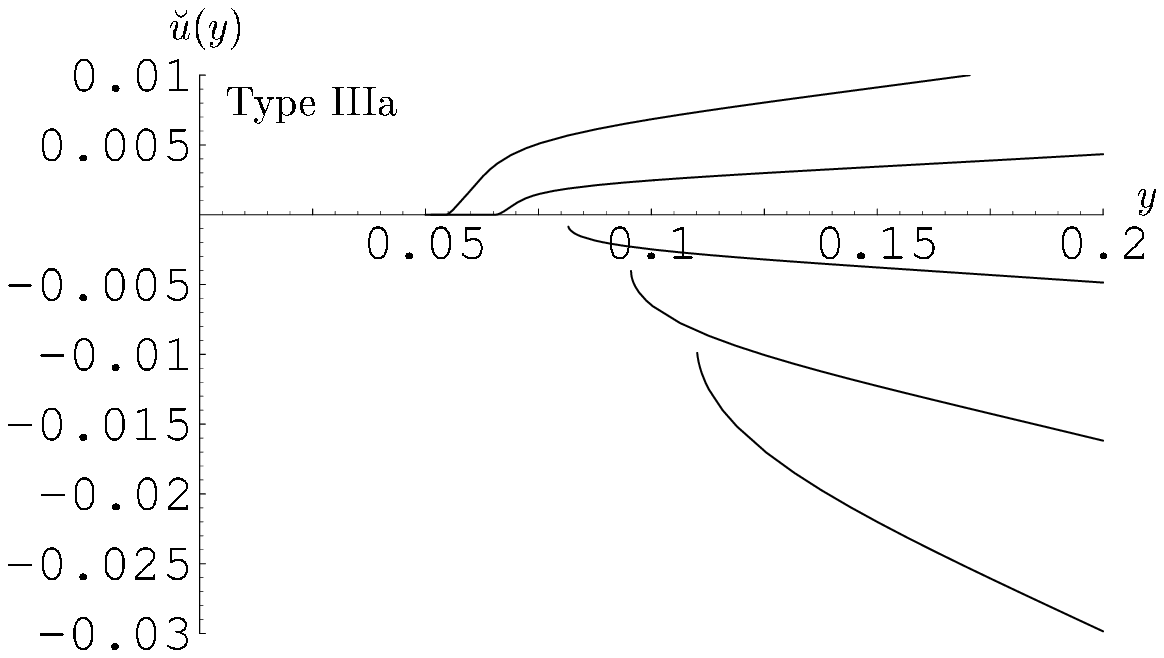}
\end{center}
\parbox[c]{\textwidth}{\caption{\label{3.vier}{\footnotesize Numerical solutions to the full flow equation with initial conditions $\lh(\yh) = 0.2$ and various values $\vh(\yh)$ given at the scale $\yh = 0.5$. These lead to trajectories of Type IIIa which terminate at $y_{\rm term} > 0$. Positive values $\vh(\yh)>0$ yield an extension of the trajectories towards smaller values $y_{\rm term}$, but in general do not prevent the termination of the trajectory. The dashed line indicates the former singularity $\lh = y/2$.}}}
\end{figure}

We find that including the effect of a running $\vh(y)$ in general does not prevent the termination of the trajectories at a finite $\y_{\rm term} > 0$. Comparing to the case of $\vh(\yh) = 0$ we find that {\it negative} values $\vh(\yh)$ lead to an even earlier termination of the trajectory. The third diagram of Fig. \ref{3.vier} further shows that, unlike in the case of the approximated flow equation \rf{3.14}, the trajectories starting with $\vh(\yh) < 0$ do not reach $\vh(y_{\rm term}) = 0$ but terminate at a nonzero value $\vh(y_{\rm term}) < 0$ already. This is due to the structure of the modified boundary $y = 2 \lh + \vh$. Here negative values $\vh$ compensate a cosmological constant that, setting $\vh=0$, would already have run into the boundary. Close to $y = y_{\rm term}$ we find that $\vh(y)$ decreases rapidly so that this compensation becomes impossible. As a consequence, the trajectory reaches the boundary with a finite (negative) value $\vh(y_{\rm term}) < 0$, contrary to the case of a positive $\vh$.  

Considering the trajectories in the region $\vh(\yh) > 0$ we see that {\it positive} values $\vh(y)$ generally lead to a decrease of $\lh(y)$ so that the corresponding trajectories terminate at a smaller value $y_{\rm term}$ than their counterpart starting with $\vh(\yh)=0$. But in general this mechanism cannot be used to prevent the termination of the trajectory since we cannot choose arbitrarily large values for the initial value of $\vh(\yh)$ due to the boundary \rf{3.11}. We therefore find that the possibility of lowering $y_{\rm term}$ by choosing $\vh(\yh) > 0$ is rather limited so that only the trajectories terminating at sufficiently small values $y_{\rm term}$ can be turned into a trajectory of Type IIa or Type Ia by taking $\vh(\yh) > 0$.

Figure \ref{3.vier} further shows that the mechanism which lowers $y_{\rm term}$ is not operative close to the termination point of the trajectory but is rather related to the general effect which $\vh(y)$ has on the running of $\lh(y)$ in the region $y > y_{\rm term}$. Close to $y \approx y_{\rm term}$ we find that $\vh(y)$ vanishes very quickly. Therefore $\vh(y)$ cannot have any healing effect on the flow of $\lh(y)$ near the boundary. 

\begin{table}
\begin{tabular}{@{\extracolsep{\fill}} ccc} 
Type & $\vh(\yh)$ chosen & Changes in the flow of $\lh(y)$ \\ \hline 
&  $\vh(\hat{y}) > \vh(\hat{y})_{\rm crit}$ & Type Ia \\
&                                           & Type IIa  \\[-1.5ex]
\raisebox{1.5ex}[-1.5ex]{\hspace*{2mm} Type Ia} & \raisebox{1.5ex}[-1.5ex]{$\vh(\hat{y}) = \vh(\hat{y})_{\rm crit}$ } & Fine-tuning of $\vh(\hat{y})$ \\
&  $\vh(\hat{y}) < \vh(\hat{y})_{\rm crit}$ & Type VIa \\[-1.5ex] 
&                                           & new solutions with $\lh(0)>0$ \\ \hline
 & $\vh(\hat{y}) > 0$ & Type Ia \\ 
\raisebox{1.5ex}[-1.5ex]{\hspace*{2mm} Type IIa} & $\vh(\hat{y}) < 0$ & Type IIIa  \\ \hline
         &                                              & generic: Type  IIIa \\[-1.5ex]  
&  $\vh(\hat{y}) > 0$ & (only solutions close to the region IIa \\[-1.5ex] 
\raisebox{1.5ex}[-1.5ex]{\hspace*{2mm} Type IIIa} & & can be converted to a Type IIa or Ia trajectory.) \\
& $\vh(\hat{y}) < 0$ & Type IIIa \\ 
\end{tabular}
\caption{\label{3.one} {\footnotesize Summary of the modifications in the RG flow of $\lh(y)$ arising from the inclusion of the $V \ln(V/V_0)$--term in the truncation.}}
\end{table}
The impact of a nonzero coupling $\vh(\yh)$ on the running of $\lh(y)$ is summarized in Table \ref{3.one} which is organized as follows. The column ``Type'' indicates the type of trajectory found when solving the flow equation \rf{3.10} with $\vh(\yh) = 0$. The column ``$\vh(\yh)$ chosen'' indicates which values of $\vh(\yh)$ lead to the modifications in the flow of $\lh(y)$ listed in the column ``Changes in the flow of $\lh(y)$''.
\end{subsubsection}
\begin{subsubsection}{Scaling laws in the IR region}
In Sec. III.A we found that the autonomous fixed point equation \rf{3.4} gives rise to a Gaussian fixed point at the origin of $u$-$\lambda$-$g$--space which should govern the scaling behavior of these coupling constants in the IR-region. In this subsection we therefore investigate if the solutions of \rf{3.10} reflect the expected scaling laws for $\vh$ and $\lh$. In this course we use the flow equation \rf{3.10} to find one representative trajectory for every Type discussed in Section III.B. In order to focus on the IR properties ($k \rightarrow 0)$ we display the solutions in double-logarithmic plots.

We start our investigation with the trajectories of Type Ia which are characterized by a negative IR value of the cosmological constant: $\lh(0) < 0$. As typical trajectories we choose the solutions with $\lh(\yh) = 0.2$ and $\vh(\yh) = 0.1$, $\vh(\yh) = -0.1$ at the starting point $\yh = 1$. They are shown in Fig. \ref{3.funf}.
\begin{figure}[t]
\renewcommand{\baselinestretch}{1}
\epsfxsize=0.49\textwidth
\begin{center}
\leavevmode
\epsffile{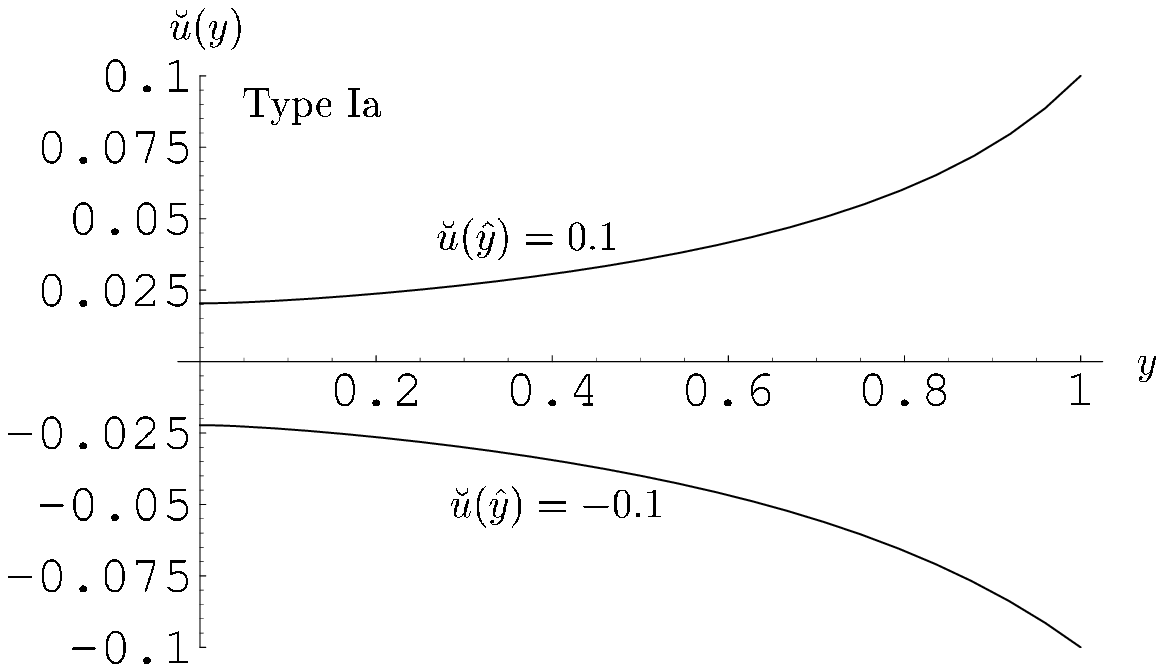}
\epsfxsize=0.48\textwidth
\epsffile{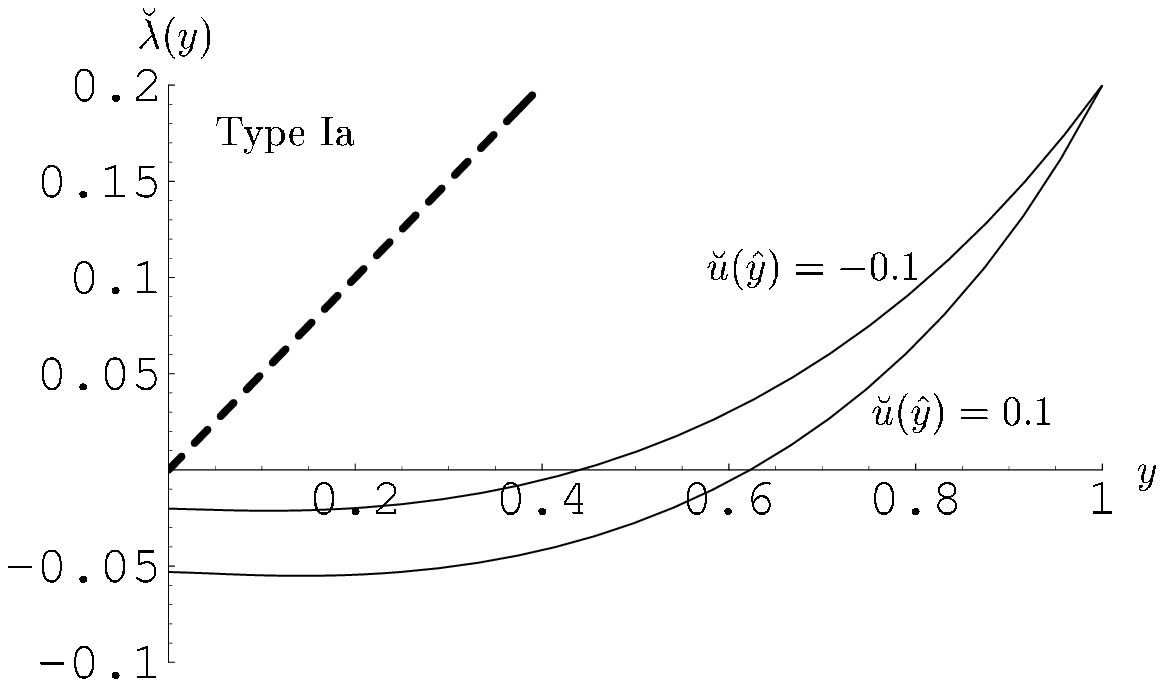}
\end{center}
\epsfxsize=0.49\textwidth
\vspace{2mm}
\begin{center}
\leavevmode
\epsffile{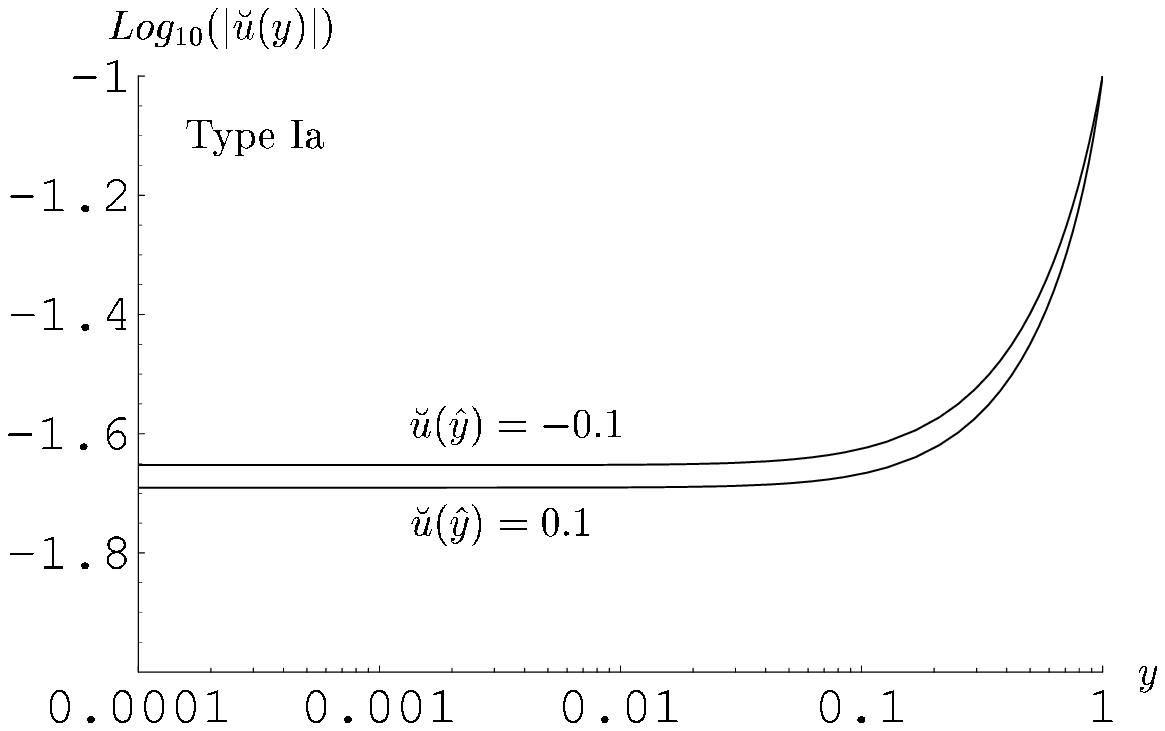}
\epsfxsize=0.48\textwidth
\epsffile{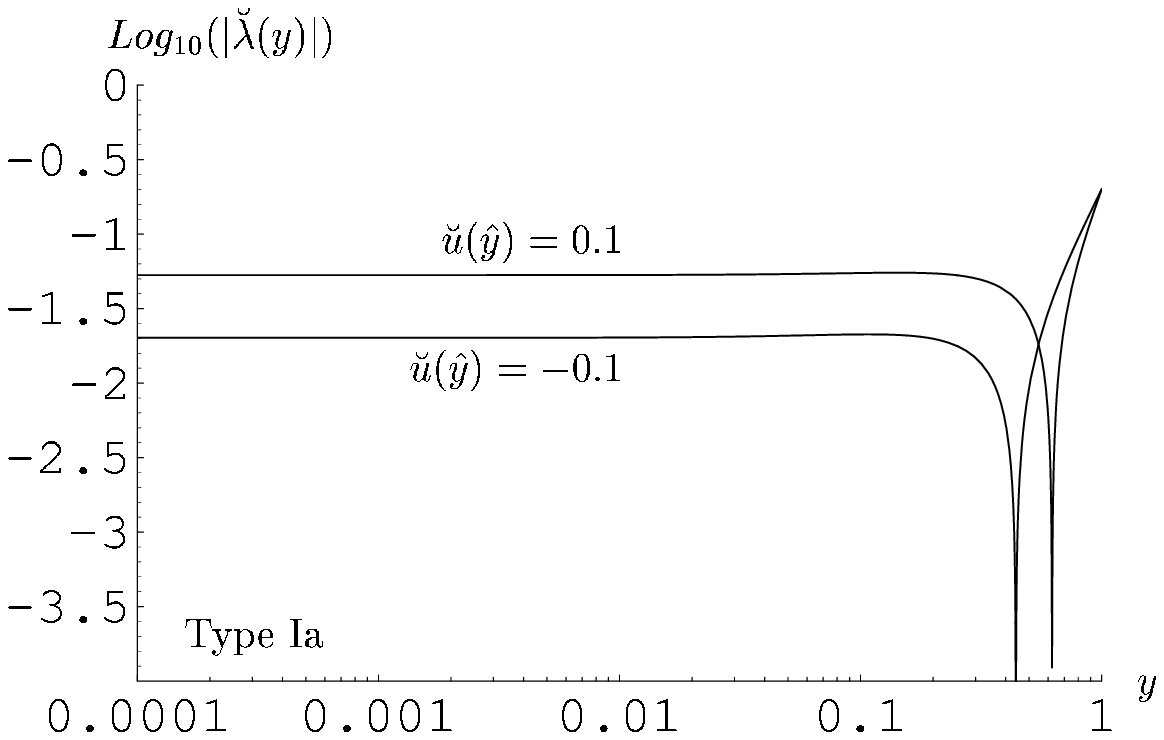}
\end{center}
\parbox[c]{\textwidth}{\caption{\label{3.funf}{\footnotesize RG flow of a typical trajectory of Type Ia arising from the initial conditions $\lh(\yh) = 0.2$ with $\vh(\yh) = 0.1$ and $\vh(\yh) = -0.1$, respectively. The dashed line indicates the boundary of $\lh$-space at $\lh = y/2$. The cusp appearing in the fourth diagram indicates that $\lh(y)$ becomes negative below a certain value $y$. In the IR-regime $(y < 0.01)$ both $\lh(y)$ and $\vh(y)$ are constant.}}}
\end{figure}

The double-logarithmic diagrams in the second line of Fig. \ref{3.funf} show that both trajectories, starting with a positive and negative value $\vh(\yh)$, have the same qualitative properties. In the IR-region $y \lesssim 0.01$ both $\lh(y)$ and $\vh(y)$ take on approximately constant values. This is exactly the scaling behavior found for the trajectories of Type Ia when considering eq. \rf{3.9} with $\alpha_1 \equiv \lb_0 < 0$. This matches the RG flow of the cosmological constant $\lh(y)$ resulting from the Einstein-Hilbert truncation. Hence the inclusion of the coupling $\vh$ in the truncation does not lead to a change in the scaling laws of the cosmological constant for trajectories of the Type Ia.

We now investigate the new trajectory class VIa which corresponds to a positive IR-value of the cosmological constant: $\lh(0) > 0$. A typical solution arising from the initial conditions $\lh(\yh) = 0.45$ and $\vh(\yh) = -0.05$ specified at $\yh = 1$ is shown in Fig. \ref{3.sechs}.
\begin{figure}[t]
\renewcommand{\baselinestretch}{1}
\epsfxsize=0.49\textwidth
\begin{center}
\leavevmode
\epsffile{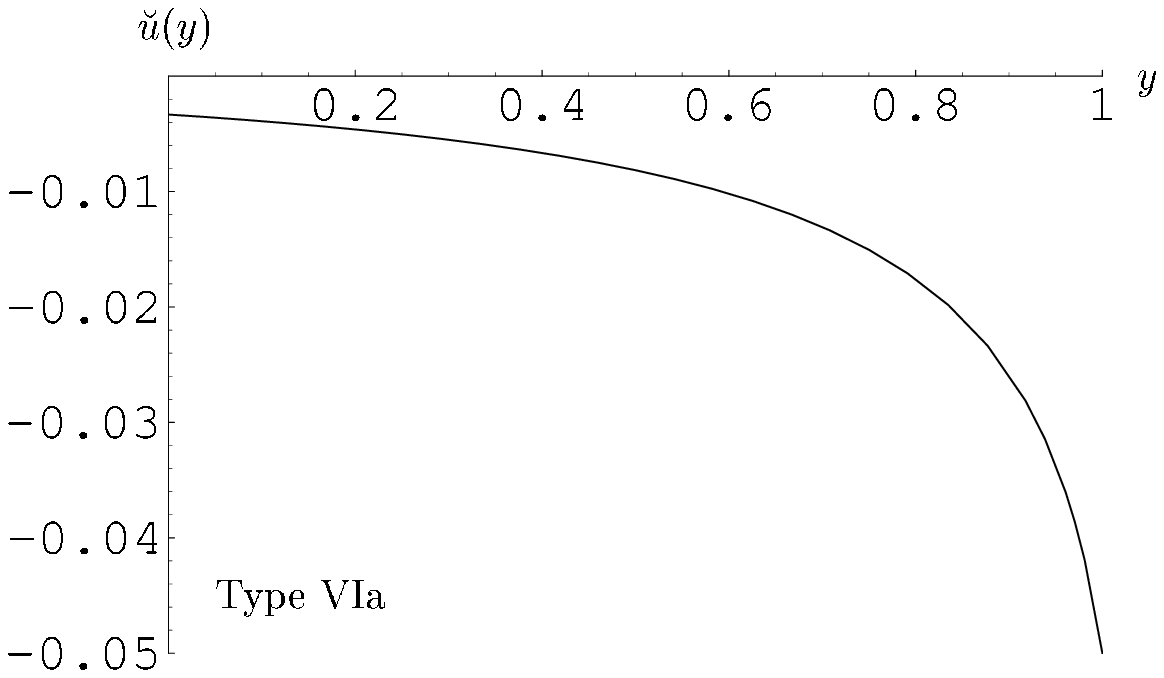}
\epsfxsize=0.48\textwidth
\epsffile{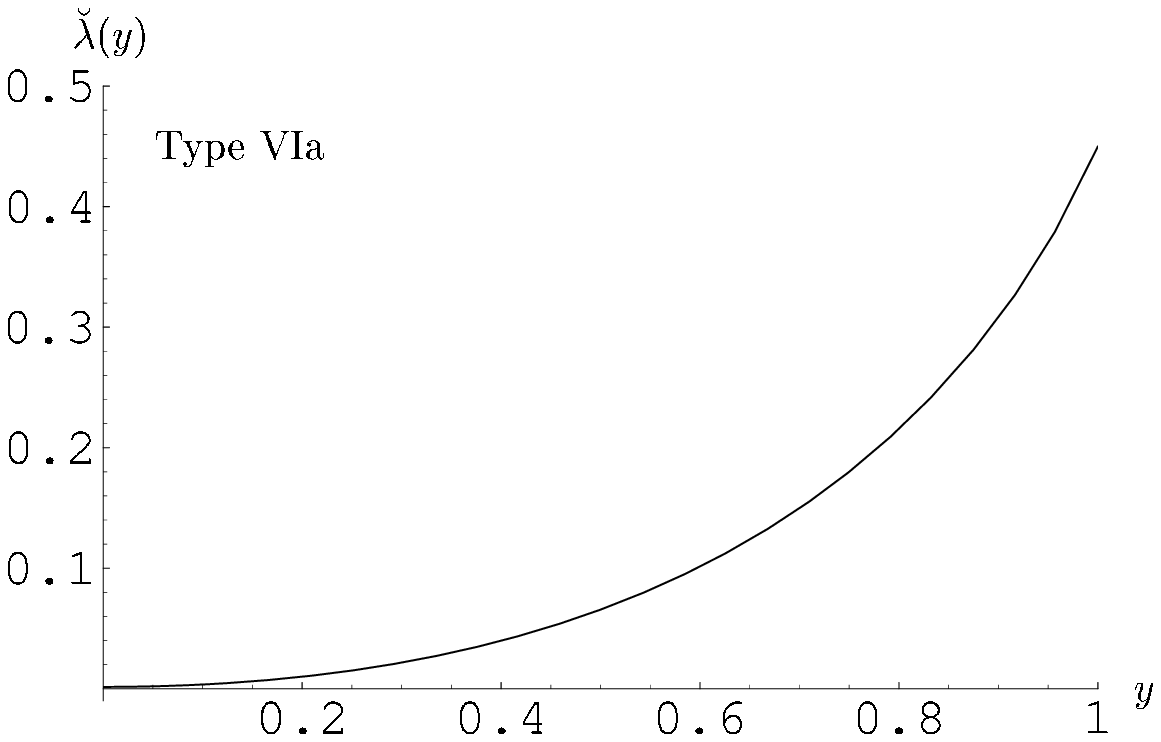}
\end{center}
\epsfxsize=0.49\textwidth
\vspace{2mm}
\begin{center}
\leavevmode
\epsffile{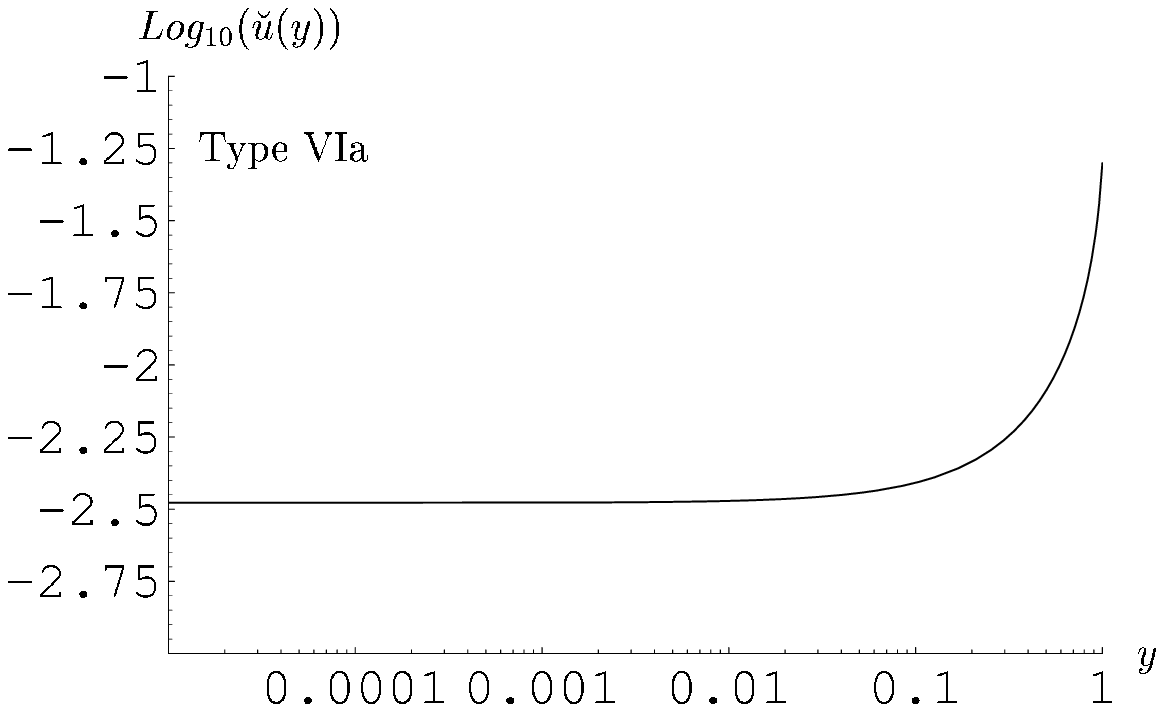}
\epsfxsize=0.48\textwidth
\epsffile{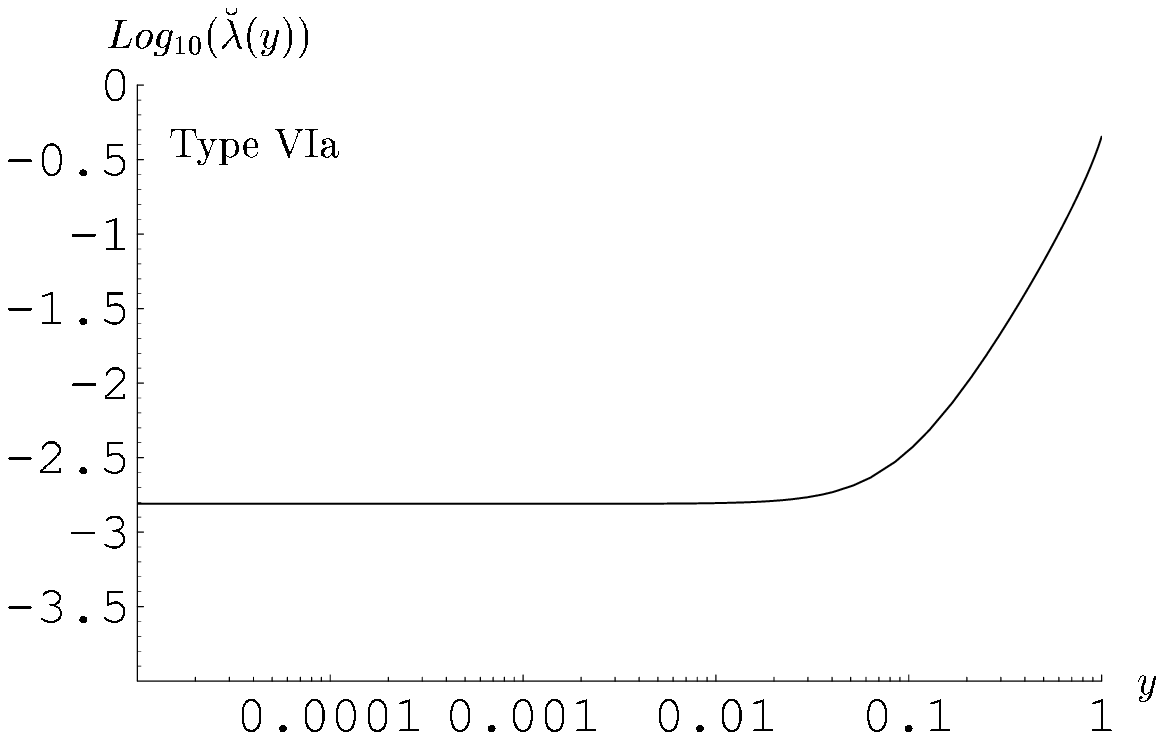}
\end{center}
\parbox[c]{\textwidth}{\caption{\label{3.sechs}{\footnotesize Typical trajectory of Type VIa with initial conditions $\lh(\yh) = 0.45$ and $\vh(\yh) = -0.05$. These lead to a positive IR value of the cosmological constant: $\lh(0)>0$. In the IR-region $(y < 0.01)$ the coupling constants $\lh(y)$ and $\vh(y)$ assume constant values.}}}
\end{figure}

Here we find that $\vh(y)$ and $\lh(y)$ again take on constant values in the IR-region. This resembles the behavior found for the trajectories of the Type Ia. The smooth curve $\log_{10}(\lh(y))$ in the fourth diagram of Fig. \ref{3.sechs} thereby indicates that $\lh(y)$ is positive in the entire region $0 \le y \le 1$.

Next we investigate the scaling properties of the trajectories of Type IIIa. These terminate at the boundary singularity of $\vh$-$\lh$--space. In this case we again fix our initial conditions at the scale $\yh = 0.5$ and choose $\lh(\yh) = 0.2$ and $\vh(\yh)=0.05$, $\vh(\yh) = -0.05$. The resulting typical trajectories are shown in Fig. \ref{3.sieben}. 
\begin{figure}[t]
\renewcommand{\baselinestretch}{1}
\epsfxsize=0.49\textwidth
\begin{center}
\leavevmode
\epsffile{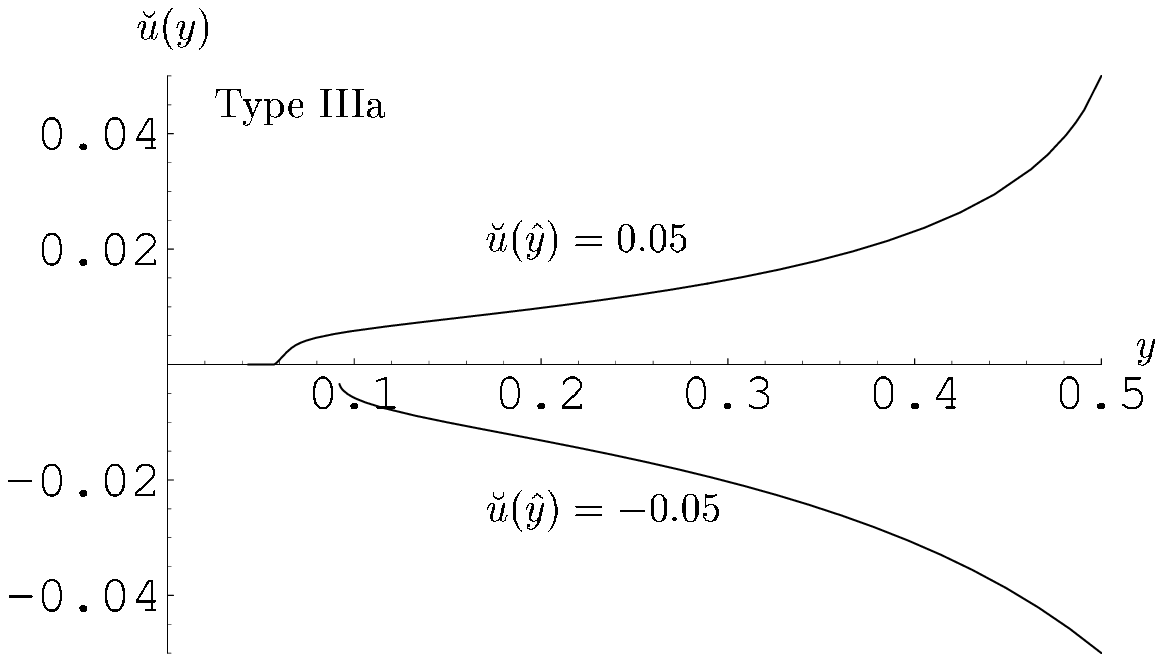}
\epsfxsize=0.48\textwidth
\epsffile{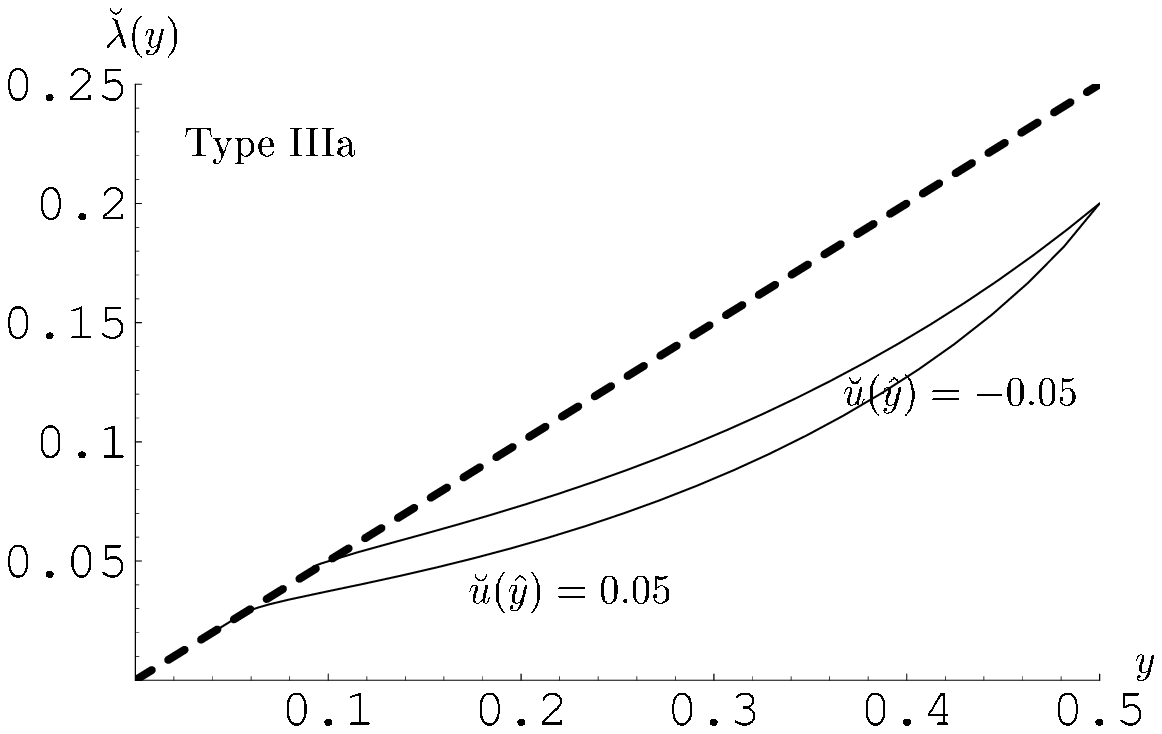}
\end{center}
\epsfxsize=0.49\textwidth
\vspace{2mm}
\begin{center}
\leavevmode
\epsffile{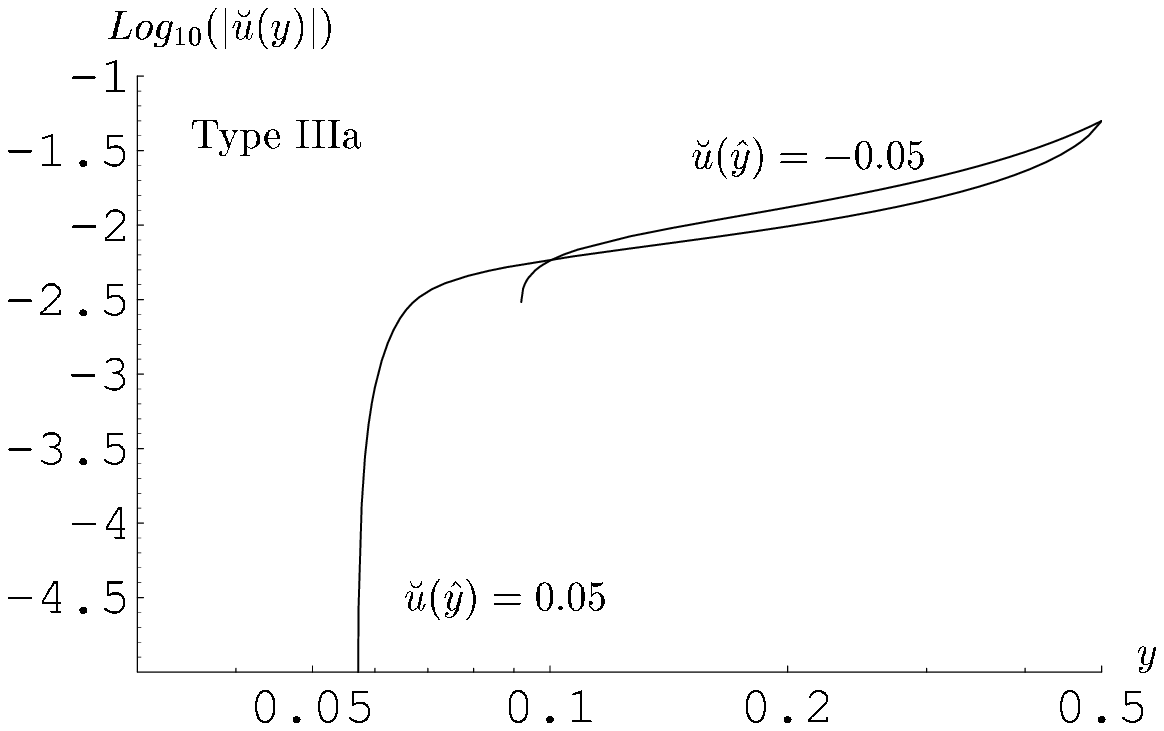}
\epsfxsize=0.48\textwidth
\epsffile{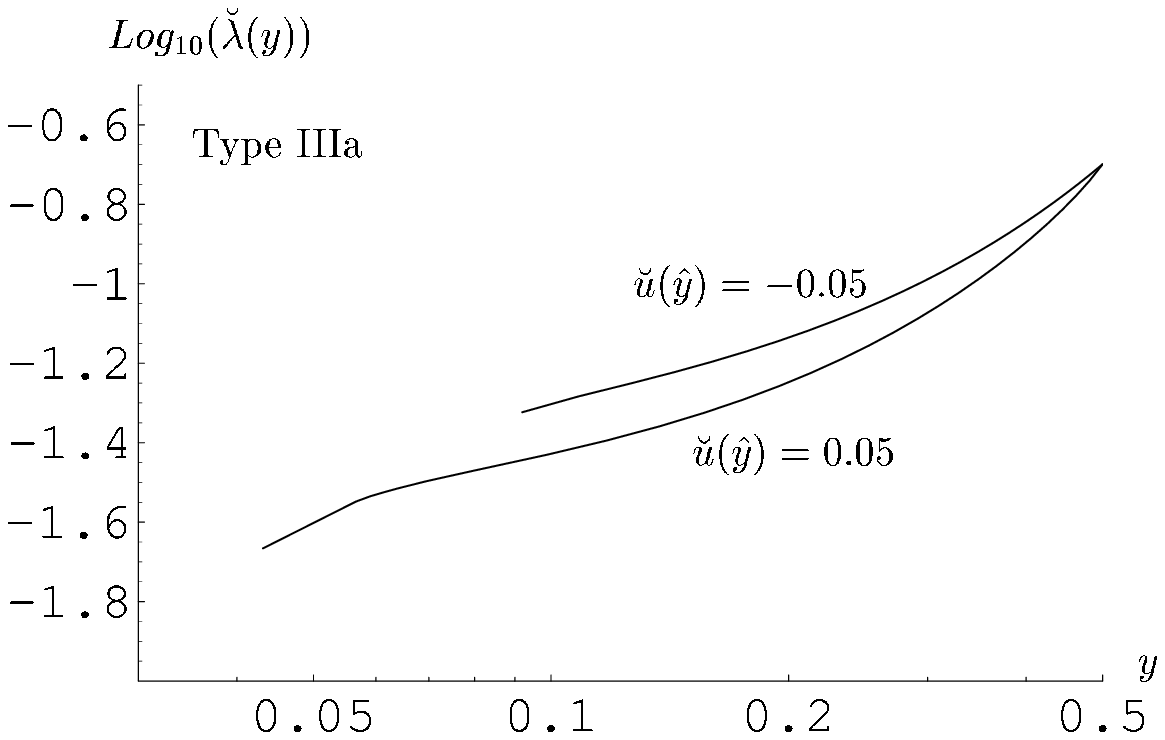}
\end{center}
\parbox[c]{\textwidth}{\caption{\label{3.sieben}{\footnotesize Typical trajectories of Type IIIa with initial conditions $\lh(\yh) = 0.2$ and $\vh(\yh) = 0.05$, $\vh(\yh) = -0.05$ given at $\yh = 0.5$. The trajectories terminate at the boundary singularity $y = 2 \lh + \vh$. Comparing to the trajectory starting with $\vh(\yh) = 0$, positive and negative values $\vh(\yh)$ lead to a decrease or increase of $y_{\rm term}$, respectively.}}}
\end{figure}

Figure \ref{3.sieben} confirms that, compared to the trajectory starting with $\vh(\yh) =0$, the initial conditions $\vh(\yh) > 0$ and $\vh(\yh) < 0$ lead to a decreased and increased value $y_{\rm term}$, respectively. The third diagram reveals that close to $y \gtrsim y_{\rm term}$ the modulus of $\vh(y)$ decreases rapidly. The third diagram thereby clearly demonstrates that a positive $\vh(\yh) > 0$  vanishes identically at $y = y_{\rm term}$, while the trajectory starting with $\vh(\yh) < 0$ ends at a nonzero value $\vh(y_{\rm term}) < 0$.

The last trajectory class found in Section III.B are the trajectories of Type IIa which are characterized by a vanishing IR value of the cosmological constant: $\lh(0)=0$. These can arise either from choosing $\vh(\yh)=0$ and fine-tuning of $\lh(\yh)$ or by fine-tuning $\vh(\yh) = \vh(\yh)_{\rm crit} < 0$ for a trajectory which was of Type Ia originally. To investigate the properties of these cases we choose the typical trajectories arising from $\lh(0) = 0$ and $\vh(0) = 0$, $\vh(0) = -0.01$\footnote{As one easily checks, these are admissible initial conditions. The $\Fbeta$-functions of \rf{3.10} are finite and well defined there.}. These are shown in Fig. \ref{3.acht}.
\begin{figure}[t]
\renewcommand{\baselinestretch}{1}
\epsfxsize=0.49\textwidth
\begin{center}
\leavevmode
\epsffile{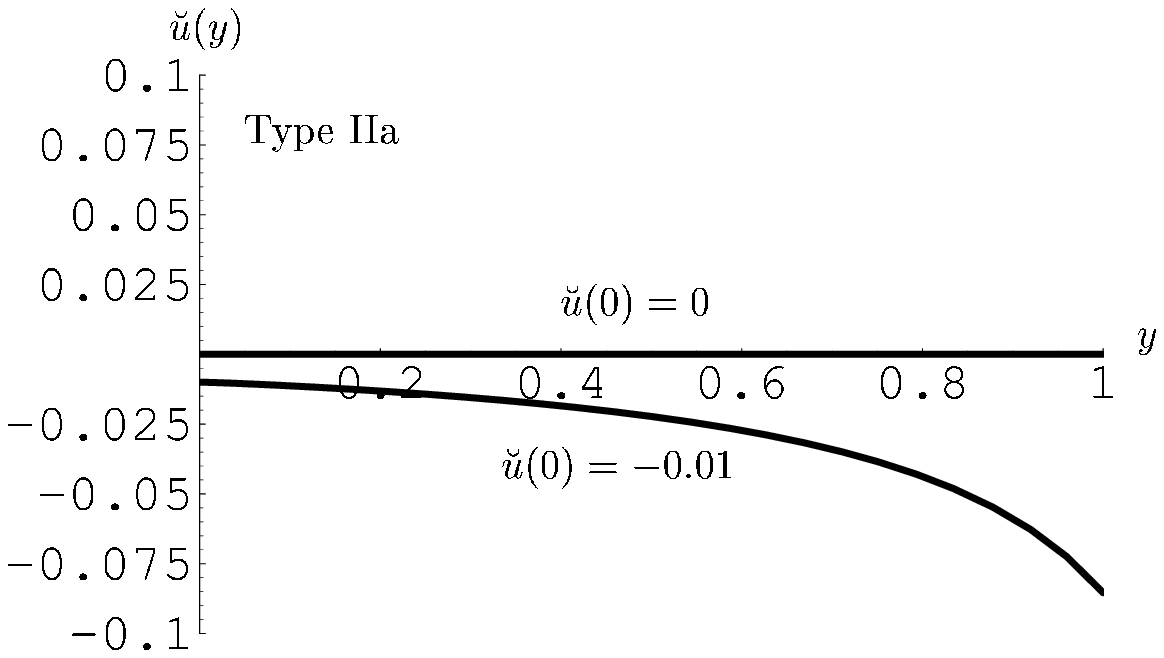}
\epsfxsize=0.48\textwidth
\epsffile{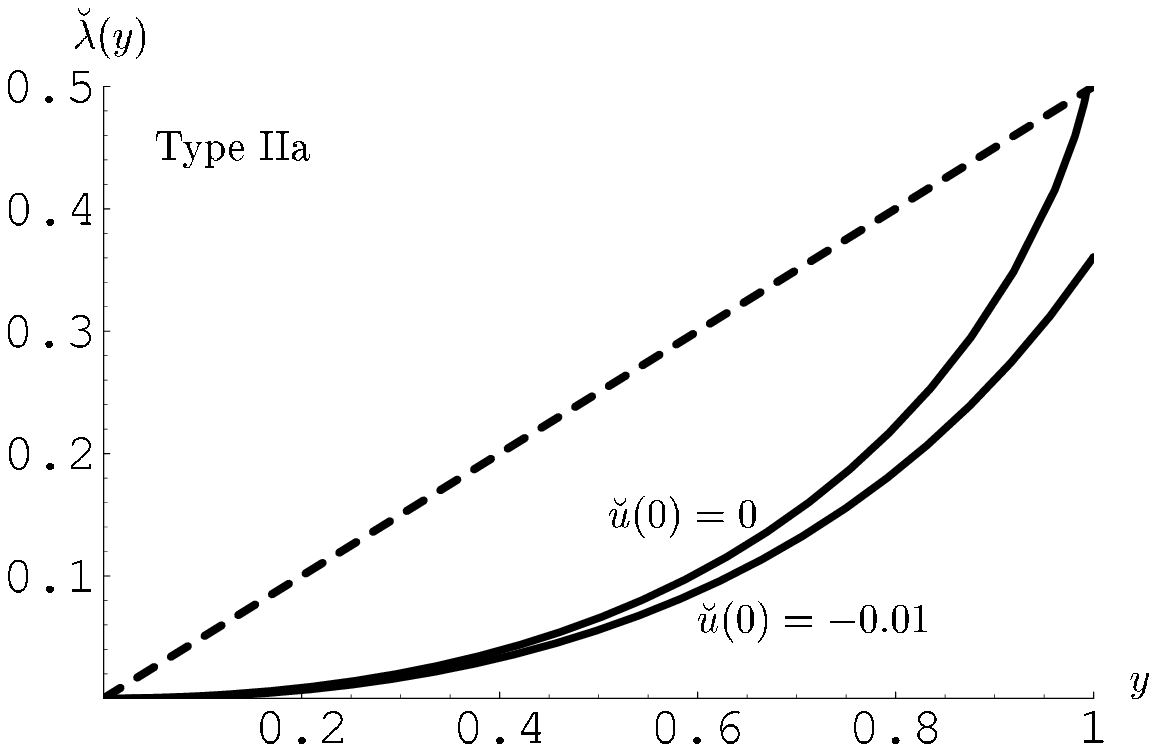}
\end{center}
\epsfxsize=0.49\textwidth
\vspace{2mm}
\begin{center}
\leavevmode
\epsffile{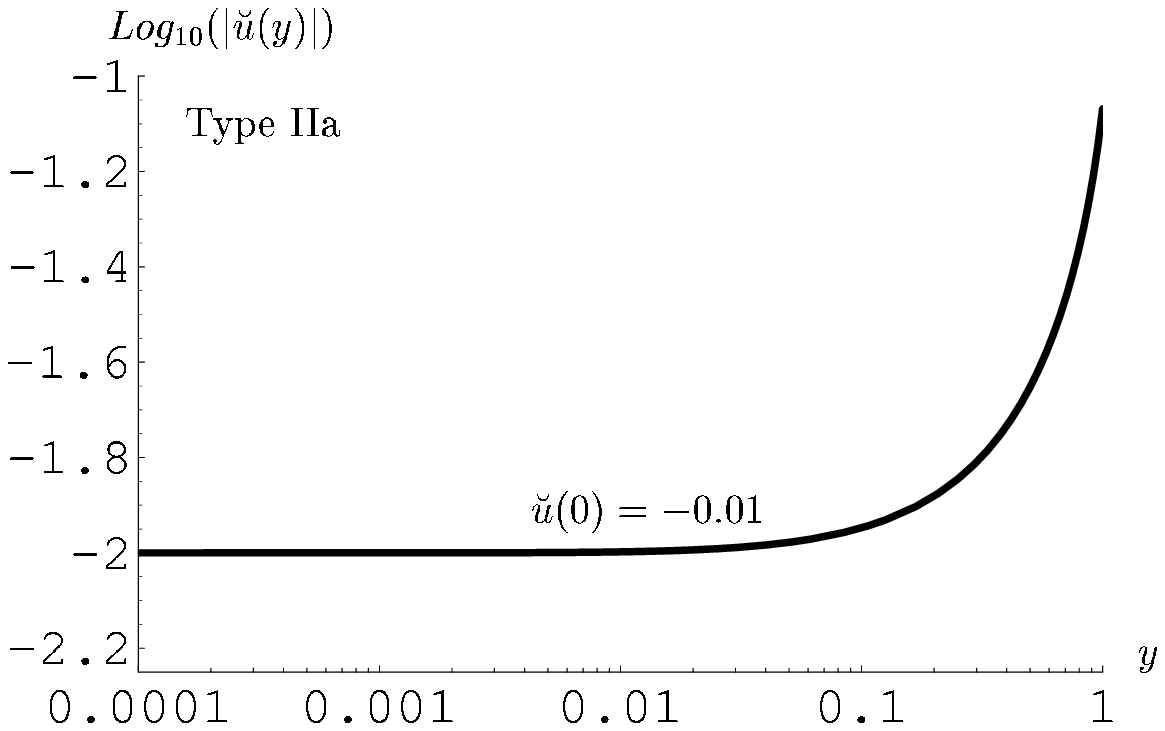}
\epsfxsize=0.48\textwidth
\epsffile{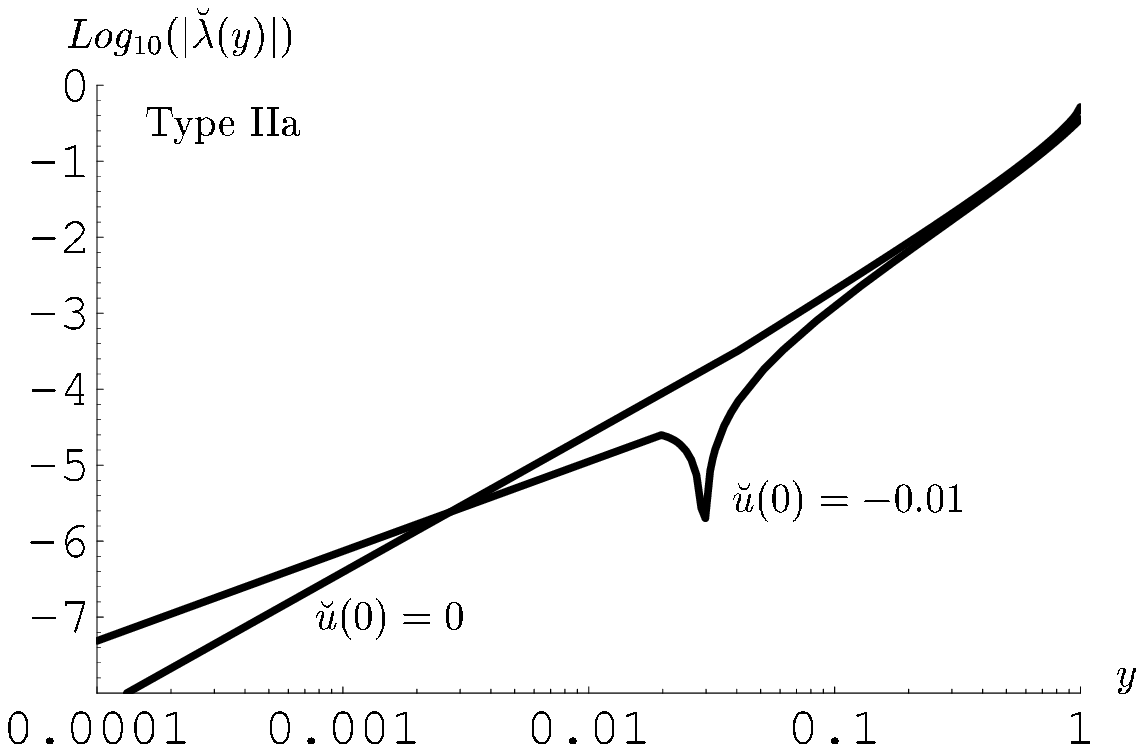}
\end{center}
\parbox[c]{\textwidth}{\caption{\label{3.acht}{\footnotesize Typical trajectories of Type IIa with initial conditions $\lh(0) = 0$ with $\vh(0) = 0$ and $\vh(0) = -0.01$, respectively. The cusp appearing in the fourth diagram showing the logarithm of the modulus of the cosmological constant indicates that the trajectory starting with $\vh(0) = -0.01$ approaches $\lh(0) = 0$ from the negative side $\lh(y) < 0$.}}}
\end{figure}

We find that the trajectories with $\vh(0) = 0$ show a monotonic increase of $\lh(y)$ with $y$. This resembles the scaling behavior for $\lh(y)$ found in the Einstein-Hilbert truncation. 

For the trajectory starting with $\vh(0) = -0.01$ the third diagram of Fig. \ref{3.acht} shows that $\vh(y)$ is approximately constant in the IR-region $y \lesssim 0.01$. The fourth diagram of Fig. \ref{3.acht} is of particular interest in this case. Here we find the typical cusp indicating that the curve $\lh(y)$ crosses the zero line. Together with the second diagram this indicates that $\lh(y)$ becomes negative below a certain value $y$. Only close to $y=0$ the curve $\lh(y)$ bends upward again so that $\lh(0) = 0$ is reached from below.

The results found in this section can be summarized as follows. For the trajectories of Type Ia and of Type IIa with $\vh(\yh) = 0$ the new coupling $\vh(y)$ does not lead to a change in the IR scaling laws of the cosmological constant found in the Einstein-Hilbert truncation. For the trajectories of Type IIa with $\vh(\yh) \not= 0$ and the Type VIa we find that $\vh(\y)$, and in the second case also $\lh(\y)$, takes on constant values in the IR region $y \lesssim 0.01$. This, and the crossing of the zero line in the case of Type IIa trajectories, indicates that these trajectories have the typical behavior of a trajectory of Type Ia whose parameter $\vh(\yh)$ has been chosen such that it runs to $\lh(0) > 0$ and $\lh(0) = 0$, respectively. For the trajectories of Type IIIa the nonlocal coupling generically does not prevent the termination of the trajectories at a finite $y_{\rm term} > 0$.
\end{subsubsection}
\end{subsection}
\mysection{Scale-dependent $S^4$ solutions}
In the previous Section we investigated the RG flow of the coupling constants $\vh$ and $\lh$ in the $V$+$V \ln V$--truncation. We found that all admissible initial conditions imposed at the scale $\yh = 1$ lead to trajectories of the classes Ia and IIa with well defined IR values for the coupling constants. Substituting these coupling constants into the truncated effective average action results in the $k$-dependent functional
\be\label{4.1}
\Gamma_k[g] = \frac{1}{16 \pi G} \, \int d^4x \sqrt{g} (-R + 2 \lb_k) \, + \, \frac{1}{16 \pi G} \, \v_k \, V  \, \ln(V / V_0)
\ee
whose $k$-dependence is explicitly known and leads to a well defined limit $\Gamma_0[g]$.

In this Section we investigate the $k$-dependent stationary points of $\Gamma_k[g]$. In a slight abuse of language we shall refer to them as ``classical solutions''. Actually we are considering a kind of ``RG improved general relativity'' here, i.e. equations of motion which explicitly depend on the scale $k$, the resolution of the ``microscope'' used. The following discussion  focuses on maximally symmetric $4$-dimensional Euclidean space-times of the type $S^4$.
\begin{subsection}{$S^4$ solutions at constant couplings}
Before investigating the scale-dependence of the stationary points of \rf{4.1} we first consider solutions to the equations of motion arising from \rf{4.1} at a fixed value of $k$. In the Introduction we saw already that the modified Einstein equations resulting from the action \rf{4.1} are given by \rf{1.2}, and that they are solved by a 4-sphere provided its radius satisfies eq. \rf{1.8}.

In this subsection we employ a graphical method in order to get a qualitative understanding of which ``on-shell'' values of the radius $r$ are possible for the various $\cf_k$-truncations.

Substituting the metric of a 4-sphere with radius $r$ into \rf{4.1} results in the effective action $\Gamma^{\rm Sphere}(r)$ which is an ordinary function of the radius $r$ of the $S^4$. Introducing the Planck length $\ell_{\rm Pl} \equiv m_{\rm Pl}^{-1} \equiv \sqrt{G}$ and the dimensionless radius $\rh \equiv r/ \ell_{\rm Pl} \equiv r \, m_{\rm Pl}$, the function $\Gamma^{\rm Sphere}(\rh)$ reads:
\be\label{4.3}
\Gamma^{\rm Sphere}(\rh) = \frac{1}{16 \pi} \left[ - 12 \, \ob_4 \, \rh^2 + 2 \, \ob_4 \, \lh \, \rh^4 + \ob_4 \, \vh \, \rh^4 \, \ln(\ob_4 \rh^4) \right] 
\ee
Here we have specified the reference volume to be $V_0 = \ell_{\rm Pl}^4$. It is easy to see that the extrema of the function $\rh \mapsto \Gamma^{\rm Sphere}(\rh)$ are precisely the solutions of eq. \rf{1.8}, i.e. they correspond to the spherical solutions of the equations of motion. This is a consequence of the ``principle of symmetric criticality'' \cite{palais}. Because of the maximal symmetry of $S^4$, the variation of the action and the restriction of the functional to $S^4$-spaces may be interchanged.

We are now interested in finding stable minima of $\Gamma^{\rm Sphere}(\rh)$, i.e. parameters $(\vh, \lh)$ for which $\Gamma^{\rm Sphere}(\rh; \lh, \vh)$ is bounded below and possesses a nontrivial minimum at $\rh > 0$. In this context it is of particular interest if there exists a ``flat limit'', a minimum of $\Gamma^{\rm Sphere}(\rh)$ which is located at very large $\rh \rightarrow \infty$. Equation \rf{1.7} then shows that for these solutions the {\it effective} cosmological constant vanishes or becomes very small without the need of vanishing coupling constants $\lh$ and $\vh$.

In order to find out which combinations of coupling constants result in stable minima of $\Gamma^{\rm Sphere}(\rh)$ we choose typical values for the coupling constants: $\lh = \pm 0.1$ and $\vh = \pm 0.1$. The resulting functions $\Gamma^{\rm Sphere}(\rh; \lh = \pm 0.1, \vh = \pm 0.1)$ are plotted in Fig. \ref{4.eins}. Here the dashed line shows $\Gamma^{\rm Sphere}(\rh)$ for the $V$+$V \ln V$--truncation. The solid line corresponds to the function
\be\label{4.4}
\Gamma^{\rm Sphere}(\rh) = \frac{1}{16 \pi} \left[ - 12 \, \ob_4 \, \rh^2 + 2 \, \ob_4 \, \lh \, \rh^4 + \ob_4^2 \, \wh \, \rh^8 \right] 
\ee
which, in the same way as above, follows from the $V$+$V^2$--truncation \rf{1.1a}.

Figure \ref{4.eins} shows that the functions $\Gamma^{\rm Sphere}(\rh)$ for the $V$+$V \ln V$-- and $V$+$V^2$--truncations possess the same qualitative properties. They lead to the same structure of minima and the same asymptotic behavior. Only the effective actions with $\vh > 0, \wh > 0$ yield a stable minimum and are bounded below. Somewhat counterintuitively, {\it decreasing} a negative $\lh$, i.e. {\it increasing} $|\lh|$, generically results in the minimum being located at {\it larger} values $\rh$. Comparing the location of the minimum for the two truncations we see that the $V$+$V \ln V$--ansatz leads to much larger values $\rh_{\rm min}$ then the $V$+$V^2$--ansatz. 
\begin{figure}[t]
\renewcommand{\baselinestretch}{1}
\epsfxsize=0.49\textwidth
\begin{center}
\leavevmode
\epsffile{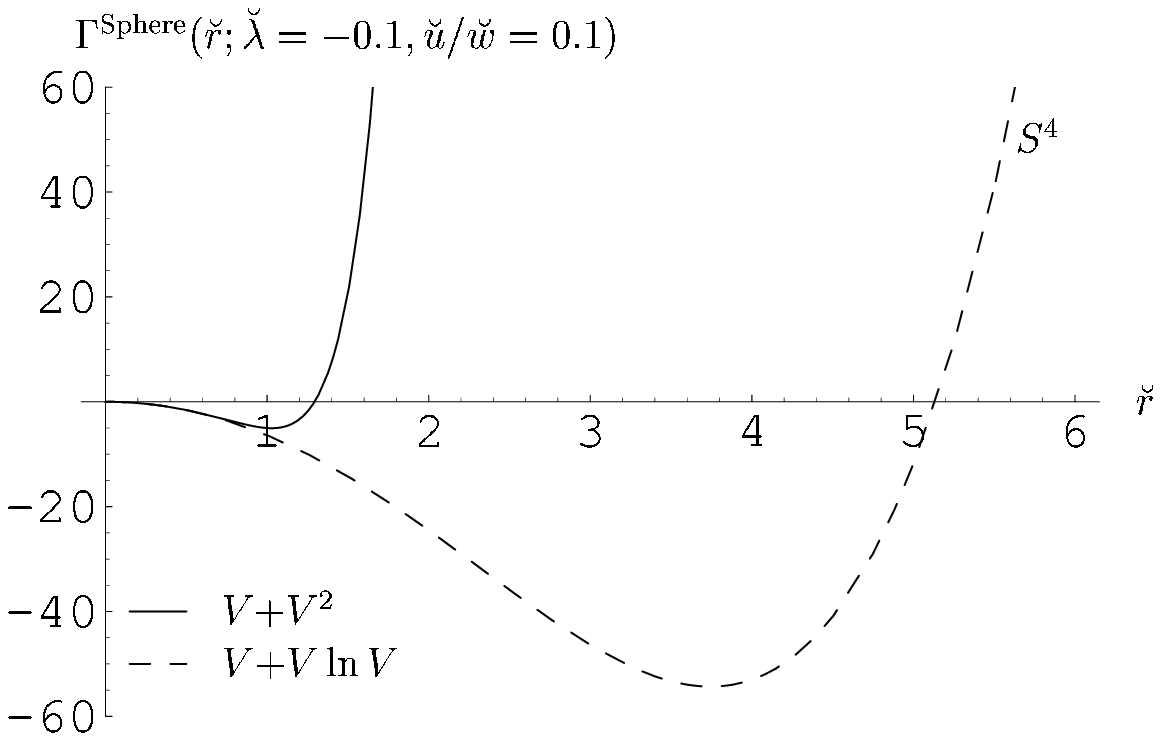}
\epsfxsize=0.48\textwidth
\epsffile{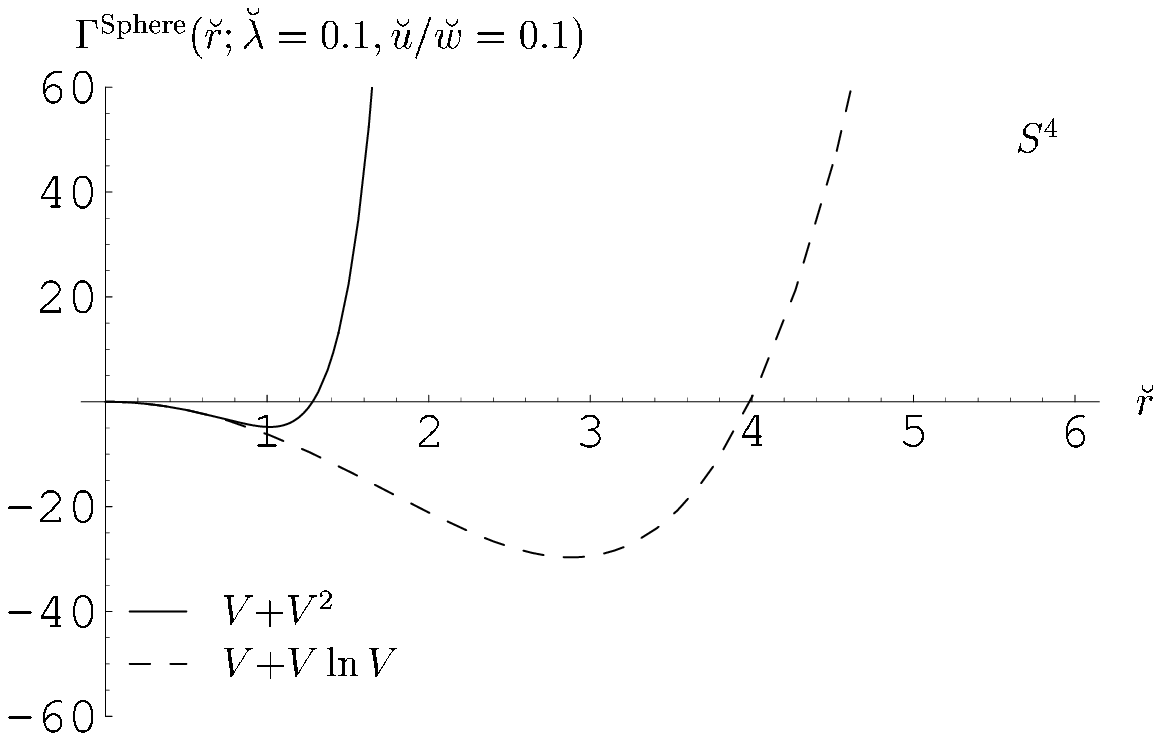}
\end{center}
\epsfxsize=0.49\textwidth
\vspace{2mm}
\begin{center}
\leavevmode
\epsffile{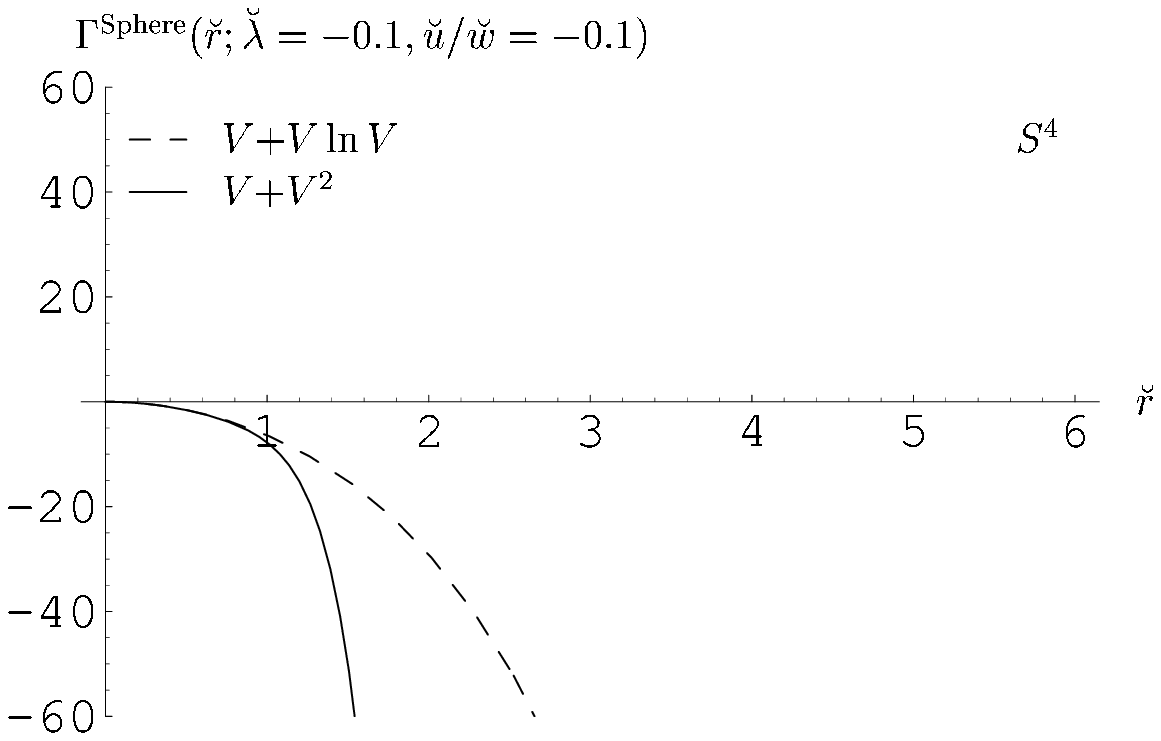}
\epsfxsize=0.48\textwidth
\epsffile{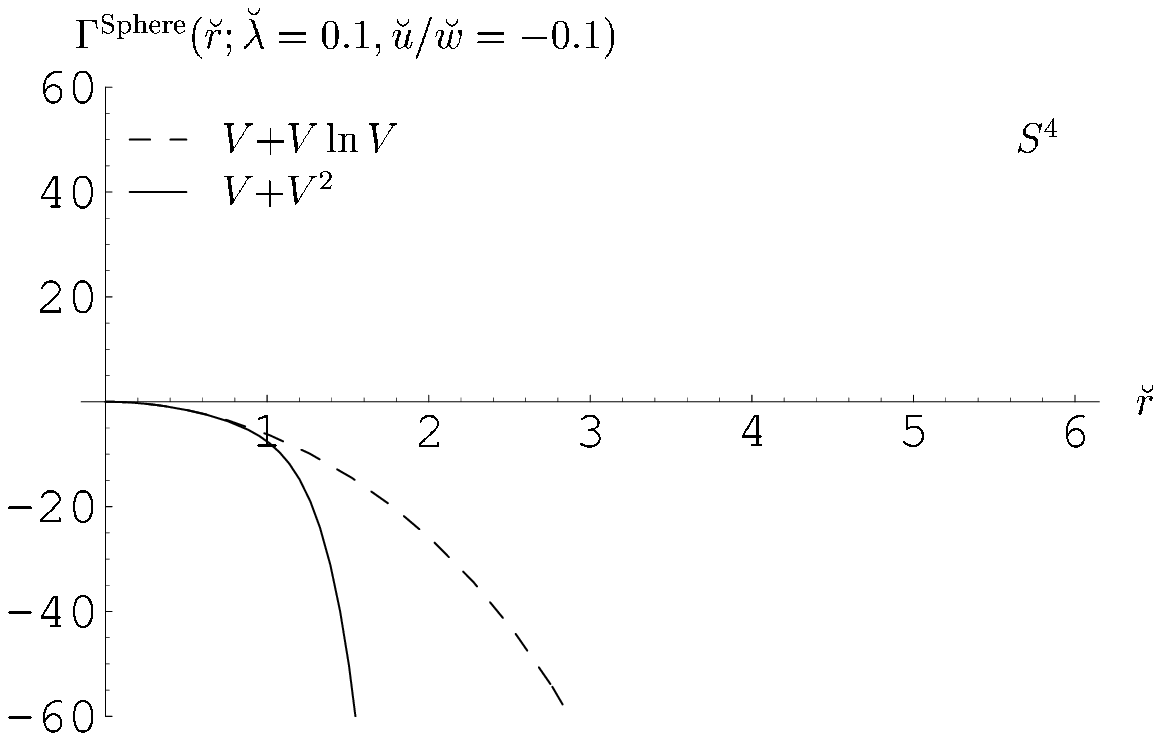}
\end{center}
\parbox[c]{\textwidth}{\caption{\label{4.eins}{\footnotesize Dependence of $\Gamma^{\rm Sphere}(\rh; \lh, \vh/\wh)$ on the radius $\rh$ for typical values of the coupling constants: $\lh = \pm 0.1$ and $\vh/\wh = \pm 0.1$ in the case of the $V$+$V \ln V$- and $V$+$V^2$-truncation, respectively. The four cases resulting from the combination of the parameters $\lh, \vh$ (dashed line) and $\lh, \wh$ (solid line) are shown in the diagrams above. Only for positive values $\vh > 0, \wh > 0$ a stable minimum occurs.}}}
\end{figure}

Considering the lower diagrams in Fig. \ref{4.eins} we find that, in the region with negative values $\vh$ and $\wh$, $\Gamma^{\rm Sphere}(\rh)$ is unbounded below. The action can be lowered to arbitrarily low negative values by increasing $\rh$. However, the corresponding ``minimum'' $\rh \rightarrow \infty$ is not a stationary point of $\Gamma^{\rm Sphere}$. The only critical point in the region $\vh,\wh < 0$ arises for negative values $\lh < 0$ and is not shown in the diagrams. For $\vh, \wh , \lh < 0$, $\Gamma^{\rm Sphere}(\rh)$ assumes a maximum at a finite value $\rh > 0$. The resulting critical point then corresponds to an unstable solution of the equations of motion.

In the following discussion we will therefore limit our investigations to the region of coupling constant space where stable minima of $\Gamma^{\rm Sphere}(\rh)$ exist. In the case of the $V$+$V \ln V$--truncation this corresponds to $\vh > 0$. The properties of the scale dependent solutions arising from this ansatz will be discussed in the next subsection. The analogous results for the $V$+$V^2$--truncation with $\wh > 0$ are summarized in the Appendix.
\end{subsection}
\begin{subsection}{Switching on the RG running}
Let us now proceed and study the $k$-dependence of the position of the stable minimum, $\rh_{\rm min}$. Inserting the $k$-dependent coupling constants $\lh \equiv \lh(y)$ and $\vh \equiv \vh(y)$ into eq. \rf{1.8} leads to the following equation for the scale dependent minimum $\rh_{\rm min} \equiv \rh_{\rm min}(y)$:  
\be\label{4.7}
\vh(y) \, \rhm^2(y) + \vh(y) \, \rhm^2(y) \, \ln \left[ \ob_4 \, \rhm^4(y) \right] + 2 \, \lh(y) \, \rhm^2(y) - 6 = 0
\ee
After substituting a trajectory $y \mapsto \left( \vh(y),\lh(y) \right)$ into \rf{4.7}, $\rh_{\rm min}(y)$ parameterizes the location of the minimum along the particular trajectory under consideration. To simplify our notation we drop the subscript ``min'' from $\rh_{\rm min}$ in the following .

The main emphasis of our studies is on the IR-value of the function $\rh(y)$ since $\rh(y \rightarrow 0)$ is directly related to the cosmological constant problem or, equivalently, the flatness problem. It is particularly important to investigate if there are trajectories along which $\rh(y)$ gives rise to almost flat solutions with $\rh(0) \gg 1$. For these solutions the cosmological constant proper, $|\lh(0)|$, might well be very large, but for solving the cosmological constant problem it is sufficient that $\lh_{\rm eff} \approx 0$. 

In order to obtain well-defined values $\rh(0)$ we restrict the investigations in this Section to trajectories of the Types Ia and IIa running inside the positive coupling region $\vh > 0$. Choosing these trajectories then guarantees that the IR-limits of $\vh(y)$ and $\lh(y)$ are well defined. The restriction to the region with $\vh > 0$ implies that $\Gamma^{\rm Sphere}(\rh)$ has a stable minimum.

In the previous Section we found that for the trajectories considered the couplings $\vh, \lh$ were bounded by their initial values: $\vh(y) < \vh(\yh)$ and $\lh(y) < \lh(\yh)$. Using this information in eq. \rf{4.7} implies that $\rh(y)$ should increase with decreasing $y$:
\be\label{4.8}
\rh(y) > \rh(\yh), \quad \forall \, y < \yh, \quad \mbox{for trajectories of Type Ia and IIa}
\ee  
Stated differently, this means that the quantum fluctuations which are taken into account via the RG running help us in making the universe large and flat.

In order to investigate this mechanism in detail we fix initial values $\left( \vh(\yh), \lh(\yh) \right)$ in the region $\vh(\yh) > 0$, $2 \lh(\yh) + \vh(\yh) \le \yh$, at $\yh = 1$, resulting in trajectories of the Types Ia and IIa. We use the flow equation \rf{3.10} to find the IR values $\vh(0)$ and $\lh(0)$ which arise from these initial values. By substituting the couplings at the Planck scale, $\left( \lh(\yh), \vh(\yh) \right)$, and in the IR, $\left( \vh(0), \lh(0) \right)$, into eq. \rf{4.7} we then determine the initial radius $\rh(\yh)$ and the evolved radius $\rh(0)$ in dependence of $\vh(\yh)$ and $\lh(\yh)$.  

The resulting radii $\rh \left(y=\yh; \vh(\yh), \lh(\yh) \right)$ and $\rh \left(y=0; \vh(\yh), \lh(\yh) \right)$ are shown in Fig. \ref{4.zwei}. (Note that Fig. \ref{4.zwei} displays the decadic logarithm of $\rh(y)$. The label ``$x$'' at the vertical axis denotes the value $\rh(y) = 10^{x}$).  
\begin{figure}[tp]
\renewcommand{\baselinestretch}{1}
\epsfxsize=0.49\textwidth
\begin{center}
\leavevmode
\epsffile{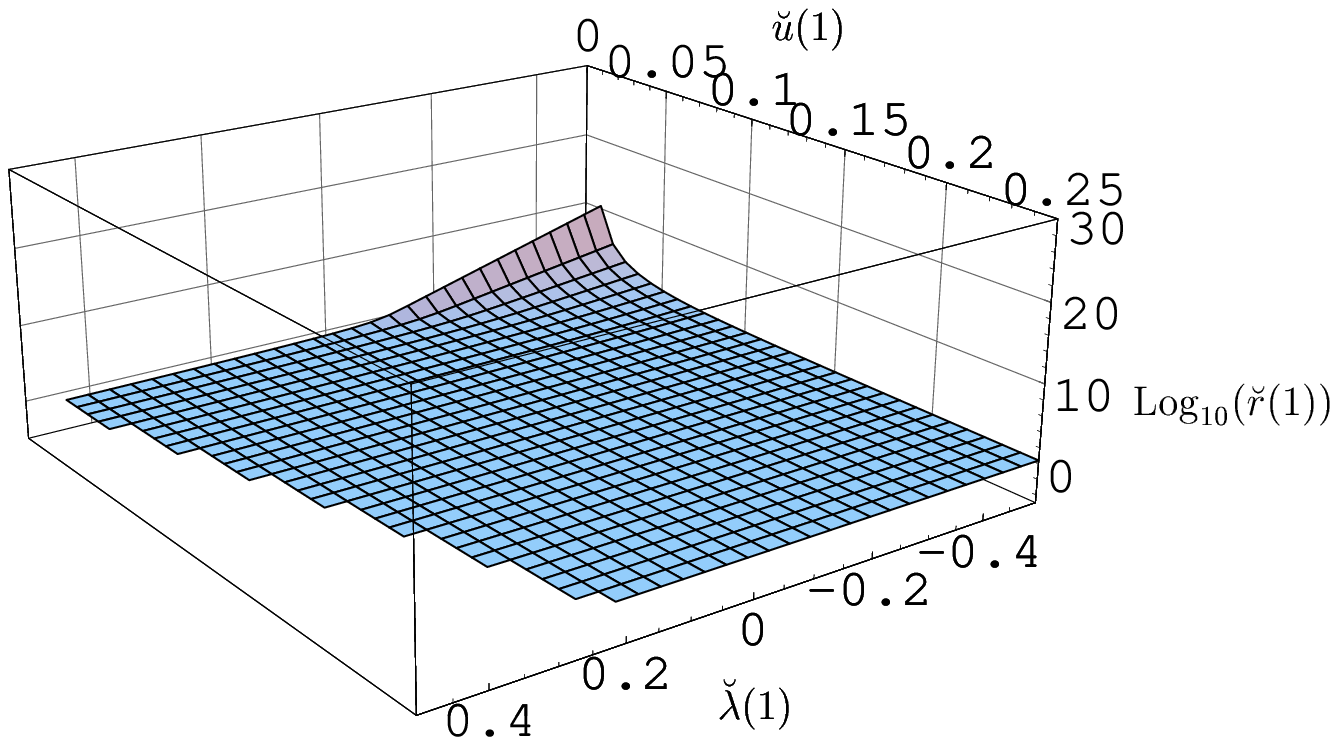}
\epsfxsize=0.48\textwidth
\epsffile{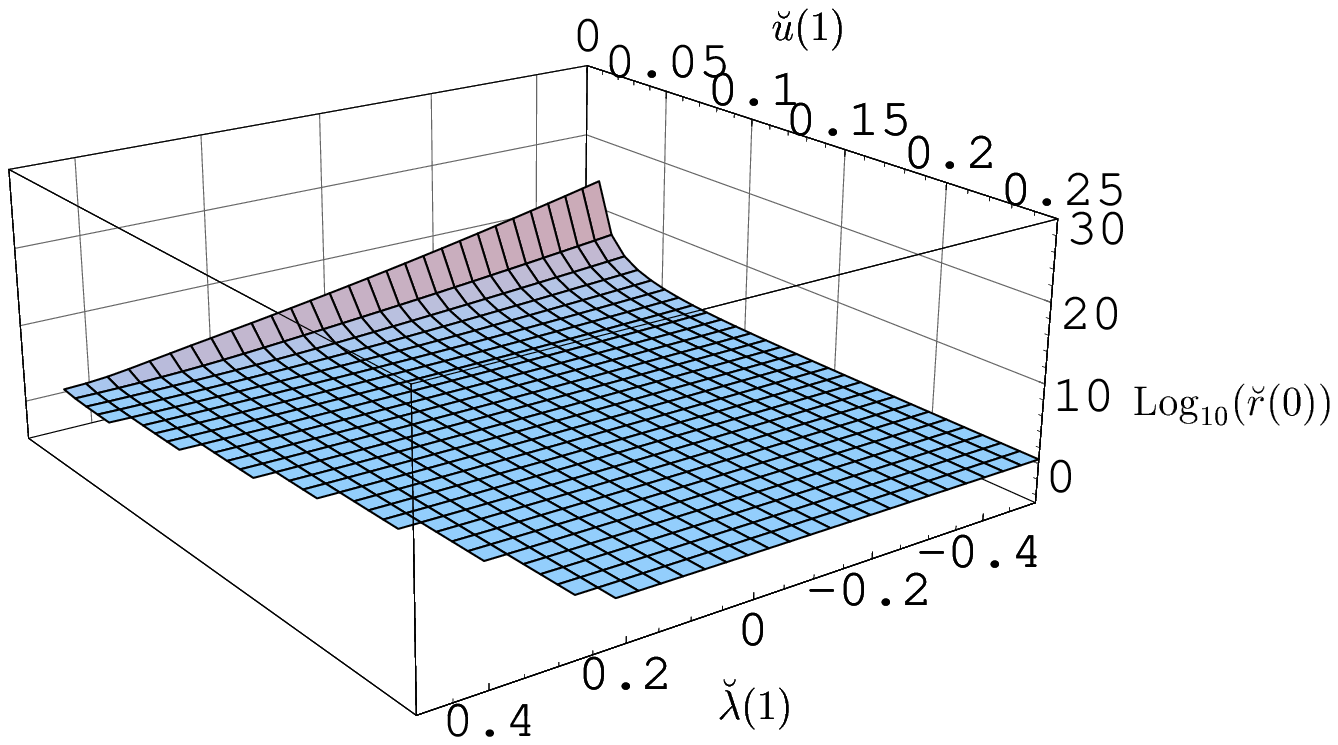}
\end{center}
\vspace{10mm}
\epsfxsize=0.99\textwidth
\begin{center}
\leavevmode
\epsffile{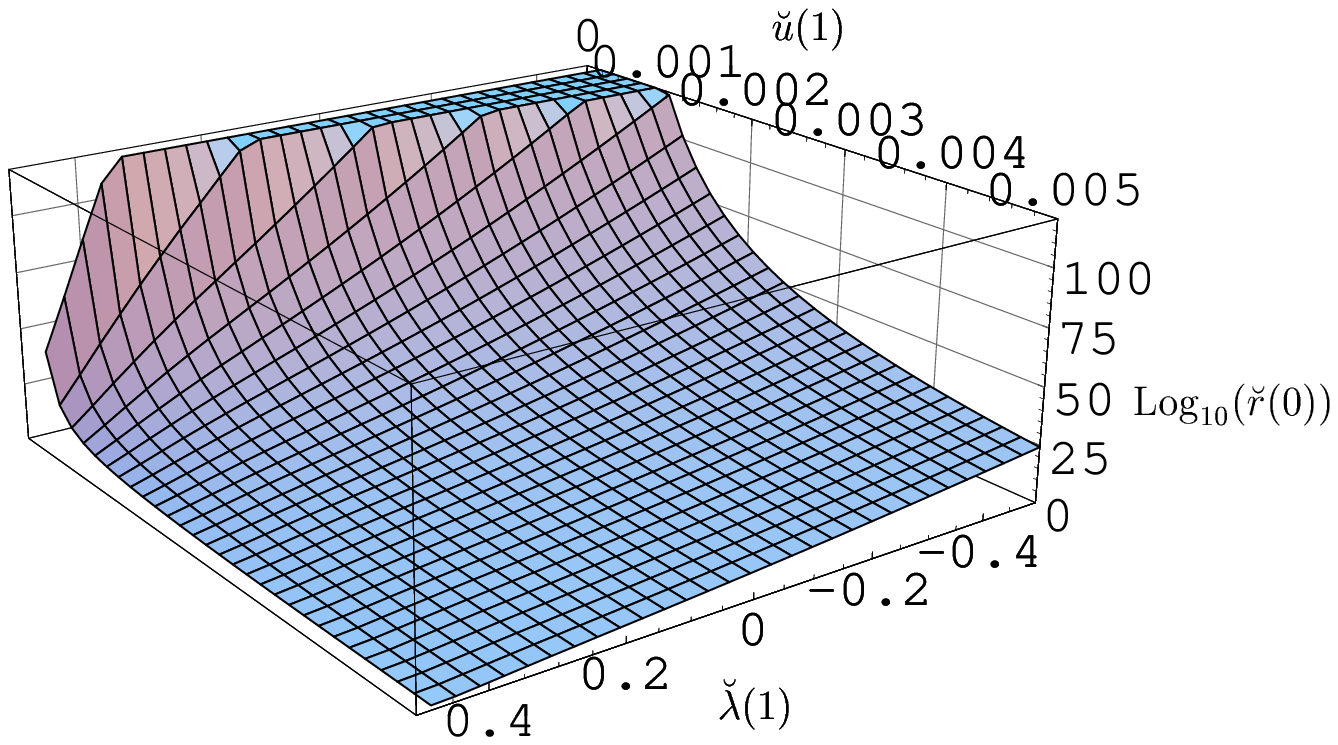}
\end{center}
\parbox[c]{\textwidth}{\caption{\label{4.zwei}{\footnotesize Dependence of $\rh \left(y=\yh, \vh(\yh), \lh(\yh) \right)$ and $\rh \left(y=0,\vh(\yh),\lh(\yh) \right)$ on the initial values $\left( \vh(\yh), \lh(\yh) \right)$ given at $\yh = 1$. The functions $\rh(y=1)$ and $\rh(y=0)$ are shown in the first and second diagram, respectively. The third diagram shows the region of initial values $\left( \vh(\yh), \lh(\yh) \right)$ which result in particularly large radii $\rh(0) > 10^{20}$.}}}
\end{figure}

The first diagram of Fig. \ref{4.zwei} shows the numerical values of $\rh \left(\yh; \vh(\yh), \lh(\yh) \right)$, the radius determined from the Planck scale parameters. The missing shaded squares correspond to initial conditions in the region $2 \lh(\yh) + \vh(\yh) \ge \yh$ which do not give rise to well-defined RG trajectories. The first diagram indicates that generic initial data $\left( \vh(\yh), \lh(\yh) \right)$ lead to $\rh(\yh) \approx O(1)$, i.e. a radius of the order of the Planck length. The only exception is the region with $\lh(\yh) < 0$ and $\vh(\yh) \ll 1$ where we find values $\rh(\yh) \gg 1$.

We then switch on the running of the coupling constants and consider the ``renormalized'' values $\rh \left(y=0; \lh(\yh), \vh(\yh) \right)$ which are shown in the second diagram of Fig. \ref{4.zwei}. The important result is that including the effect of the RG flow results in a considerable extension of the $\left( \vh(\yh), \lh(\yh) \right)$-region in which large radii, $\rh(0) \gg 1$, occur. In particular we see that this region now also extends to $\lh(\yh) > 0$. The most important property of this diagram is that there exists an extended domain of initial data $\left( \vh(\yh), \lh(\yh) \right)$ which result in large nearly flat space-times with $\rh(0) \gtrsim 10^{20}$, say.

The region $\lh(\yh) < 0.5, 0 < \vh(\yh) < 0.005$ in which $\rh(0)$ is particularly large is displayed in detail in the third diagram of Fig. \ref{4.zwei}.

This diagram shows that there exists a 2-dimensional region of initial values $\left( \vh(\yh), \lh(\yh) \right)$, indicated by the flat top, whose trajectories give rise to ``macroscopic'' radii $\rh(0) \gtrsim 10^{125}$, i.e. $r(0) \gtrsim 10^{125} \ell_{\rm Pl}$. Since these macroscopic radii occur for an extended domain of initial data, there is no need to fine-tune initial conditions in order to obtain $\rh(0) \gtrsim 10^{125}$. The only requirement in this case is that the initial value of $\v_k$ is at least 3 orders of magnitude smaller than $\mp^2$. Moreover, $\vh(\yh)$ may be arbitrarily small: all initial values $0 < \vh(\yh) < 10^{-3}$ result in a radius $\rh(0) > 10^{125}$.

The third diagram of Fig. \ref{4.zwei} further shows that the region in coupling constant space where these macroscopic radii occur is {\it not} located close to the boundary $2 \lh(y) + \vh(y) = y$ where the use of our truncation could be problematic.

Even though the Euclidean space-times $S^4$ considered here cannot be compared directly to the Robertson-Walker space-times of Lorentzian signature which are relevant to cosmology it is nevertheless plausible to compare the two pertinent length scales. Loosely speaking, the radius of the $S^4$ is analogous to the Hubble radius $r_{\rm H}$ in the Robertson-Walker case. If we recall that in the present universe $r_{\rm H} \approx 10^{60} \ell_{\rm Pl}$ it is clear that values as large as $\rh(0) \approx 10^{125}$ or $r \approx 10^{125} \ell_{\rm Pl}$ are by far sufficient for a dynamical solution of the cosmological constant problem on the basis of the ``RG improved Taylor-Veneziano mechanism''. 

In order to understand the appearance of the macroscopic radii analytically, we return to eq. \rf{4.7}. For the large values of $\rh$ found in Fig. \ref{4.zwei} we can neglect the ``$-6$'' there. The resulting equation can easily be solved for $\rh$: 
\be\label{4.10}
\rh = (\ob_4)^{-1/4} \, \mbox{exp} \left[ - \, \frac{\lh}{2 \, \vh} \, - \, \frac{1}{4} \right ], \qquad  \left( \rh \gg 1 \right).
\ee
Obviously there is an exponential correlation between the coupling constants $\vh, \lh$ and the radius $\rh$. Hence a small change in $\vh, \lh$ will have an exponentially large effect on $\rh$. Equation \rf{4.10} implies that in order to obtain $\rh(0) \approx 10^{125}$ it is sufficient that $|\lh(0)/\vh(0)| \approx 250$ with $\lh(0)/\vh(0) < 0$. Since $\Gamma^{\rm Sphere}(r)$ possesses a stable minimum for positive values $\vh > 0$ only, this leads to the condition that in order to make this mechanism work {\it $\lh(0)$ has to be negative}. Hence the trajectories of Type Ia are the natural candidates for obtaining large spherical solutions with $\rh(0) \gg 1$.

To investigate the $y$-dependence of $\rh(y)$ along a typical trajectory of Type Ia we choose the trajectory arising from the initial conditions $\lh(\yh) = 0.3$ and $\vh(\yh) = 0.0002$. The trajectory as well as the functions $\rh(y), \log_{10}(\rh(y))$, and the exponent appearing in eq. \rf{4.10} are shown in Fig. \ref{4.drei}. 
\begin{figure}[p]
\renewcommand{\baselinestretch}{1}
\footnotesize 
\epsfxsize=0.49\textwidth
\begin{center}
\leavevmode
\epsffile{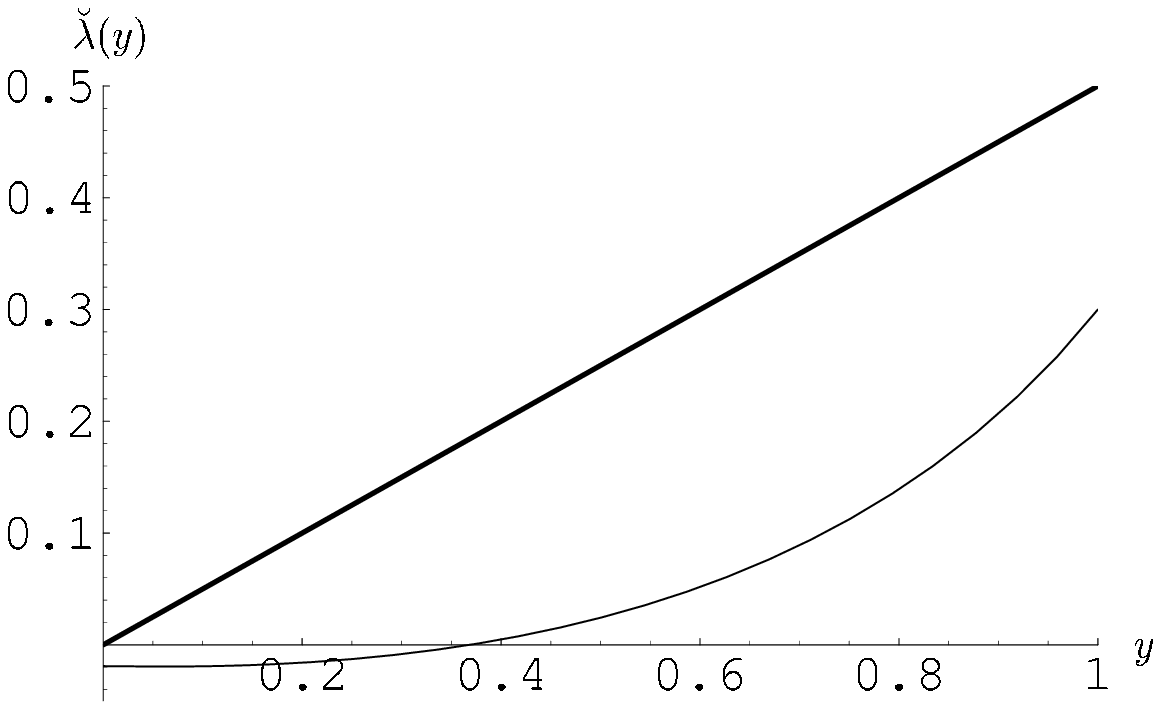}
\epsfxsize=0.48\textwidth
\epsffile{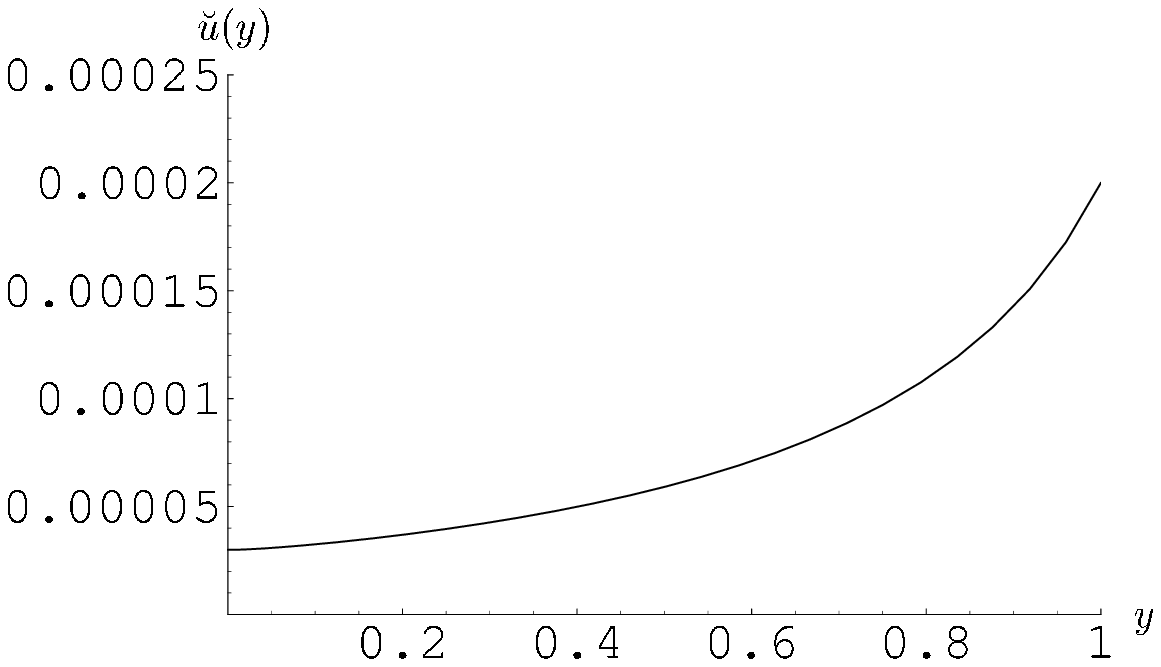}
\end{center}
\epsfxsize=0.49\textwidth
\vspace{2mm}
\begin{center}
\leavevmode
\epsffile{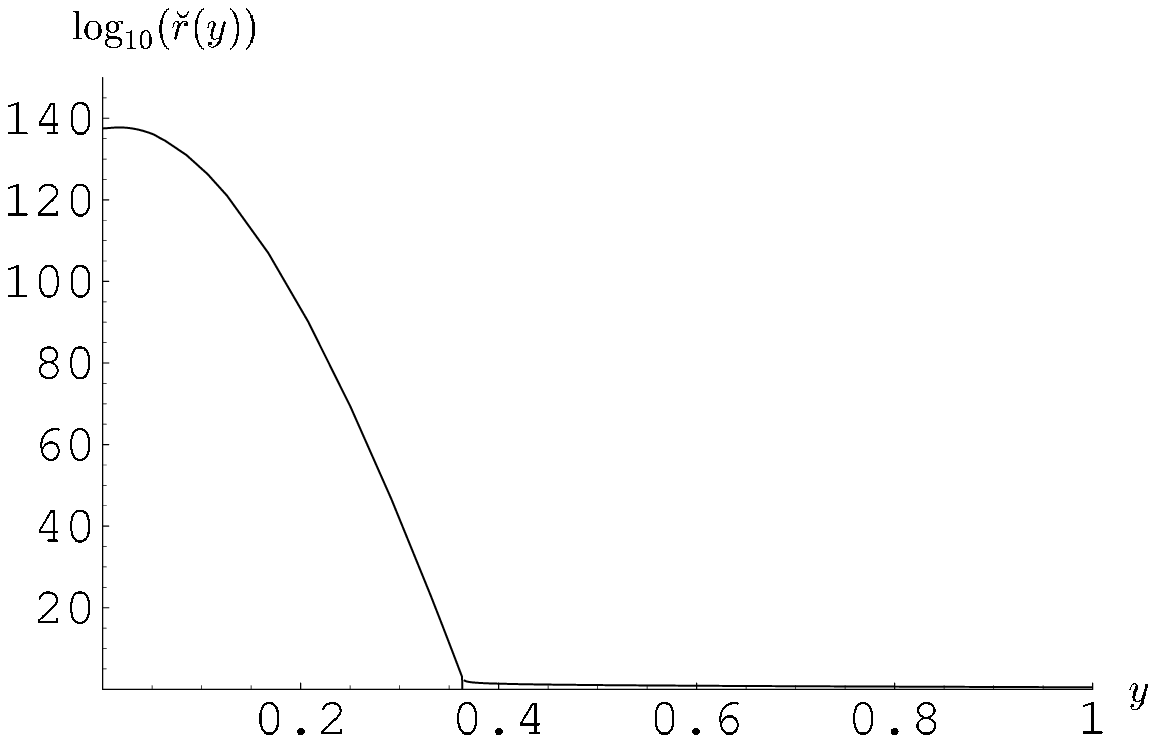}
\epsfxsize=0.48\textwidth
\epsffile{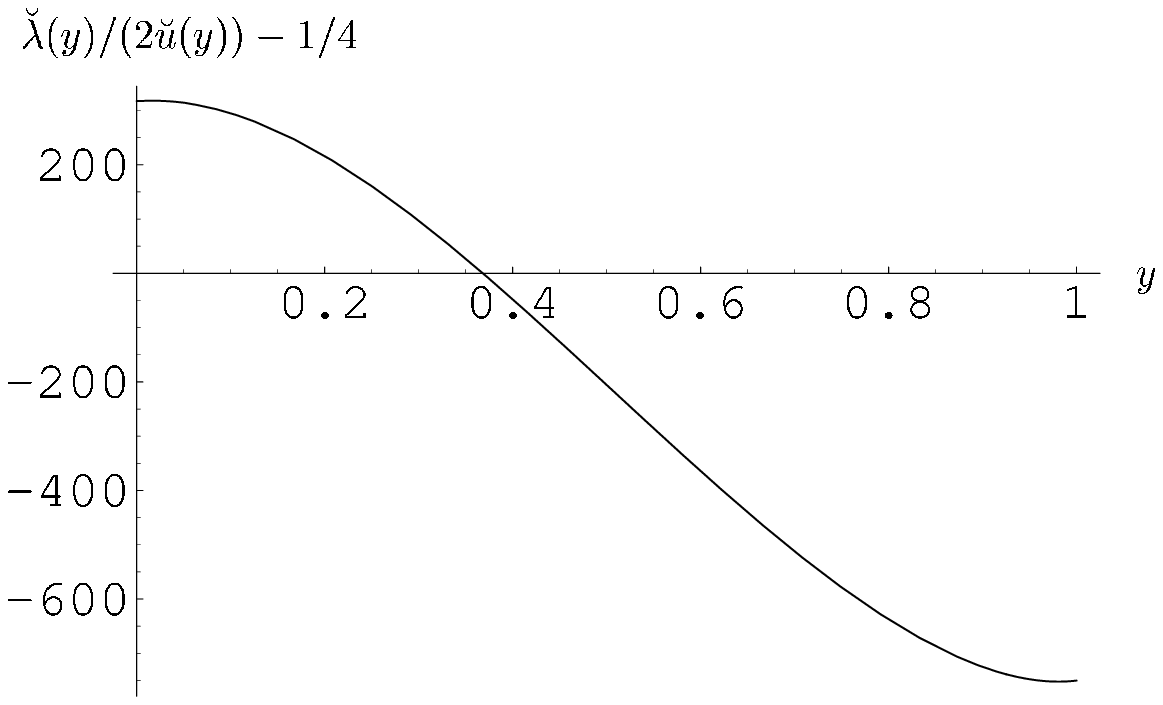}
\end{center}
\epsfxsize=0.49\textwidth
\vspace{2mm}
\begin{center}
\leavevmode
\epsffile{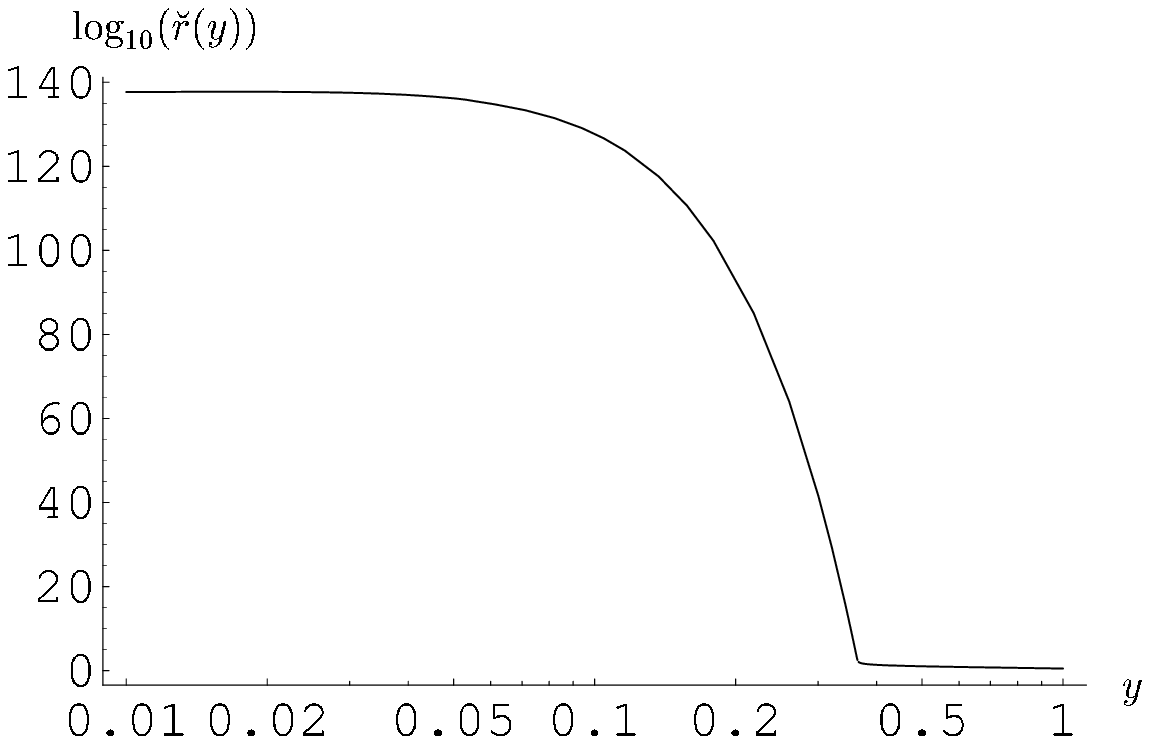}
\epsfxsize=0.48\textwidth
\epsffile{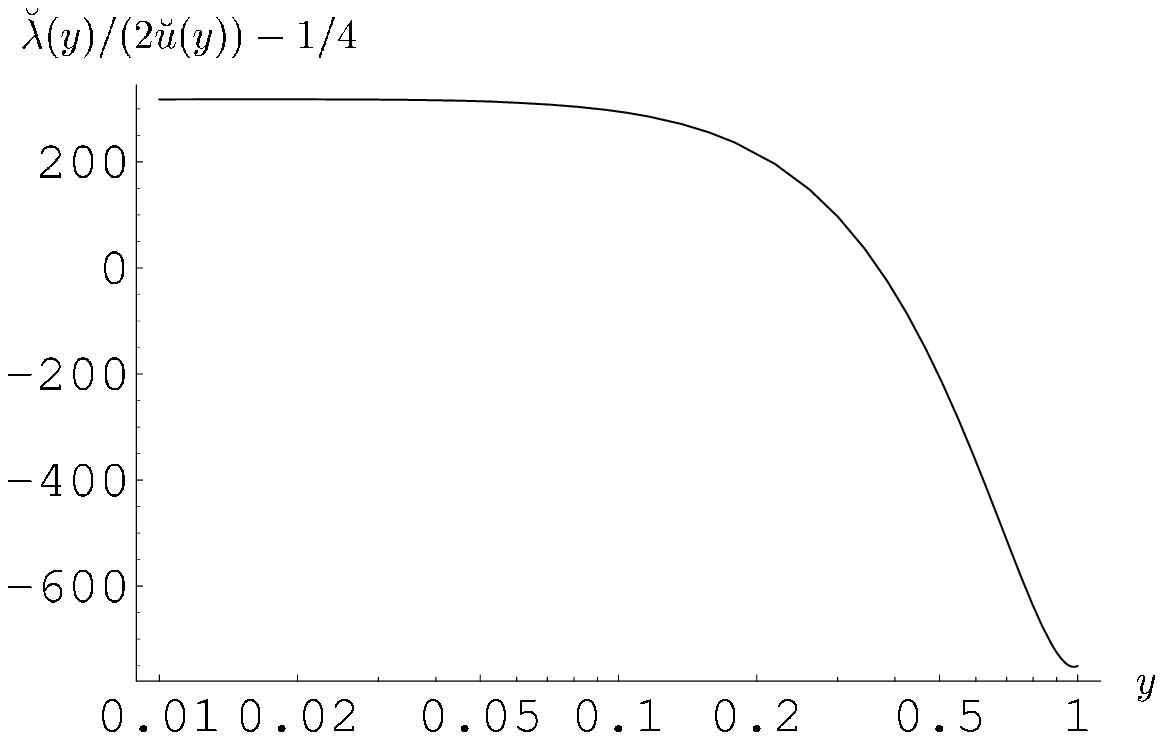}
\end{center}
\parbox[c]{\textwidth}{\caption{\label{4.drei}{\footnotesize The functions $\lh(y), \vh(y), \rh(y)$ of a typical trajectory of Type Ia, starting from initial conditions $\vh(\yh) = 0.0002$ and $\lh(\yh) = 0.3$. In the region where $\lh(y) > 0$ the radius $\rh(y)$ is approximately constant and of order 1. $\lh(y)$ becoming negative induces a ``phase transition'' or ``inflation'' in the sense that $\rh(y)$ increases rapidly towards a macroscopic value $\rh(y) > 10^{125}$. In the IR region $(y < 0.01)$, $\rh(y)$, is constant again.}}}
\end{figure}

Here we see that for large $y$-values where $\lh(y)$ is positive the radius $\rh(y)$ is almost constant and of order 1. The argument of the exponential in eq. \rf{4.10} is negative in this case. The approximation \rf{4.10} would lead to values $\rh \ll 1$. It is certainly not applicable in this regime since the exact numerical solution of \rf{4.7} shows that $\rh(y) \approx 1$.

Lowering $y$, the behavior changes when $\lh(y)$ turns negative. In this case the argument of the exponential in eq. \rf{4.10} turns positive and we find that $\rh(y)$ grows rapidly assuming a macroscopic value $\rh(y) \gg 1$.

In the IR region $y \lesssim 0.01$ both $\lh(y)$ and $\vh(y)$ take on constant values. As a result, $\rh(y)$ ``freezes out'' and keeps its macroscopic value.

Taking the trajectory from Fig. \ref{4.drei} as an example we find a typical ``magnification factor'' of 
\be\label{4.10a}
\frac{r(k=0)}{r(k=\hk)} \approx 10^{138}. 
\ee
This huge number indicates that the RG running is of crucial importance for the actual size of the universe. 

In summary we find that the transition from $\rh(y) \approx O(1)$ to very large radii $\rh(y) \gg 1$ is induced by the cosmological constant proper, $\lh(y)$, crossing zero and turning negative. On the other hand, looking at eq. \rf{1.7} it is clear that the {\it effective} cosmological constant is positive along the entire trajectory. In the region where $\lh(y) > 0$ the small radii $\rh(y) \approx O(1)$ correspond to $\lambda_{\rm eff}(y) \approx \mp^2$. At low scales, $\lh(y)$ turning negative induces a ``quenching'' of $\lambda_{\rm eff}$ so that, in the IR region, the effective cosmological constant takes on very small positive values, $\lambda_{\rm eff}(y) \ll \mp^2$. 

Thus we find that including the nonlocal invariant $V \ln(V/V_0)$ with a small coupling $\vh(\yh)$ into the effective action indeed results in a tiny, positive IR value of $\lambda_{\rm eff}$ at the endpoint of any Type Ia trajectory. While the quenching of $\lambda_{\rm eff}$ due to the nonlocal invariant takes place at the classical level already, the inclusion of the RG running leads to a tremendous amplification of this effect. Taking these quantum effects into account, the same initial data specified at the Planck scale lead to by far larger universes than expected classically.
\end{subsection}
\mysection{The modified Gaussian fixed point}
We now return to the partial differential equation \rf{2.31b} which describes the scale-dependence of $f_k(\vartheta) \equiv k^{d-2} \, \cf_k(V k^d), \vartheta \equiv V \, k^d$. Using this equation for the full nonlinear function $f_k$ we shall study the properties of the modified Gaussian fixed point (MGFP) in the now infinite dimensional truncation subspace. Thereby we generalize the methods used to investigate the fixed point properties in Section III.A to the infinite dimensional setting of ``fixed functions''. 

In order to remove any explicit $k$-dependence from \rf{2.31b} we have to introduce the dimensionless Newton constant $g(k) \equiv G \, k^{d-2}$ satisfying the trivial equation $\partial_t g = (d-2)g$ since $G$ is assumed constant. This leads to the following autonomous system of equations:  
\be\label{5.3}
\partial_t \, g = \Fbeta_g(g, f), \qquad \partial_t \, f_k(\vartheta) = \Fbeta_f(g, f) 
\ee
The $\Fbeta$-functions are:
\bea\label{5.4}
\Fbeta_g(g, f) &=& (d-2) \, g \\[1.5ex]
\label{5.5}
\Fbeta_f(g, f) &=& - \left\{ (2-d) \, f_k(\vartheta) + d \, \vartheta \, f_k^{\prime}(\vartheta) \right\} \\[1.5ex]
\nonumber && + 8 \, \pi \, g \, \bigg\{ 
(4 \pi)^{-d/2} \, \vartheta \, \left( \frac{d \, (d+1)}{2} \, \p{1}{d/2}{-2 \, f_k^{\prime}(\vartheta)}{} - 2 \, d \, \p{1}{d/2}{0}{} \right) \hspace{1cm} \\ 
\nonumber && \qquad \qquad - \, \frac{1}{1 - 2 \, f_k^{\prime}(\vartheta)} + \frac{1}{1 - 2 \, f_k^{\prime}(\vartheta) - \frac{2d}{d-2} \, f_k^{\prime \prime}(\vartheta) \, \vartheta } 
\bigg\}
\eea
(The prime denotes a derivative with respect to $\vartheta$.)

The fixed point condition for these evolution equations, $\Fbeta_g(g^*, f^*) = 0, \Fbeta_f(g^*, f^*) = 0,$ only admits the trivial solution $g^* = 0$ which corresponds to the MGFP. Substituting $g = g^* = 0$ into $\Fbeta_f(g^*, f^*) = 0$ we find the following simple condition for the vanishing of the second $\Fbeta$-function:
\be\label{5.7}
(2-d) \, f^*(\vartheta) + d \, \vartheta \, f^{* \prime}(\vartheta) = 0
\ee
This equation is easily integrated. We find that the fixed point is characterized by $g^* = 0$ together with
\be\label{5.8}
f^*(\vartheta) = c \, \vartheta^{(d-2)/d}
\ee
Here $c$ denotes an arbitrary constant of integration which actually parameterizes a one-parameter family of fixed points.

Specializing to $d=4$ dimensions eq. \rf{5.8} reads
\be\label{5.9}
f^*(\vartheta) = c \sqrt{\vartheta}.
\ee
On the level of the dimensionful function $\cf_k$ this result corresponds to
\be\label{5.10}
\cf^* = c \, \sqrt{V} = c \, \left( \int \, d^4x \sqrt{g} \right)^{1/2}.
\ee
At the fixed point, $\Gamma_k[g]$ has the somewhat peculiar form
\be\label{5.11}
\Gamma^*[g] = \frac{1}{16 \pi G} \, \int d^4x \, \sqrt{g} (-R) + \frac{c}{8 \pi G} \, \left( \int d^4x \sqrt{g} \right)^{1/2}
\ee
Here $c$ acts as an a priori undetermined coupling constant of zero canonical dimension which multiplies the dimensionless $\sqrt{V}/G$. We observe that the cosmological constant vanishes at $\Gamma^*$.

The fixed points \rf{5.11} are scale invariant in the sense that the associated modified Einstein equation does not contain any dimensionful coupling constant:
\be\label{5.11a}
R_{\mn} - \frac{1}{2} \, g_{\mn} \, R = - \, \frac{1}{2} \, \frac{c}{\sqrt{V[g]}} \, g_{\mn}
\ee
As a consequence, eq. \rf{5.11a} cannot fix the ``size'' of the universe. If we insert a $4$-sphere, for example, its radius drops out, and one finds that spheres of any radius are solutions to eq. \rf{5.11a} provided that the parameter $c$ assumes the special value $c = 12 \,  \pi \, \sqrt{2/3}$. 

Note that trajectories starting without a nonlocal ``seed'', i.e. with $f_k(\vartheta) \propto \vartheta$, cannot be in the basin of attraction of the $\sqrt{V}$-fixed points because the flow \rf{2.33} preserves the form $f_k(\vartheta) \propto \vartheta$.

Let us now investigate the stability properties of the fixed point \rf{5.8}. We linearize $g = g^* + \delta g = \delta g$ and $f_k(\vartheta) = f^*(\vartheta) + \delta f_k(\vartheta)$, and make the following ansatz for the small perturbations:
\be\label{5.11b}
\delta g = \epsilon \exp(-\theta t) \, y_g, \quad \delta f_k(\vartheta) = \epsilon \exp(-\theta t) \, Y(\vartheta)
\ee
Here $\epsilon$ is an infinitesimal parameter, and $\theta$ will be the stability coefficient (critical exponent) associated with the scaling field $\left( y_g, Y(\vartheta) \right)$. (Recall that $\theta$ is defined such that $\theta < 0$ corresponds to eigendirections which are attractive in the IR, i.e. for $t = \ln(k/\hk) \rightarrow -\infty$, while directions with $\theta > 0$ are IR repulsive.) The infinite dimensional stability matrix of the MGFP has the structure
\be\label{5.13}
{\bf B}_{\rm MGFP} = \left[ \begin{array}{ll} \frac{\partial \Fbeta_g}{\partial g} & \frac{\partial \Fbeta_g}{\partial \epsilon}\Big|_{\epsilon=0} \\[1.2ex] \frac{\partial \Fbeta_f}{\partial g} & \frac{\partial \Fbeta_f}{\partial \epsilon}\Big|_{\epsilon=0} \end{array}\right]_{\rm MGFP}
\ee
Its eigenvectors are infinite ``columns'' $(y_g, Y(\vartheta))^{\sf T}$. Because $\Fbeta_g$ is independent of $f$ and of $\epsilon$ therefore, $B_{\rm MGFP}$ is a lower triangular matrix:
\be\label{5.14}
{\bf B}_{\rm MGFP} = \left[ \begin{array}{cl} (d-2) & 0 \\[1.2ex] \frac{\partial \Fbeta_f}{\partial g}  & \frac{\partial \Fbeta_f}{\partial \epsilon}\Big|_{\epsilon=0} \end{array}\right]_{\rm MGFP}
\ee

For the stability coefficient $\theta = 2-d$ there is an obvious eigenvector with $y_g \not= 0$ and an accompanying $Y(\vartheta) \not=0$. In terms of the rescaled function $X(\vartheta) \equiv  Y(\vartheta) / y_g$, the form of $Y(\vartheta)$ can be determined by solving the following differential equation:
\be\label{5.14b}
d \, \vartheta \, \frac{d}{d \vartheta} \, X(\vartheta) = \frac{\partial \Fbeta_f}{\partial g} \bigg|_{\rm MGFP}
\ee
In $d=4$, and for $c \not= 0$, $\frac{\mbox{$\partial$} \Fbeta_f}{\mbox{$\partial g$}} \bigg|_{\rm MGFP} $ is given by
\be\label{5.14c}
\frac{\partial \Fbeta_f}{\partial g} \bigg|_{\rm MGFP} = 8 \pi + O(\sqrt{\vartheta})
\ee
Substituting this result, eq. \rf{5.14b} is easily integrated and yields
\be\label{5.14d}
X(\vartheta) = 2 \pi \, \ln(\vartheta) + \mbox{const} + O(\sqrt{\vartheta})
\ee
The $O(\sqrt{\vartheta})$ terms are regular and vanish in the IR ($k \rightarrow 0$) where, for $V$ fixed, $\vartheta = V \, k^4 \rightarrow 0$. We now use this result to write down the trajectory $\Gamma_k = \Gamma^* + \delta \Gamma_k$ due to the eigenvector $\left( y_g, Y(\vartheta) \right)^{\sf T}$. We make the assumption that it approximates a trajectory of the full nonlinear system which gets close to the MGFP {\it in the infrared}, i.e. for $k \rightarrow 0$. Expanding for small $k $, $\delta \Gamma_k$ is given by:
\bea\label{5.14e}
\delta \Gamma_k &=& \frac{\epsilon^{\prime} \hk^{-2}}{16 \pi G^2} \, \bigg\{ 16 \pi G \, \ln(k) + \int \! d^4x \, \sqrt{g} \,  R - 2 \, c \, \sqrt{V} + 4 \pi G \, \ln(V) \\
\nonumber && \qquad \qquad \; + \, \mbox{const} \, + \mbox{terms vanishing for $k \rightarrow 0$} \bigg\}
\eea
Here $\epsilon^{\prime} \equiv \epsilon \, y_g$, and $\hk$ is the fixed reference scale from the definition of the renormalization group time $t \equiv \ln(k/\hk)$. Remarkably, apart from a field independent term which we can ignore, all terms in $\delta \Gamma_k$ either remain constant or vanish in the IR limit $k \rightarrow 0$. Hence, at the level of dimensionful quantities, the eigendirection corresponding to $\theta = 2 - d$ is not repulsive.
 
The eigenvectors of the matrix ${\bf B}_{\rm MGFP}$ with $\theta \not= 2 - d$ are of the form $(0, Y)^{\sf T}$. In this case the function $Y$ is determined by the entry
\be\label{5.14a}
\frac{\partial \Fbeta_f}{\partial \epsilon} \bigg|_{\epsilon = 0} \equiv \frac{\partial}{\partial \epsilon} \, \Fbeta_f \bigg( g=0, f^*(\vartheta) + \epsilon \, Y(\vartheta) \bigg) \bigg|_{\epsilon = 0}
\ee
The function $Y$ has to satisfy
\be\label{5.15}
\left\{ d \, \vartheta \, \frac{\partial}{\partial \vartheta} + (2-d) \right\} Y(\vartheta) = \theta \, Y(\vartheta)
\ee
This differential equation has a solution for every real $\theta$:
\be\label{5.16}
Y(\vartheta) = c_{\theta} \, \vartheta^{( d - 2 + \theta)/d} \equiv Y_{\theta}(\vartheta)
\ee
This result is rather unusual, both because the spectrum of ${\bf B}_{\rm MGFP}$ is {\it continuous} and because there are infinitely many relevant and irrelevant eigenvectors. (In conventional field theories the spectrum is discrete typically and there are only a few relevant directions.) While it is true that there are infinitely many repulsive directions in the ``theory space'' of dimensionless couplings one should bear in mind that the relation between dimensionless and dimensionful coupling constants involves explicit powers of $k$. In order to obtain the trajectory $\Gamma_k = \Gamma^* + \delta \Gamma_k$ corresponding to a specific eigendirection one has to combine those explicit powers of $k$ with the factor $\exp(- \theta t) = (\hk/k)^\theta$ coming from \rf{5.11b}. It is quite remarkable that in the case at hand all factors of $k$ cancel precisely so that $\delta \Gamma_k$ actually does not depend on $k$. 

In 4 dimensions the eigenvectors are $Y_{\theta}(\vartheta) = c_{\theta} \, \vartheta^{(2 + \theta)/4}$. Reintroducing dimensionful variables this becomes at the level of $\Gamma_k$,
\be\label{5.18}
\delta \Gamma_k[g] = \frac{1}{8 \pi G} \, \epsilon \,c_{\theta} \, \hk^{\theta} \, V^{(2 + \theta)/4}
\ee
Obviously the RHS of \rf{5.18} is completely independent of $k$. Thus, at the linearized level, the perturbations \rf{5.18} neither grow nor decay, and their actual stability properties can be inferred from the higher orders in $\epsilon$ only.

Combining \rf{5.14e} and \rf{5.18} we can say that there exists no linearized trajectory $\Gamma^* + \delta \Gamma_k$ which, for $k \rightarrow 0$, is repelled by the MGFP. While this ``taming'' of the IR behavior due to the $\sqrt{V}$-fixed point is not yet a solution to the cosmological constant problem is represents significant progress compared to the usual situation where the $\lb_k$--direction at the GFP is strongly repulsive. According to eq. \rf{3.9}, the dimensionful cosmological constant runs proportional to $k^4$ near the GFP; the analogous perturbation of the $\sqrt{V}$-fixed point is completely independent of $k$, however. (See eq. \rf{5.18} for $\theta = 2$.)

An investigation of the nonlinear stability properties of the NGFP and the determination of its basin of attraction would require solving the nonlinear partial differential equation \rf{5.3} which is beyond the scope of the present paper. Instead we shall study a simplified 2-dimensional flow in the next section.
\mysection{The $V$+$\sqrt{V}$--Truncation}
In the previous Section we found that $\Gamma^*[g]$ contains an invariant proportional to $\sqrt{V}$. Motivated by this result we now analyze a 2-dimensional truncation of $\Gamma_k$ which includes this structure. We investigate the RG flow arising in the ``$V$+$\sqrt{V}$--truncation''
\be\label{6.1}
\Gamma_k[g; \gb] = \frac{1}{16 \pi G} \int d^dx \, \sqrt{g} \, \left( -R + 2 \lb_k \right) + \frac{1}{8 \pi G} \, \ub_k \sqrt{V} + \mbox{classical gauge fixing}
\ee
This ansatz for $\Gamma_k$ contains a running cosmological constant $\lb_k$ and the coupling of the $\sqrt{V}$--term, $\ub_k$.
\begin{subsection}{Projecting the flow equation}
In order to derive the flow equation for the running couplings $\lb_k$ and $\ub_k$ we make use of the partial differential equation \rf{2.31} which describes the RG behavior of a general nonlinear function $\cf_k(V)$. Comparing eq. \rf{6.1} to the ansatz including the $\cf_k$-term in eq. \rf{2.5}, we find that the $V$+$\sqrt{V}$--truncation corresponds to choosing
\be\label{6.2}
\cf_k \left( V \, k^d \right) = \ub_k \, \sqrt{V} + \lb_k \, V = \ub_k \, k^{-d/2} \, \sqrt{\vartheta} + \lb_k \, k^{-d} \, \vartheta = \cf_k \left( \vartheta \right)
\ee
Substituting this ansatz into eq. \rf{2.31} its LHS becomes
\be\label{6.3}
\frac{1}{8 \pi G} \, \left( \partial_t \ub_k \, \sqrt{V} + \partial_t \lb_k \, V \right)
\ee
In order to determine the flow equation for $\lb_k$ and $\ub_k$ it therefore suffices to determine the coefficients of the terms proportional to $\sqrt{V}$ and $V$ appearing on the RHS of \rf{2.31}.

Calculating the first and second derivative of $\cf_k(\vartheta)$ with respect to its argument,
\be\label{6.4}
\cf_k^{\prime}\left( \vartheta \right) = \frac{ \ub_k \, k^{-d/2}}{2 \, \sqrt{\vartheta}} + \lb_k \, k^{-d}, \qquad \cf_k^{\prime\prime} \left( \vartheta \right) = - \, \frac{\ub_k \, k^{-d/2}}{4 \, \vartheta^{3/2}},
\ee
we see that, unlike in the case of the $V$+$V \ln V$-- and $V$+$V^2$--truncation, these expressions contain inverse powers  $V$. As we will see shortly, it is due to this new feature that the $V$+$\sqrt{V}$--truncation does {\it not} give rise to a boundary singularity at $\lambda = 1/2$.

Substituting the derivatives \rf{6.4}, the RHS of \rf{2.31} yields: 
\bea\label{6.5}
\nonumber &&(4 \pi)^{-d/2} \, k^d \, V \left\{ \frac{d(d+1)}{2} \, \p{1}{d/2}{- \, \frac{2 \, \lb_k}{k^2} - \, \frac{\ub_k}{k^2 \, \sqrt{V}}}{} - 2 d \, \p{1}{d/2}{0}{} \right\} \\
&& - \frac{ \sqrt{V} }{ - \, \ub_k \, k^{-2} + \sqrt{V} \, \left( 1 - 2 \lb_k \, k^{-2}\right) } + \frac{\sqrt{V} }{\frac{4-d}{2 (d-2)} \, \ub_k \, k^{-2} + \sqrt{V} \, \left( 1 - 2 \, \lb_k \, k^{-2}\right) }
\eea

To determine the terms proportional to $\sqrt{V}$ and $V$ we expand this expression in a power series in $\sqrt{V}$ at $V=0$. Let us briefly comment on this expansion. We begin with the case $\ub_k \not= 0$. Using the definition of $\p{p}{n}{w}{}$ in \rf{2.28} we find 
\bea\label{6.6}
\nonumber \p{1}{d/2}{ - \, \frac{2 \, \lb_k}{k^2} - \, \frac{\ub_k}{k^2 \, \sqrt{V}} }{} &=& \frac{1}{\Gamma(d/2)} \int^{\infty}_{0} \!\! dz z^{d/2-1} \frac{ R^{(0)}(z) \, - \, z\, R^{(0)\prime}(z)}{\left[ z +  R^{(0)}(z) - \, \frac{2 \, \lb_k}{k^2} - \, \frac{\ub_k}{k^2 \, \sqrt{V}} \right]} \\ 
\nonumber &=& \sqrt{V} \, \frac{1}{\Gamma(d/2)} \int^{\infty}_{0} \!\! dz z^{d/2-1} \, \frac{ R^{(0)}(z) \, - \, z\, R^{(0)\prime}(z)}{\left[ \sqrt{V} \, \left( z +  R^{(0)}(z) - \, \frac{2 \, \lb_k}{k^2} \right) - \, \ub_k \, k^{-2} \right]} \\ 
&=& \tilde{c} \, \frac{k^2}{\ub_k} \, \sqrt{V} + \mbox{higher powers of $\sqrt{V}$}
\eea
Here $\tilde{c}$ denotes a finite constant that depends on the particular choice of $R^{(0)}(z)$. We see that the leading term in the expansion of $\p{1}{d/2}{ - \, \frac{2 \, \lb_k}{k^2} - \, \frac{\ub_k}{k^2 \, \sqrt{V}} }{} $ is of order $\sqrt{V}$. When inserted into \rf{6.5} it produces only a term $\propto V^{3/2}$ outside our truncation subspace and therefore does not contribute to the flow equation. The expansion of the penultimate term in eq. \rf{6.5} is straightforward and yields
\be\label{6.7}
\frac{ \sqrt{V} }{ \ub_k \, k^{-2} - \sqrt{V} \, \left( 1 - 2 \lb_k \, k^{-2}\right) } = \frac{k^2}{\ub_k} \, \sqrt{V} + \frac{k^4}{\ub_k^2} \, \left( 1 - 2 \, \lb_k \, k^{-2} \right) \, V + O\left(\sqrt{V}^{~3} \right)
\ee
Expanding the last term in eq. \rf{6.5} we need to distinguish the cases $d=4$ and $d \not= 4$. In $d=4$ dimensions the term  becomes independent of $V$ and lies outside the truncation subspace. For $d \not= 4$ the expansion yields
\bea\label{6.8}
\nonumber \frac{\sqrt{V} }{\frac{4-d}{2 (d-2)} \, \frac{\ub_k }{k^{2}} + \sqrt{V} \, \left( 1 - 2 \, \frac{\lb_k}{ k^{2}}\right) } &=& \\ 
&& \hspace{-4cm} \frac{2(d-2) \, k^2}{(4-d) \, \ub_k} \, \sqrt{V} - \frac{4(d-2)^2 \, k^4}{(4-d)^2 \, \ub_k^2} \, \left( 1 - 2 \, \frac{\lb_k}{k^{2}} \right) \, V + O\left(\sqrt{V}^{~3} \right),  \qquad \mbox{for $d \not= 4$}
\eea
Introducing the dimensiondependent coefficients
\be\label{6.9}
c_1(d) = \left\{ \begin{array}{cl} 1 & \quad \mbox{for $d=4$} \\ \frac{d}{4-d} & \quad \mbox{for $d\not=4$} \end{array} \right., \qquad c_2(d) = \left\{ \begin{array}{cl} 1 & \quad \mbox{for $d=4$} \\ \frac{d(8-3d)}{(4-d)^2} & \quad \mbox{for $d\not=4$} \end{array} \right.
\ee
the RHS of eq. \rf{2.31} takes on the following form:
\bea\label{6.10}
\nonumber && c_1(d) \, \frac{k^2}{\ub_k} \, \sqrt{V} + \left\{ -2 d \, (4 \pi)^{-d/2} \, k^d \, \p{1}{d/2}{0}{} + c_2(d) \, \frac{k^2}{\ub_k^2} \, \left( k^2 - 2 \lb_k \right) \right\} \, V  \\ 
&& \qquad \qquad + \mbox{terms outside the truncation subspace} 
\eea
Comparing the coefficients of $\sqrt{V}$ and $V$ in the eqs. \rf{6.3} and \rf{6.10} then yields the flow equation for $\lb_k$ and $\ub_k$:
\bea\label{6.11}
\nonumber \partial_t \, \ub_k &=& 8 \pi G \, c_1(d) \, \frac{k^2}{\ub_k} \\
\partial_t \lb_k &=& 8 \pi G \, \left\{ -2d \, (4 \pi)^{-d/2} \, k^d \, \p{1}{d/2}{0}{} + c_2(d) \, \frac{k^2}{\ub_k^2} \, \left( k^2 - 2 \lb_k\right)\right\}
\eea

It is important to note that the above derivation is valid only if $\ub_k \not= 0$. If $\ub_k = 0$ for some value of $k$, the expansions \rf{6.6}, \rf{6.7}, and \rf{6.8} break down, and the $V$-dependence of \rf{6.5} changes abruptly. As a consequence, $\ub_k$ will continue to vanish at all lower scales, and the evolution of $\lb_k$ is governed by the decoupled equation\footnote{For $d=4$, this equation has been studied extensively in \cite{myself}, where it has been obtained by switching off the running of $G_k$ in the $\Fbeta$-functions derived in the Einstein-Hilbert truncation.}
\be\label{6.11a}
\partial_t \lb_k = (4 \pi)^{1-d/2} \, G \, k^d \, \left[ d(d+1) \, \p{1}{d/2}{-2\lb_k/k^2}{} - 4d \, \p{1}{d/2}{0}{} \right]
\ee

The abrupt change of the $V$-dependence at $\ub_k=0$ suggests already that the $V$+$\sqrt{V}$--truncation is probably not very reliable for small values of $|\ub_k|$, and that the flow tends to create additional invariants different from $\sqrt{V}$. In fact, we are actually interested in RG trajectories describing universes which become large in the IR $(V \rightarrow \infty)$ and which have small nonlocalities. Under these conditions $\p{1}{d/2}{-\frac{2 \lb_k}{k^2} - \frac{\ub_k}{k^2 \sqrt{V}}}{}$ equals approximately $\p{1}{d/2}{-2 \lb_k / k^2}{}$ which is independent of $V$. It appears to be of order $\sqrt{V}$ only if we choose a basis of $\cf_k$'s consisting of powers of $\sqrt{V}$ and project on it at $\ub \not=0$.

The parameter space of the $V$+$\sqrt{V}$--truncation is the $\lb$-$\ub$--plane with the $\ub = 0$--line removed. According to the second equation of \rf{6.11} the $\Fbeta$-function of $\lb_k$ diverges on this line. Interestingly enough, there is no boundary singularity at $\lambda = 1/2$.
\end{subsection}
\begin{subsection}{The solution of the flow equation}
We now solve the system of differential equations \rf{6.11} analytically. We see that the flow equation of $\ub_k$ is independent of $\lb_k$ and therefore decouples. Using $\partial_t = k \frac{\partial}{\partial k}$ it is easily integrated and yields
\be\label{6.12}
\ub_k = \pm \, \sqrt{ \ub_{\hk}^2 - 8 \pi G \, c_1(d) \, \left( \hk^2 - k^2 \right) }
\ee
As usual $\hk$ is the scale where we impose the initial value, $\ub_{\hk}$. Later on we shall set $\hk = \mp$. Next we substitute \rf{6.12} into the flow equation for the cosmological constant:
\be\label{6.13}
k \, \frac{\partial}{\partial k} \, \lb_k = 8 \pi G \, \left[ -2d \, (4 \pi)^{-d/2} \, k^d \, \p{1}{d/2}{0}{} + \frac{c_2(d) \, k^2 \, \left( k^2 - 2 \lb_k \right) }{ \ub_{\hk}^2 - 8 \pi G \, c_1(d) \, \left( \hk^2 - k^2 \right) } \right]
\ee
This differential equation can also be integrated analytically. For general $d$ it leads to a rather complicated expression. Setting $d=4$ the general solution simplifies considerably. It consists of two branches distinguished by the sign of the square root in \rf{6.12}: 
\bea\label{6.14}
\nonumber \ub_k &=& \pm \, \sqrt{ \ub_{\hk}^2 - 8 \pi G  \, \left( \hk^2 - k^2 \right) } \\ 
\lb_k &=& - \, \frac{8 G^2 \, \p{1}{2}{0}{} \left( 2 k^6 + \hk^6 - 3 \hk^2 k^4 \right) + 6 G \left( \hk^4 - k^4\right) - 3 \, \lb_{\hk} \, \ub_{\hk}^2}{3  \left[ \ub_{\hk}^2 - 8 \pi G  \, \left( \hk^2 - k^2 \right) \right] }
\eea
This result gives a complete description of the RG flow of the $V$+$\sqrt{V}$--truncation in 4 dimensions. $\lb_{\hk}$ and $\ub_{\hk}$ are the initial values of $\ub_k$ and $\lb_k$ specified at the scale $\hk$. The relevant branch is determined by the sign of $\ub_{\hk}$. 

In order to illustrate the properties of the solution \rf{6.14} we introduce the following ``$\mp$--scaled'' quantities:
\be\label{6.15}
\uh(y) \equiv \ub_k, \quad \lh(y) \equiv \frac{\lb_k}{m_{\rm Pl}^2}, \quad y \equiv \kh^2 = \frac{k^2}{m_{\rm Pl}^2}
\ee
In terms of these couplings the eqs. \rf{6.14} read:
\bea\label{6.16}
\uh(y) &=& \pm \sqrt{\uh(\yh = 1)^2 - 8 \, \pi \left( 1 - y \right)} \\
\label{6.17}
\lh(y) &=& - \, \frac{8 \, \p{1}{2}{0}{} \left( 2 \, y^3 - 3 \, y^2 + 1 \right) + 6 \, \left( 1 - y^2 \right) - 3 \, \lh(\yh = 1) \, \uh(\yh = 1)^2}{3 \, \left[ \uh(\yh=1)^2 - 8 \pi \, \left( 1 - y \right) \right] }
\eea
Here we have identified the initial scale with $\mp$ by setting $\yh = \hk^2/\mp^2 = 1$.

Figure \ref{6.eins} visualizes the RG flow of $\uh(y)$ for various positive and negative initial values $\uh(\yh)$.  
\begin{figure}[t]
\renewcommand{\baselinestretch}{1}
\epsfxsize=0.49\textwidth
\begin{center}
\leavevmode
\epsffile{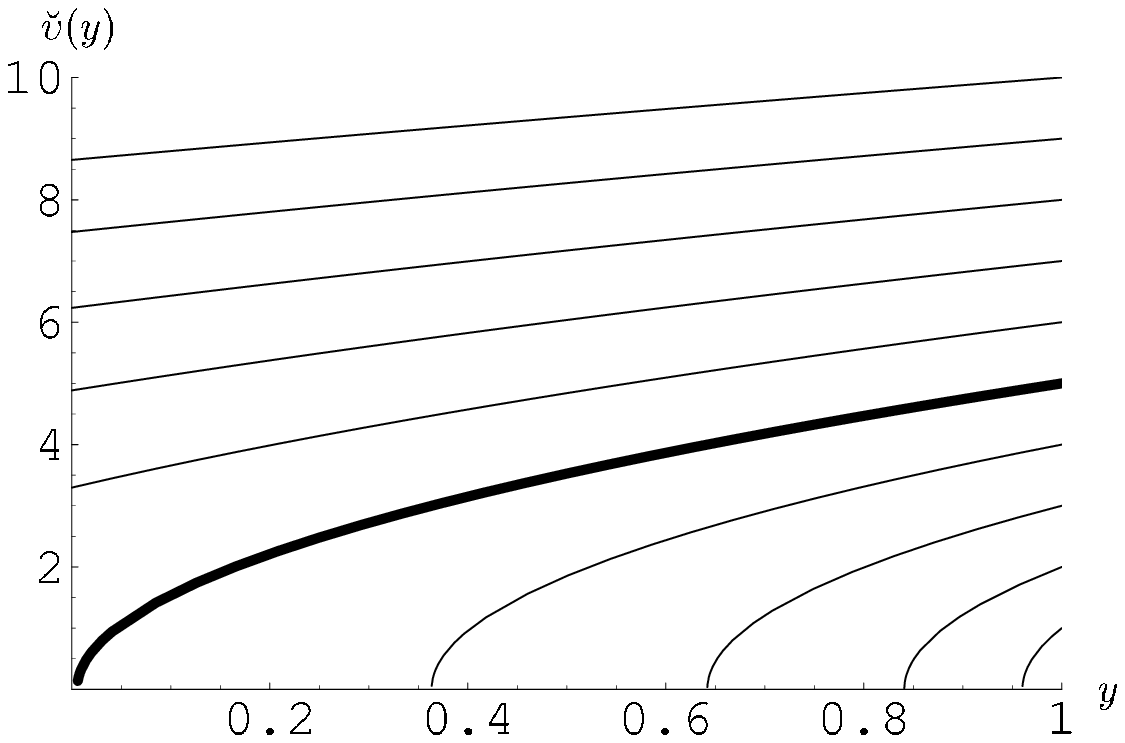}
\epsfxsize=0.48\textwidth
\epsffile{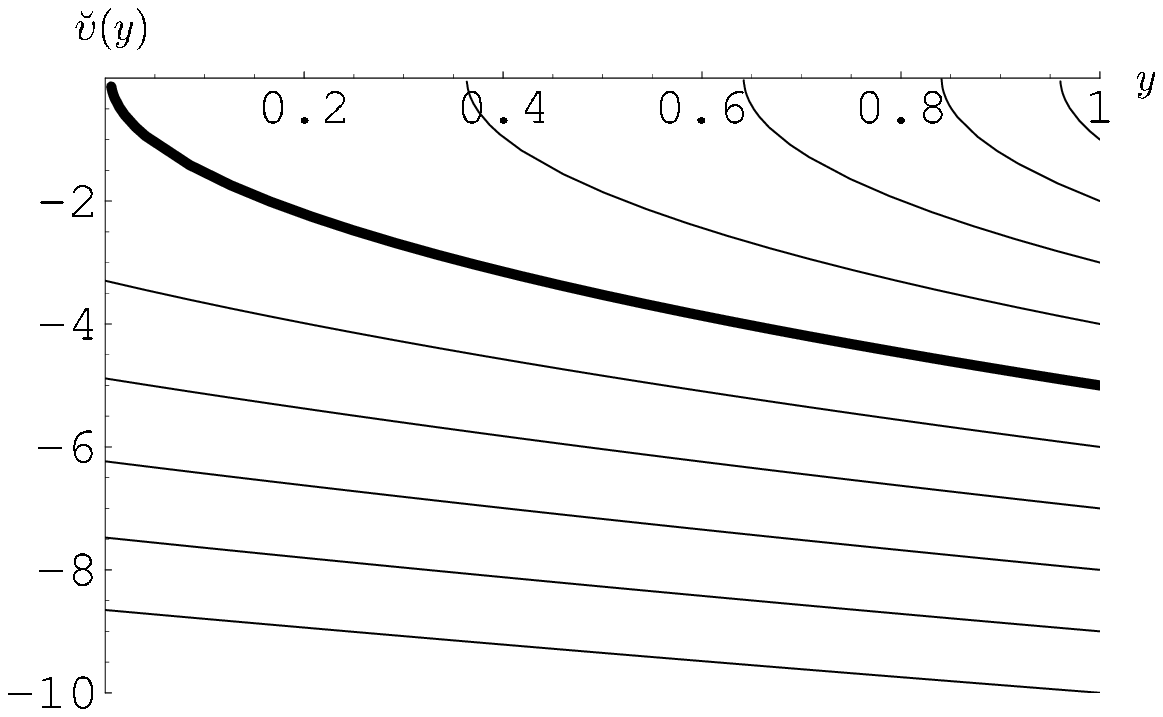}
\end{center}
\parbox[c]{\textwidth}{\caption{\label{6.eins}{\footnotesize Graphical illustration of $\uh(y)$ given by eq.  \rf{6.16} with selected positive and negative initial values $\uh(\yh)$ in the left and right diagram, respectively. The trajectories of Class II starting with $|\uh(\yh)| \ge |\uh(\yh)_{\rm crit}|$ can be continued to $y=0$. Those of Class I with $|\uh(\yh)| < |\uh(\yh)_{\rm crit}|$ run into the singularity at $\uh = 0$ and terminate at finite values $y_{\rm term}>0$. The bold line indicates the trajectory separating the trajectories of Class I and II.}}}
\end{figure}
Here we see that for all trajectories $|\uh(y)|$ is bounded above by its initial value $|\uh(\yh)|$. Figure \ref{6.eins} further shows that the trajectories of the $V$+$\sqrt{V}$--truncation can be separated into two classes. The ones starting below a critical value, $|\uh(\yh)| < |\uh(\yh)_{\rm crit}| \equiv \sqrt{8 \pi \yh}$, run into the singularity at $\uh = 0$ and terminate at    
\be\label{6.18}
y_{\rm term} = \yh - \frac{\uh^2_{\yh}}{8 \pi}.
\ee
These trajectories will be labeled as trajectories of Class I. The trajectories starting with initial values $|\uh(\yh)| \ge |\uh(\yh)_{\rm crit}|$ can all be continued down to $y=0$. These will be labeled as trajectories of Class II. Depending on their initial value $\uh(\yh)$ these trajectories can end at any IR-value $\uh(0)$.  

Let us now discuss the RG evolution of the cosmological constant. Figure \ref{6.zwei} shows $\lh(y)$ along typical trajectories of Class I, starting with $\uh(\yh) = 4$ and various positive and negative initial values $\lh(\yh)$. 
\begin{figure}[t]
\renewcommand{\baselinestretch}{1}
\epsfxsize=0.60\textwidth
\begin{center}
\leavevmode
\epsffile{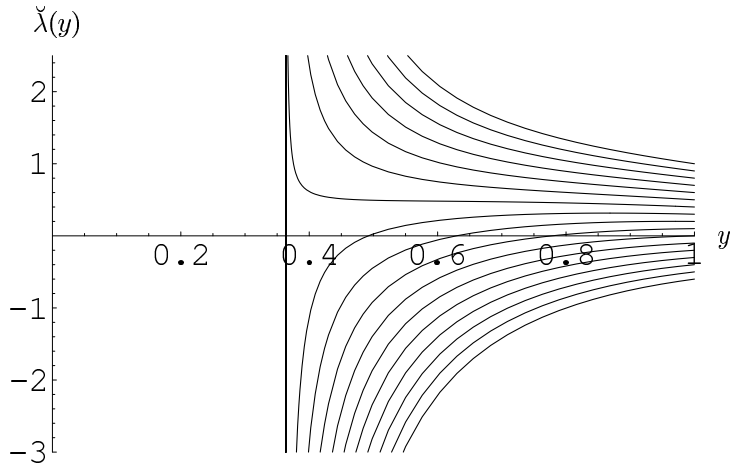}
\end{center}
\parbox[c]{\textwidth}{\caption{\label{6.zwei}{\footnotesize Graphical illustration of $\lh(y)$ for typical Class I trajectories, starting from $\uh(\yh) = 4$ and various positive and negative values $\lh(\yh)$. At $y = y_{\rm term}$, $\lh(y)$ diverges independently of the initial value $\lh(\yh)$, causing the termination of the RG trajectory at a finite $y_{\rm term} > 0$.}}}
\end{figure}
Here we find that, as the trajectories approach $y_{\rm term}$ where $\uh(y_{\rm term})=0$, the values of $\lh(y)$ diverge. This divergence causes the termination of the trajectory at the finite scale $y_{\rm term}$ because $|\lh(y_{\rm term})| = \infty$ does not allow us to continue the evolution of $\lb_k$ with eq. \rf{6.11a}.

\begin{figure}[t]
\renewcommand{\baselinestretch}{1}
\epsfxsize=0.80\textwidth
\begin{center}
\leavevmode
\epsffile{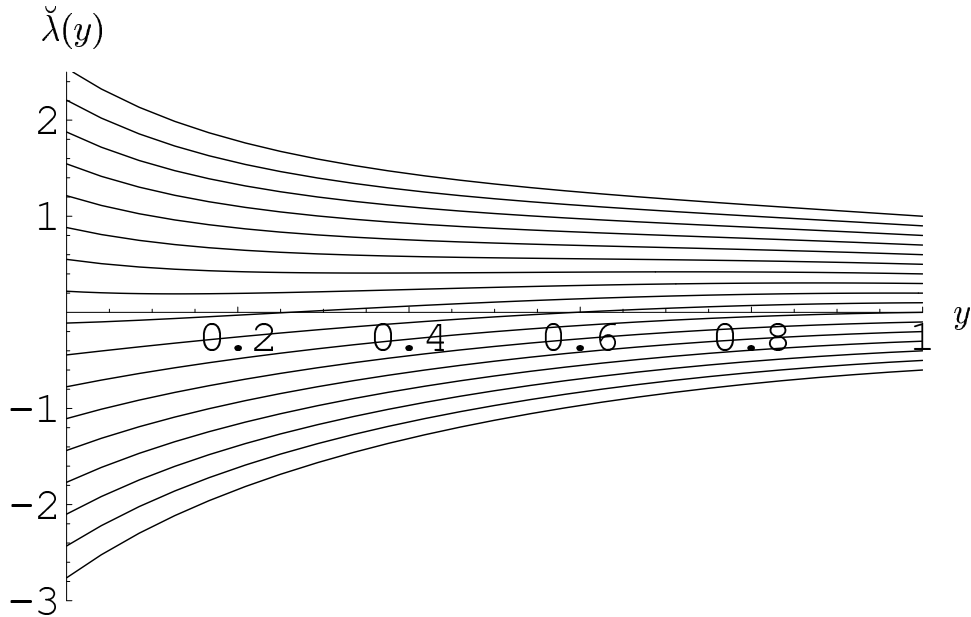}
\end{center}
\parbox[c]{\textwidth}{\caption{\label{6.drei}{\footnotesize Graphical illustration of $\lh(y)$ given by eq. \rf{6.17} for typical Class II trajectories, starting from $\uh(\yh) = 6$ and various positive and negative values $\lh(\yh)$. All trajectories can be continued to $y=0$. Depending on the initial value $\lh(\yh)$ they yield positive, vanishing or negative IR-values $\lh(0)$. There is no boundary singularity preventing the trajectories from reaching positive IR-values $\lh(0) > 0$.}}}
\end{figure}
Figure \ref{6.drei} shows a typical set of trajectories of Class II starting with $|\uh(\yh)| > |\uh(\yh)_{\rm crit}|$. In the plot we have chosen $|\uh(\yh)| = 6$ and various positive and negative initial values $\lh(\yh=1)$. Figure \ref{6.drei} illustrates that all trajectories of this class can be continued down to $y=0$. The important new feature found here is that, depending on the initial value $\lh(\yh)$ of the trajectory, we find negative, zero, as well as positive IR-values for the cosmological constant. 

Probably the terminating Class I trajectories are not described reliably by the truncation, at least not close to $y_{\rm term}$ where $\uh$ approaches zero. On the other hand, the trajectories of Class II which stay away from the problematic neighborhood of $\uh = 0$ have a chance of being realistic. The $V$+$\sqrt{V}$--truncation is interesting from the point of view that, contrary to the other 2-parameter truncations, it has no problems in producing positive renormalized cosmological constants $\lh(0)$.

The fixed points \rf{5.11} are characterized by a vanishing cosmological constant and an arbitrary (nonzero) prefactor of $V^{1/2}$. Since, in the $V$+$\sqrt{V}$--truncation, $\lh(0)$ can have any value it is obvious that in this truncation the trajectories are not attracted towards the projection of the MGFP onto the truncation subspace. We do not know whether this reflects a genuine property of the MGFP, or whether it is due to our 2-dimensional approximation of $\cf_k$--space.
\end{subsection}

\mysection{Discussion and Conclusions}
In this paper we used the exact RG equation of Quantum Einstein Gravity \cite{ERGE} in order to study the scale dependence of nonlocal effective actions of the form $\cf_k(V)$ where $V$ is the Euclidean space-time volume. Such investigations are both interesting in their own right and they are important for an understanding of quantum gravity at large distances. One of the physical motivations of the present work is the conjecture that strong IR quantum effects might provide a solution to the cosmological constant problem. The Einstein-Hilbert truncation is too simple, however, to encapsulate such effects; in order to obtain a small value of the renormalized cosmological constant one has to fine-tune the initial point of the RG trajectory to be extremely close to the separatrix.

In Section III we presented a detailed investigation of the RG flow for an $\cf_k$ of the form $V$+$V \ln V$, and in the Appendix we performed a similar analysis for $V$+$V^2$. These specific choices were motivated by the fact that these actions had been discussed before in the context of wormhole physics. Nevertheless our investigation has nothing to do with wormholes directly; it is supposed to apply at length scales much larger than the wormhole size where an effective description is possible \cite{TVI,coleman,TVII}. The wormholes might be needed to provide the ``seeds'' of the nonlocalities, though.

The classification of the RG trajectories resulting from the new nonlocal invariants leads to almost the same classes of trajectories as in the Einstein-Hilbert truncation. We also found that the $V \ln(V/V_0)$ and $V^2$-terms do in general not prevent the termination of the trajectories of Type IIIa in the boundary singularity. A new feature in the RG flow are the trajectories of Type VIa which appeared in the case of the $V$+$V \ln V$--truncation. They yield a {\it positive} value of the renormalized cosmological constant $\lb_0$ which could not happen in the Einstein-Hilbert truncation. The existence of these trajectories is due to a modification of the boundary singularity where, in the case of the $V$+$V \ln V$--truncation, positive values of $\lb_0$ can be compensated by a negative coupling $\bar{u}$. 

A rather impressive and potentially very important property of the nonlocal actions becomes obvious if one looks for their (maximally symmetric) stationary points. We considered $S^4$-type solutions of the modified Einstein equations and determined their radius as a function of the parameters in $\Gamma_k$. It is remarkable that there exist solutions whose curvature is not, as usual, proportional to $\lb$ but rather is the smaller the {\it larger} the absolute value of the (proper) cosmological constant $\lb$ is. 

We used our results for the running of $\lb_k$ and the nonlocality parameters in order to determine the impact the leading renormalization effects have on the radius. In the case of the $V \ln V$--truncation we found that by including the RG evolution of the parameters between $k = m_{\rm Pl}$ and $k=0$, the $S^4$ resulting from generic Planck-size initial values (fixed at $k = m_{\rm Pl}$) has a radius which is many orders of magnitude larger than it would be classically. This ``inflation'' due to the RG running of the couplings helps in understanding how Planck-size bare parameters in an effective action valid at $k = m_{\rm Pl}$ can give rise to large and almost flat universes. 

Our most important results concerning this mechanism are displayed in Figs. \ref{4.zwei} and \ref{4.drei}. They show that an arbitrary trajectory of Type Ia with a negative $\lb_0$ of the order of $m_{\rm Pl}^2$ leads to macroscopic $S^4$--solutions if a small coupling constant $0 < \bar{u} \le 10^{-3} m_{\rm Pl}^2$ is included. From these solutions we obtain a very tiny, positive IR value of the {\it effective} cosmological constant which could be in agreement with the experimental bounds. It is this effective cosmological constant which is responsible for the curvature of space-time. Since it is extremely small for an entire class of trajectories, this mechanism does not need a fine-tuning of the initial conditions. Hence this ``RG improved Taylor-Veneziano mechanism'' provides a very promising approach to explain a small positive cosmological constant in a natural way.

We found that the ``inflation'' of Planck length universes to a macroscopic size is not a general property of all $\cf_k(V)$-truncations. While the qualitative features of the $V$+$V \ln V$- and the $V$+$V^2$--truncations are quite similar, the ``magnification factor'' $r(k=0)/r(k = \mp)$ one can achieve with the latter is not much larger than unity.

Another scenario one could think of and which also would solve the fine-tuning problem of the cosmological constant is to assume that, in the IR, $\Gamma_k$ is attracted by a fixed point $\Gamma^*$ which allows the space-time to become (almost) flat. Even within our simple class of truncations we found a first hint showing that a mechanism of this sort is indeed possible in principle. We saw that in this truncation subspace the RG flow has a line of fixed points of the type $\int \, d^4x \, \sqrt{g} \, R + c \, V^{1/2}$. The associated modified Einstein equations are scale invariant and the curvature of their solutions is completely unrelated to the parameters in the bare action.

Motivated by the structure of the modified GFP with its characteristic $\sqrt{V}$--dependence we determined in Section VI the 2-dimensional RG flow of the $V$+$\sqrt{V}$--truncation. It does not suffer from the notorious boundary singularity at $\lambda = 1/2$ and has no problems in achieving positive renormalized cosmological constants.

While these results are already very encouraging it is clear that they cannot yet be used in order to construct realistic cosmologies. The invariants $\cf_k(V)$ make sense only for Euclidean space-times and not for the Lorentzian ones of the Robertson-Walker type, say.\footnote{However, it could perhaps be possible to compare our results to simulations of Euclidean simplicial quantum gravity \cite{amb-book,amb-rev}.} In the future one of the main tasks will be to extend the RG analysis to nonlocal invariants which have a Lorentzian interpretation, and then to check whether the resulting modifications of general relativity and cosmology are phenomenologically acceptable. This includes questions of causality, and in particular the need of making the nonlocal effects at large distances compatible with the experimental bounds. The general framework of the effective average action and its RG equation should work also in the Lorentzian case, albeit applied to a different type of ``theory space'' \cite{ERGE}.

\vspace{1.2cm}
\noindent
Acknowledgement: We would like to thank A.Bonanno, W.Dittrich, H.Gies, O.Lauscher, D.Litim, R.Percacci, G.Veneziano and C.Wetterich for helpful discussions.
 
\begin{appendix}
\mysection{The $V$+$V^2$--truncation}
\renewcommand{\theequation}{\Alph{section}.\arabic{equation}}
In this Appendix we compare the $V$+$V^2$--truncation to the $V$+$V \ln V$--truncation discussed in the main part of the paper. We start by stating the fixed point properties. Then we briefly summarize the properties of the RG flow found by studying the numerical solutions of the flow equation. We then determine the dependence of the radius of $S^4$--solutions on the running coupling constants along trajectories of Type Ia and IIa. As in the case of the $V$+$V \ln V$--truncation, we employ the sharp cutoff with shape parameter $s = 1$ in all calculations.

\begin{subsection}{The Gaussian fixed point}
The starting point of our investigation of the $V$+$V^2$--truncation is the differential equation \rf{2.33}. This equation describes the RG flow of the dimensionful coupling constants $\lb_k$ and $\wb_k$ associated to the invariants $V$ and $V^2$ in an effective theory of gravity below the Planck scale. 

Introducing the dimensionless coupling constants $g(k) \equiv k^{d-2} G, \lambda(k) \equiv k^{-2} \lb_k$ and $w(k) \equiv k^{-(d+2)} \wb_k$ we write eq. \rf{2.33} in an autonomous way,
\be\label{A.1}
\partial_t \, g = \Fbeta_g(\lambda, g, w), \qquad \partial_t \, \lambda = \Fbeta_\lambda(\lambda, g, w), \qquad \partial_t \, w = \Fbeta_w(\lambda, g, w),
\ee
where the $\Fbeta$-functions are given by
\bea\label{A.2}
\nonumber \Fbeta_g(\lambda, g, w) &=& (d-2) \, g \\[1.5ex]
\nonumber \Fbeta_\lambda(\lambda, g, w) &=& -2 \lambda + 
(4 \pi)^{1-d/2}\,g\, \left\{ d \, (d+1) \, \p{1}{d/2}{-2 \lambda}{} - 4 \, d \, \p{1}{d/2}{0}{} \right\} \\
\nonumber && + 16 \pi \frac{d}{d-2} \, g \, w \, \frac{1}{1 - 2 \lambda} \\[1.5ex]
\nonumber \Fbeta_w(\lambda, g, w) &=& -(d+2) w + (4 \pi)^{1-d/2} \, 4 \, d \, (d+1) \, g \, w \, \p{2}{d/2}{-2 \lambda}{} \\
&& + 128 \pi \frac{d(3d-4)}{(d-2)^2} \, g \, w^2 \, \frac{1}{(1 - 2 \lambda)^3}
\eea

The only solution to the fixed point equation $\Fbeta_i(g^*, \lambda^*, w^*) = 0 \;\; \forall \, i \, \in \{ g, \lambda, w \}$ is the Gaussian fixed point $g^* = 0, \lambda^* = 0, w^* = 0$. From the pertinent stability matrix ${\bf B}_{ij} = \partial_j \Fbeta_i$, $i,j \, \in  \{g, \lambda, w \}$, we find the following stability coefficients (critical indices) and eigenvectors satisfying ${\bf B} V^I = - \theta_I V^I$:
\bea\label{A.5}
\nonumber \theta_1 = +2\qquad &\mbox{with}& \qquad V^1 = \left(  1, \quad 0, \quad 0 \right)^{\sf T} \\[1.5ex]
\nonumber \theta_2 = -(d-2) \qquad &\mbox{with}& \qquad V^2 = \left( (4 \pi)^{1-d/2} \, (d-3) \, \p{1}{d/2}{0}{}, \quad 1, \quad 0 \right)^{\sf T} \\[1.5ex]
\theta_3 = d + 2 \qquad &\mbox{with}& \qquad V^3 = \left( 0, \quad 0,  \quad 1 \right)^{\sf T} 
\eea

These stability coefficients and eigenvalues can be used to write down the linearized RG flow of the dimensionful coupling constants in the vicinity of the GFP: 
\bea\label{A.6}
\nonumber G_k &=& G \\
\nonumber \lb_k &=& \lb_0 + (4 \pi)^{1-d/2} \, (d-3) \, G \, k^d \, \p{1}{d/2}{0}{} \\
\w_k &=& \w_0 
\eea
As in the case of the $V$+$V \ln V$--truncation the coupling constants take on constant but in general non-zero values as $k \rightarrow 0$. The numerical values $G, \lb_0$ and $\wb_0$  depend on the RG trajectory chosen and are not determined by the properties of the fixed point. Hence the GFP in the $V$+$V^2$--truncation, too, does not provide a solution to the cosmological constant problem by achieving $\lb_0 = 0$ automatically.
\end{subsection}
\begin{subsection}{The RG Flow}
Let us now illustrate the different classes of solutions found by numerically solving the flow equation. Introducing the dimensionless scale $y \equiv \kh^2 \equiv k^2/m_{\rm Pl}^2$ and setting $d=4$, eq. \rf{2.33} takes the form:
\bea\label{A.7}
\nonumber \frac{d \, \lh(y)}{d \, y} &=& \frac{1}{2 \pi} \, y \, \left\{ -5 \ln(1-2\lh(y)/y) + \varphi_2 \right\} + 16 \pi \, \frac{\wh(y)}{(y - 2 \lh(y))^2} \\[1.5ex] \label{A.7a}
\frac{d \, \wh(y)}{d \, y} &=& \frac{10}{\pi} \, y \, \frac{\wh(y)}{(y - 2 \lh(y))} + 512 \pi \, \frac{\wh(y)^2}{(y - 2 \lh(y))^3}
\eea
For the sharp cutoff with $s=1$ the constant $\varphi_2$ has the value $\varphi_2 \equiv 2 \zeta(3)$. In order to systematically investigate the properties of the RG flow we first analyze the approximation of \rf{A.7} arising from setting $\lh = \mbox{const}$. This leads to the following decoupled flow equation for $\wh(y)$:
\be\label{A.8}
\frac{d \, \wh(y)}{d \, y} = \frac{10}{\pi} \, y \, \frac{\wh(y)}{(y - 2 \lh)} + 512 \pi \, \frac{\wh(y)^2}{(y - 2 \lh)^3}
\ee
\begin{figure}[t]
\renewcommand{\baselinestretch}{1}
\epsfxsize=0.60\textwidth
\begin{center}
\leavevmode
\epsffile{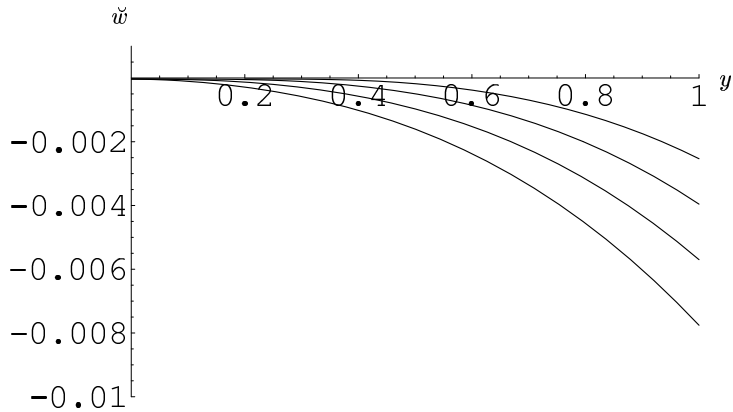}
\end{center}
\parbox[c]{\textwidth}{\caption{\label{A.eins}{\footnotesize Functions $\wh_1(y) \equiv 0$ and $\wh_2(y)$ along which the RHS of eq. \rf{A.8} vanishes. The parameter $\lh$ is chosen as $\lh = 0.1, 0, -0.1$ and $-0.2$, respectively. These curves separate regions $\wh(y) > \wh_1(y), \wh_1(y) \ge \wh(y) \ge \wh_2(y)$, and $\wh(y) < \wh_2(y)$, where the RHS of \rf{A.8} is positive, negative and positive, respectively. Decreasing values of $\lh$ thereby lead to decreasing values $\wh_2(y)$.}}}
\end{figure}

Due to the quadratic nature of its RHS eq. \rf{A.8} gives rise to two curves $y \mapsto \wh_1(y) \equiv 0$ and $y \mapsto \wh_2(y) < 0$ on which $\frac{d \wh(y)}{d y}$ vanishes. We note that $y \mapsto \wh_2(y)$ is no trajectory arising as a solution of \rf{A.8}. For the parameters $\lh = 0.1, 0, -0.1$, and $-0.2$ these curves are shown in Fig. \ref{A.eins}. They divide the $\wh$-$y$--plane into three regions, $\wh(y) > \wh_1(y)$, $\wh_1(y) \ge \wh(y) \ge \wh_2(y)$, and $\wh(y) < \wh_2(y)$ in which the RHS of \rf{A.8} is positive, negative, and positive, respectively.

To illustrate the properties of $\wh(y)$ in these regions we solve eq. \rf{A.8} numerically, choosing $\lh = -0.1$ and various positive and negative initial values $\wh(\yh)$ given at $\yh = 1$. The resulting trajectories are shown in Fig. \ref{A.zwei}.
\begin{figure}[t]
\renewcommand{\baselinestretch}{1}
\epsfxsize=0.60\textwidth
\begin{center}
\leavevmode
\epsffile{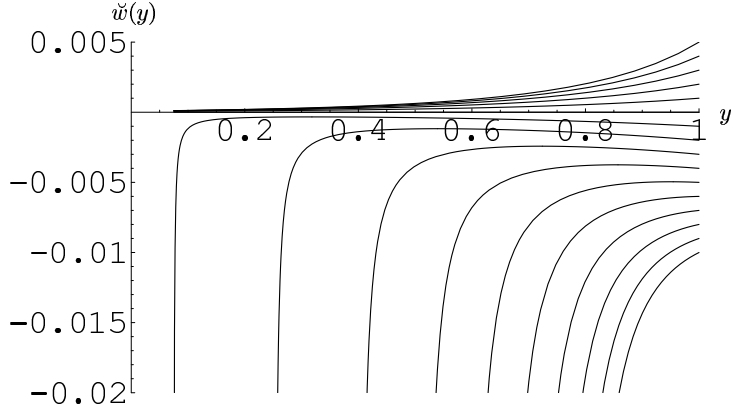}
\end{center}
\parbox[c]{\textwidth}{\caption{\label{A.zwei}{\footnotesize Numerical solutions of the approximate flow equation \rf{A.8} with $\lh = -0.1$ and various positive and negative initial values $\wh(\yh)$. The trajectories in the region $\wh(\yh) > 0$ are stable and yield $\wh(y) < \wh(\yh)$ while these starting with $\wh(\yh) < 0$ generally become unstable when reaching $\wh(y) < \wh_2(y)$. For these trajectories we find $\wh(y) \rightarrow -\infty$ at a finite value $y = y_{\rm term} > 0$.}}}
\end{figure}
This figure illustrates the following properties:
\begin{itemize}
\item In the region $\wh(y) > \wh_1(y)$ where the RHS of eq. \rf{A.8} is positive, $\wh(y)$ is well behaved and monotonically decreasing for decreasing $y$. All trajectories in this region can be continued to $y=0$.
\item For $\wh(\yh) = \wh_1(\yh) \equiv 0$ the RHS of eq. \rf{A.8} vanishes identically for all $y \le \yh$, i.e. the trajectory starting with $\wh(\yh) = 0$ has $\wh(y) = 0$ for all values $y \le \yh$. Hence $\wh = 0$ is a stability axis of the RG flow leading to a separation between the RG trajectories starting with positive and negative $\wh(\yh)$, respectively.
\item Trajectories starting in the region $\wh_1(\yh) > \wh(\yh) > \wh_2(\yh)$ in which the RHS of eq. \rf{A.8} is positive yield decreasing values $|\wh(y)| < |\wh(\yh)|$ for $y < \yh$ up to the point where the trajectories reach $\wh(y) = \wh_2(y) \equiv 0$.
\item For the trajectories in the region $\wh(y) < \wh_2(y)$ we find that $\wh(y)$ is rapidly decreasing. These trajectories terminate at a finite $y_{\rm term} > 0$, with $\wh(y \rightarrow y_{\rm term}) \rightarrow -\infty$.
\end{itemize}

Hence we see that for the trajectories in the region $\wh(y) > 0$ the coupling $\wh(y)$ is ``stable'' and the corresponding trajectories lead to well defined values $\wh(0) > 0$ while in general the trajectories in the negative coupling region run into the region $\wh(y) < \wh_2(y)$ where $\wh(y)$ becomes ``unstable'' and terminates at $y_{\rm term} > 0$.
\begin{figure}[t]
\renewcommand{\baselinestretch}{1}
\epsfxsize=0.49\textwidth
\begin{center}
\leavevmode
\epsffile{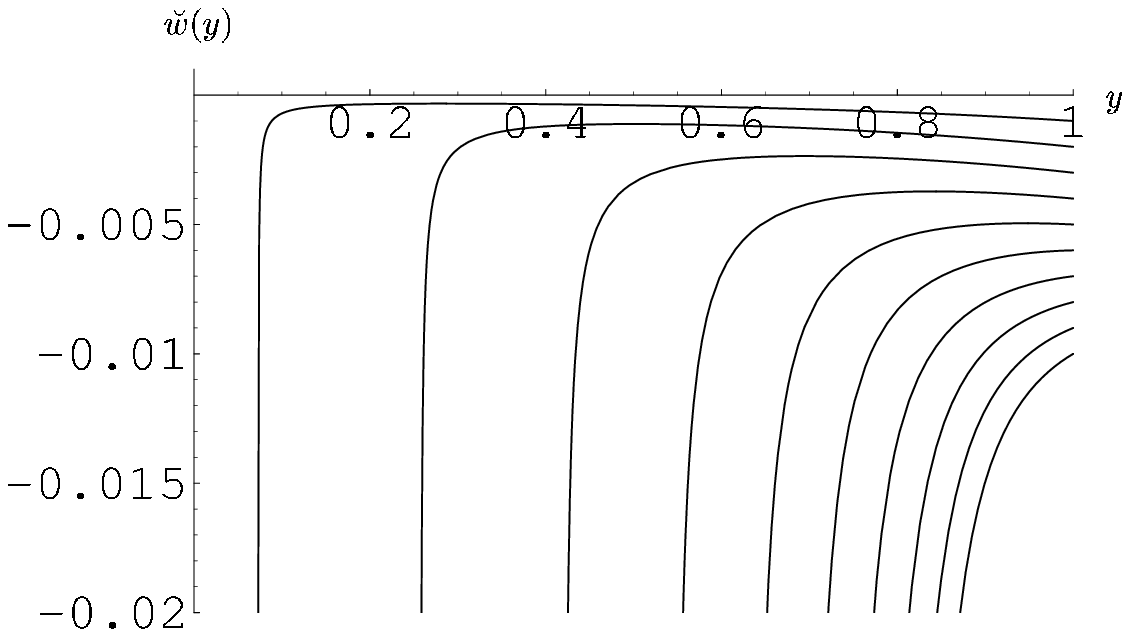}
\epsfxsize=0.48\textwidth
\epsffile{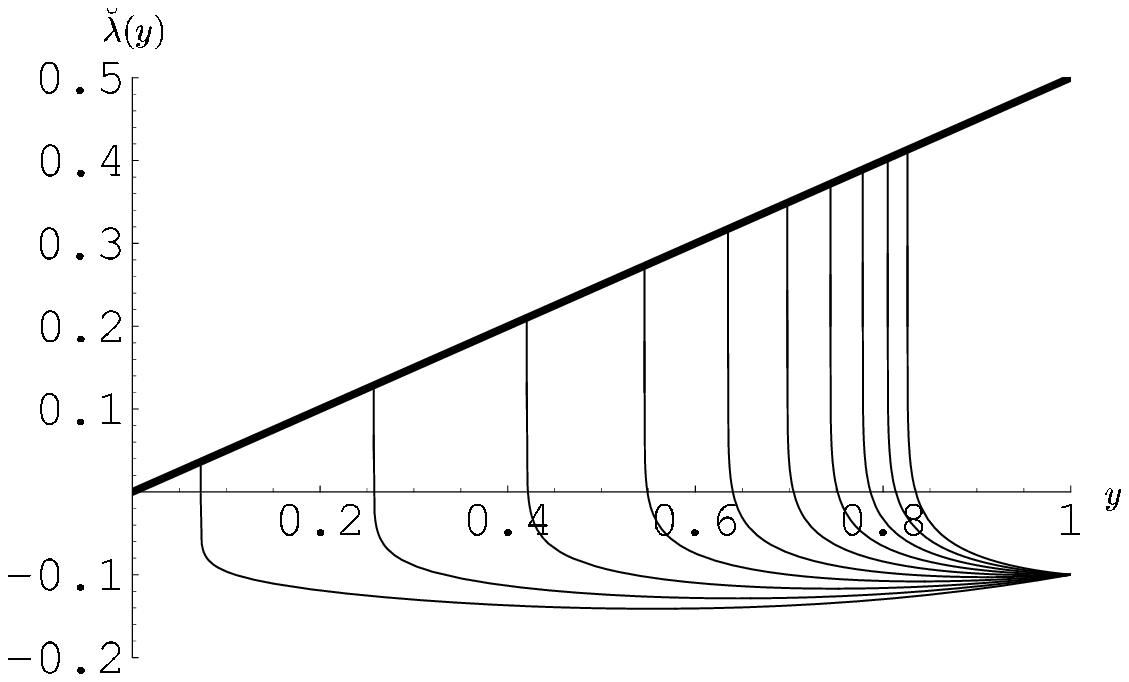}
\end{center}
\parbox[c]{\textwidth}{\caption{\label{A.drei}{\footnotesize Numerical solutions of the full flow equation in the region $\wh(\hat{y}) < 0$, illustrated by trajectories starting with $\lh(\yh) = -0.1$ and various negative values $\wh(\yh) < 0$. In the region $\wh(y) < \wh_2(y, \lh(y))$ all trajectories lead to diverging values $\wh(y \rightarrow y_{\rm term}) \rightarrow -\infty, \lh( y_{\rm term} ) = y_{\rm term}/2 $. The bold line in the second diagram illustrates the boundary singularity $\lh = y/2$.}}}
\end{figure}

Let us now return to the exact equation \rf{A.7a}. We observe that, unlike in the case of the $V$+$V \ln V$--truncation, including the nonlocal term does not lead to a modification of the boundary of coupling constant space. It is located at $\lh = y/2$. Furthermore we find that $\wh = 0$ still is a stability plane in $\wh$-$\lh$-$y$--space which separates the trajectories in the regions $\wh > 0$ and $\wh < 0$. An important consequence of this separation is that, if a trajectory starts out with a zero coupling  $\wh(\yh) = 0$, the RG flow of the $V$+$V^2$--truncation will not lead to the dynamical generation of this coupling. We will use the separation of the coupling constant space to investigate the properties of the trajectories in these regions separately.

As typical trajectories of the region $\wh(\yh) < 0$ we consider solutions starting with $\lh(\yh) = -0.1$ and various negative values $\wh(\yh)$. The resulting trajectories are shown in Fig. \ref{A.drei}.

These trajectories reflect the behavior already seen in the case of the decoupled flow eq. \rf{A.8}. As long as $\wh_1 \left(y, \lh(y) \right) > \wh(y) > \wh_2 \left(y, \lh(y) \right)$, where now the zeros of the RHS of \rf{A.7a} also depend on $\lh(y)$ along the trajectory, $\wh(y)$ is stable and bounded, $|\wh(y)|<|\wh(\yh)|$. Leaving this region of stability we again find that $\wh(y)$ rapidly diverges, $\wh(y) \rightarrow -\infty$. In this course the cosmological constant $\lh(y)$ is driven into the boundary of the coupling constant space, $\lh(y) \rightarrow y/2$, so that the corresponding trajectories terminate at a finite value $y_{\rm term} > 0$.

Regarding the RG flow in the positive coupling region $\wh \ge 0$, we find that the flow equation \rf{A.7} gives rise to trajectories of the Types Ia, IIa and IIIa. The trajectories of the new Type VIa found in the $V$+$V \ln V$--truncation are absent due to the unaltered boundary singularity $\lh = y/2$ which prevents trajectories from reaching positive values $\lh(0) > 0$.

Typical trajectories of the Types Ia and IIa are displayed in Figs. \ref{A.vier} and \ref{A.funf}. 
\begin{figure}[t]
\renewcommand{\baselinestretch}{1}
\epsfxsize=0.49\textwidth
\begin{center}
\leavevmode
\epsffile{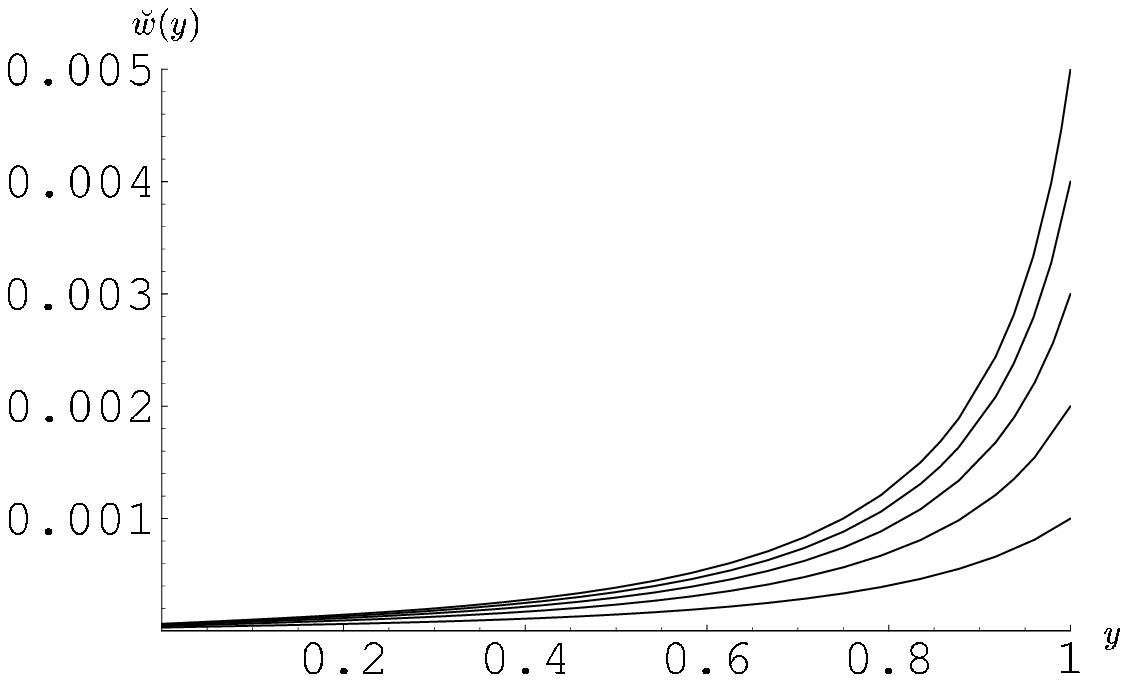}
\epsfxsize=0.48\textwidth
\epsffile{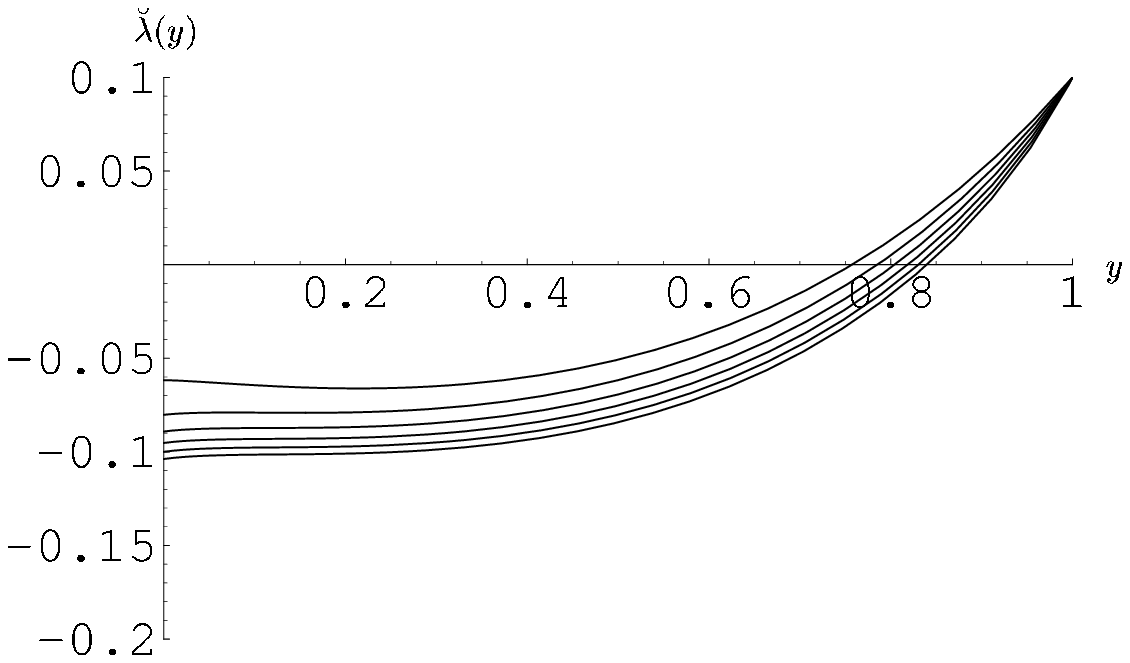}
\end{center}
\parbox[c]{\textwidth}{\caption{\label{A.vier}{\footnotesize Typical Type Ia-solutions of the full flow equation starting with $\lh(\yh) = 0.1$ and various positive values $\wh(\yh) > 0$. For the trajectories starting with $\wh(\yh) = 0$ $(\wh(\yh) = 0.005)$ we obtain the largest (smallest) IR value of the cosmological constant, i.e. a {\it positive} initial value $\wh(\yh)$ drives $\lh(0)$ away from zero.}}}
\end{figure}

Figure \ref{A.vier} shows the impact of a positive initial value $\wh(\yh)$ on $\lh(y)$ along some typical trajectories starting with $\lh(\yh) = 0.1$ and various values $\wh(\yh) > 0$. Here we find the following properties:
\begin{itemize}
\item For all trajectories shown, $\wh(y)$ decreases with decreasing $y$. All trajectories starting with $\wh(\yh) > 0$ lead to non-vanishing IR values $\wh(0) > 0$.
\item Compared to the trajectory starting at $\wh(\yh) = 0$, a nonzero initial value $\wh(\yh)$ leads to smaller values $\lh(y)$, i.e. {\it positive} values $\wh(\yh)$ drive the cosmological constant away from zero.
\item A coupling $\wh(\yh) > 0$ changes the slope $\frac{d \lh(y)}{d y}$ when we approach the IR ($y \rightarrow 0$). The trajectory with zero coupling $\wh(\yh)$ curves upward, $\frac{d \lh(y)}{d y} < 0$, while for the trajectories with $\wh(\yh) > 0$ we find positive slopes $\frac{d \lh(y)}{d y} > 0$. 
\end{itemize}
\begin{figure}[t]
\renewcommand{\baselinestretch}{1}
\epsfxsize=0.49\textwidth
\begin{center}
\leavevmode
\epsffile{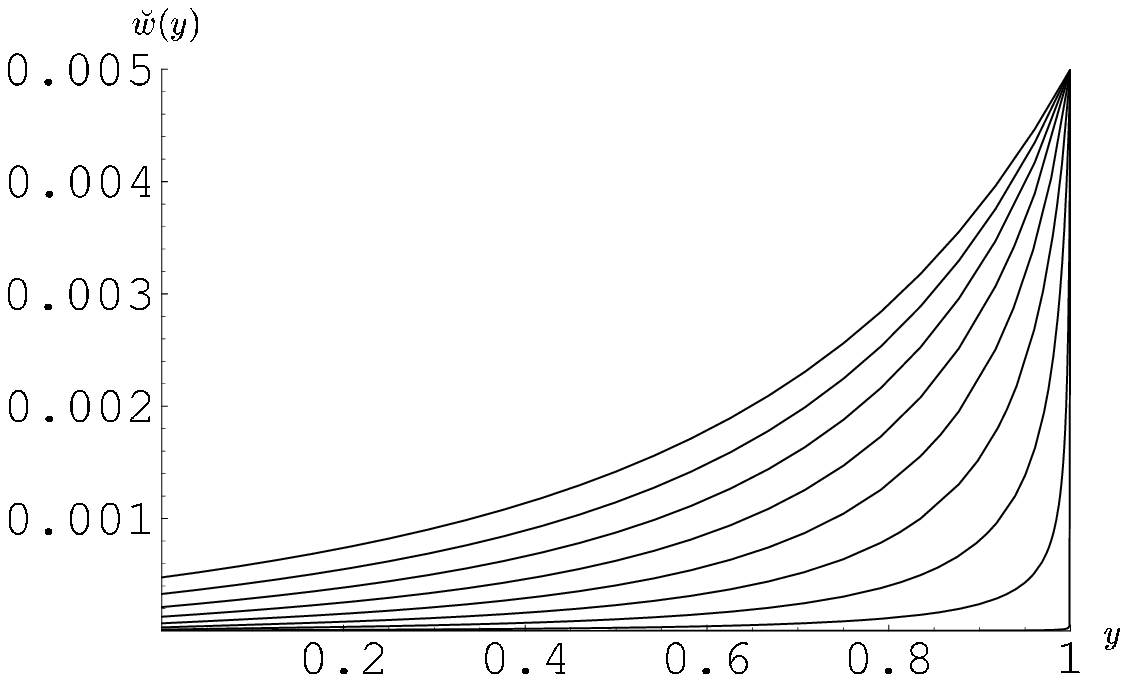}
\epsfxsize=0.48\textwidth
\epsffile{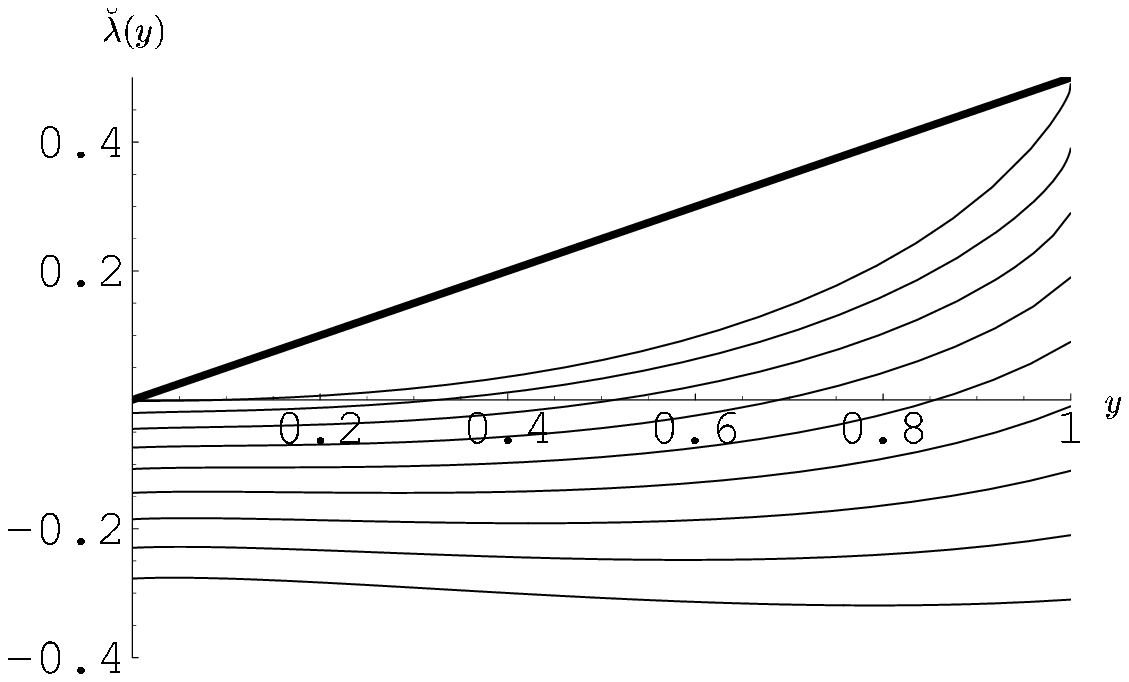}
\end{center}
\parbox[c]{\textwidth}{\caption{\label{A.funf}{\footnotesize Numerical solutions of the full flow equation starting with $\wh(\yh) = 0.005$ and various values $\lh(\yh)$. Trajectories starting close to the boundary $\lh = y/2$ yield a rapid decrease of $\wh(y)$ at $y \approx \yh$. Nevertheless all trajectories displayed lead to non-vanishing values $\wh(0) > 0$. The bold straight line in the second diagram indicates the boundary $\lh = y/2$.}}}
\end{figure}

Figure \ref{A.funf} shows the impact of different initial values $\lh(\yh)$ on the flow of $\wh(y)$. As an example we use the trajectories starting with $\wh(\yh) = 0.005$ and various values $\lh(\yh)$. We find that larger initial values $\lh(\yh)$ lead to smaller, but non-vanishing, IR values $\wh(0)$. Even the rapid decrease of $\wh(y)$ at $y \lesssim \yh$ found for the trajectory starting close to the boundary $\lh = y/2$ does not result in $\wh(0) = 0$.

The properties of the trajectories of Type IIIa are illustrated in the Figs. \ref{A.sechs} and \ref{A.sieben}. As typical trajectories we choose the solutions starting at $\lh(\yh) = 0.2$ and various values $\wh(\yh) > 0$ specified at $\yh = 0.5$.

\begin{figure}[t]
\renewcommand{\baselinestretch}{1}
\epsfxsize=0.49\textwidth
\begin{center}
\leavevmode
\epsffile{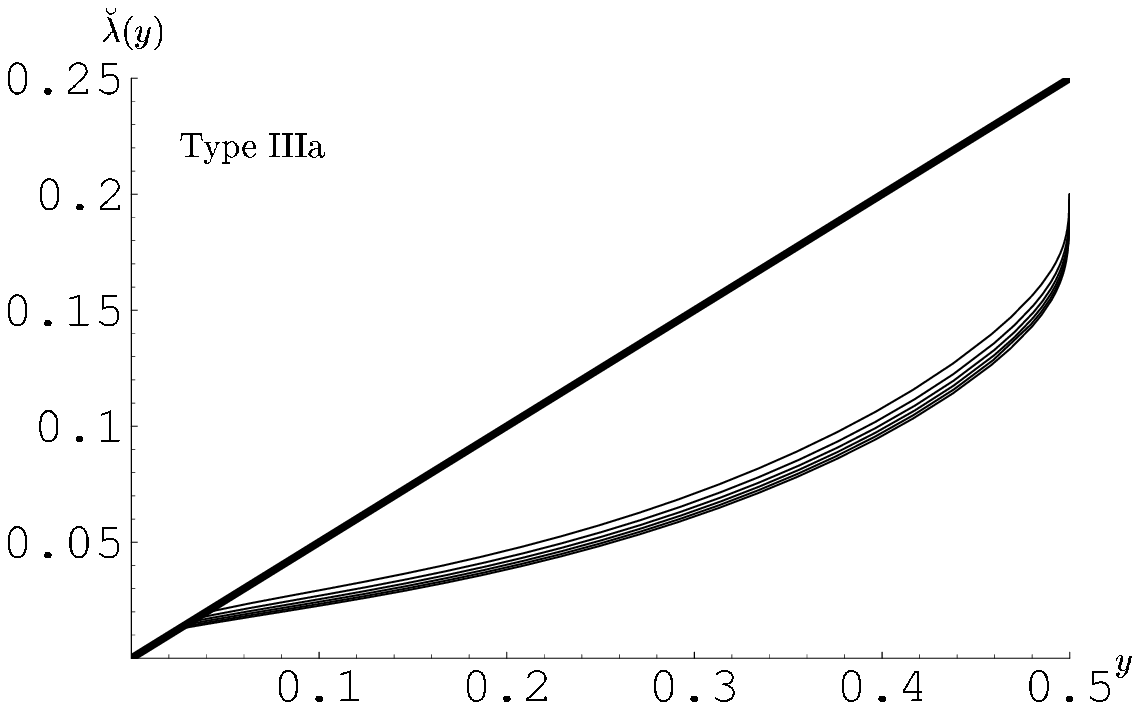}
\epsfxsize=0.48\textwidth
\epsffile{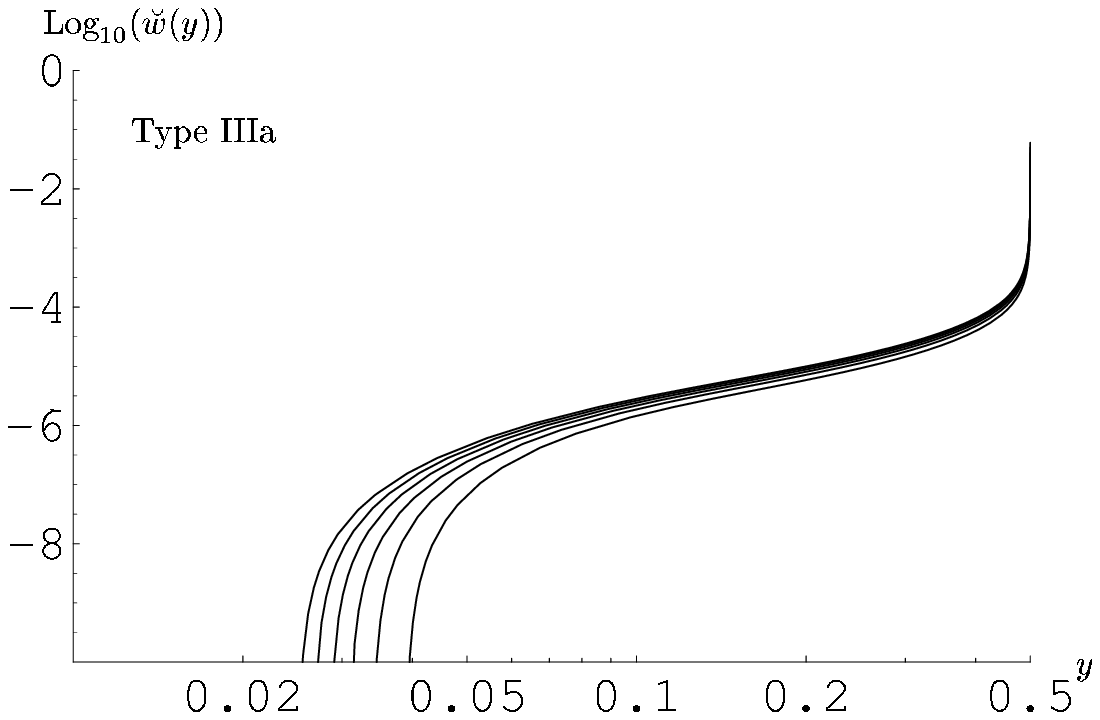}
\end{center}
\parbox[c]{\textwidth}{\caption{\label{A.sechs}{\footnotesize Typical Type IIIa trajectories with initial values $\lh(\yh) = 0.2$ and various positive values $\wh(\yh)$ given at the scale $\yh = 0.5$ The bold straight line indicates the boundary singularity $\lh = y/2$. Increasing values of $\wh(\yh)>0$ lead to smaller values $\lh(y)$ and hence to a decreasing value of $\y_{\rm term}$.}}}
\end{figure}

We see that, as in the case of the $V$+$V \ln V$--truncation, the new coupling $\wh(\yh) > 0$ does not prevent the trajectories from running into the boundary singularity $\lh = y/2$; hence they terminate at $y_{\rm term} > 0$. As in the $V$+$V \ln V$--truncation, the new coupling vanishes identically as the trajectory approaches the boundary, $\wh(y_{\rm term}) = 0$, and therefore has no ``healing effect'' on the running of $\lh(y)$.
\begin{figure}[th]
\renewcommand{\baselinestretch}{1}
\epsfxsize=0.65\textwidth
\begin{center}
\leavevmode
\epsffile{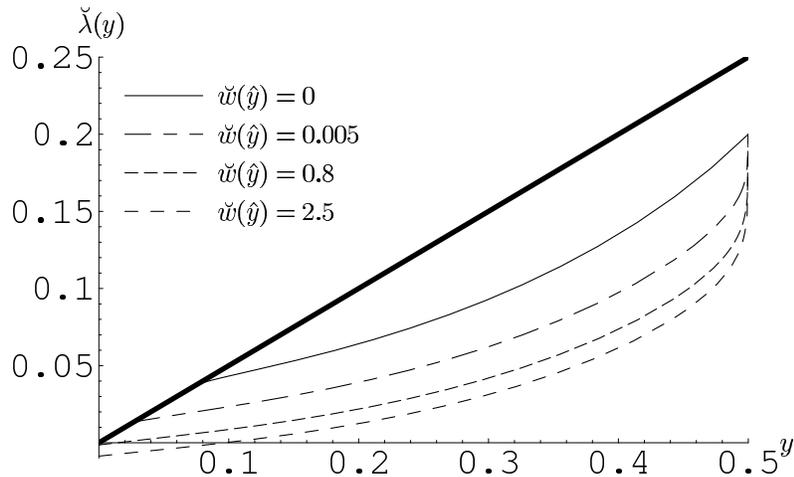}
\end{center}
\parbox[c]{\textwidth}{\caption{\label{A.sieben}{\footnotesize The effect of increasing the value $\wh(\yh)$ on a typical trajectory of Type IIIa starting with $\lh(\yh) = 0.2, \wh(\yh) = 0$ at $\yh = 0.5$. The bold straight line indicates the boundary $\lh = y/2$. Increasing $\wh(\yh)$ leads to a decrease of the IR value of the cosmological constant. For sufficiently large values $\wh(\yh)$, this effect can be used to turn a trajectory of Type IIIa into a trajectory of the Type IIa or Ia.}}}
\end{figure}

Figure \ref{A.sieben} further illustrates that, compared to the trajectory starting with $\wh(\yh) = 0$, positive values $\wh(\yh)$ lead to a decrease of $\lh(y)$ in the IR. Since we can choose arbitrarily large values $\wh(\yh)$ this mechanism can be used to turn trajectories of Type IIIa into a trajectory of Type IIa and Ia by choosing $\wh(\yh) = \wh(\yh)_{\rm crit}$ and $\wh(\yh) > \wh(\yh)_{\rm crit}$, respectively. For the ``fine-tuned'' Type IIa trajectories the IR value of $\wh(y)$ vanishes identically.
\begin{table}
\begin{tabular}{@{\extracolsep{\fill}} ccc} 
Type & $\wh(\yh)$ chosen & Changes in the flow of $\lh(y)$ \\ \hline 
& & Termination of the trajectory \\[-1.5ex]
\raisebox{1.5ex}[-1.5ex]{\hspace*{2mm} arbitrary} & \raisebox{1.5ex}[-1.5ex]{$\wh(\yh) < 0$} & at $y = y_{\rm term} > 0$ with $\wh(y_{\rm term}) \rightarrow -\infty$  \\ \hline \hline
 & & Type Ia \\[-1.5ex]
\raisebox{1.5ex}[-1.5ex]{\hspace*{2mm} Type Ia} & \raisebox{1.5ex}[-1.5ex]{$\wh(\yh) > 0$} & Change in the slope of $\lh(y)$ \\[-1.5ex]
& & in the region $y \lesssim 0.01$ from $d \lh / dy < 0$ to  $d \lh / dy > 0$ \\ \hline
& & Type Ia \\[-1.5ex]
\raisebox{1.5ex}[-1.5ex]{\hspace*{2mm} Type IIa} & \raisebox{1.5ex}[-1.5ex]{$\wh(\yh) > 0$} & $\wh(\yh) > 0$ drives $\lh(0)$ away from zero. \\ \hline
& $\wh(\yh) < \wh(\yh)_{\rm crit}$ & Type IIIa \\
\raisebox{1.5ex}[-1.5ex]{\hspace*{2mm} Type IIIa} & & Type IIa \\[-1.5ex]
& \raisebox{1.5ex}[-1.5ex]{$\wh(\yh) = \wh(\yh)_{\rm crit}$ } & Fine-tuning of $\wh(\yh)$ \\
& $\wh(\yh) > \wh(\yh)_{\rm crit}$ & Type Ia \\ 
\end{tabular}
\caption{\label{A.one} \footnotesize Summary of the effect of $\wh(y)$ on the RG flow of the cosmological constant $\lh(y)$.}
\end{table}

The effects of a nonzero coupling $\wh(\yh)$ on the RG flow of the cosmological constant are summarized in Table \ref{A.one} which is analogous to Table \ref{3.one}. The column ``Type'' determines the Type of trajectory which is obtained by setting $\wh(\yh) = 0$. The column ``$\wh(\yh)$ chosen'' indicates which values of $\wh(\yh)$ then give rise to the changes of the RG flow of the cosmological constant outlined in the column ``Changes in the flow of $\lh(y)$''. 
\end{subsection}
\begin{subsection}{Classical $S^4$ solutions}
We now study the scale dependence of the radius of the classical $S^4$--solutions in the $V$+$V^2$--truncation. The modified Einstein equations resulting from our ansatz
\be\label{A.9}
\Gamma[g] = \frac{1}{16 \pi G} \, \int d^4x \sqrt{g} (-R + 2 \lb) \, + \, \frac{1}{16 \pi G} \bar{w} \left( \int d^4x \sqrt{g} \right)^2,
\ee
are of the form \rf{1.4} again, with the effective cosmological constant given by
\be\label{A.9a}
\lambda_{\rm eff} = \lb + \w V
\ee
A 4-sphere of radius $r$ is a solution provided \rf{1.7} with \rf{A.9a} is satisfied, i.e. if
\be\label{A.9b}
\ob_4 \, \w \, r^6 + \lb \, r^2 -3= 0
\ee

If one inserts the metric of an $S^4$ into the functional \rf{A.9} and introduces the dimensionless radius $\rh \equiv r / \ell_{\rm Pl}$ one obtains the function $\Gamma^{\rm Sphere}(\rh)$ of eq. \rf{4.4}. Its extrema are given by the solutions of \rf{A.9b}. In dimensionless variables, and now with $y$-dependent coupling constants, it reads 
\be\label{A.11}
\ob_4 \, \wh(y) \, \rh(y)^6 + \lh(y) \, \rh(y)^2 - 3 = 0
\ee

In Fig. \ref{4.eins} we saw that a stable minimum of $\Gamma^{\rm Sphere}(\rh)$ only occurs for trajectories in the positive coupling region $\wh(y) > 0$. This is exactly the region of coupling constant space where we found ``stable'' trajectories of the Types Ia, IIa and IIIa. As in the case of the $V$+$V \ln V$--truncation we shall now compute the ``running $S^4$-radius'' by solving \rf{A.11} for $\rh(y)$ with the running coupling constants of the Type Ia and IIa trajectories inserted. First we focus on the dependence of the ``initial radius'' at the Planck scale, $\rh \left(y=1; \wh(\yh), \lh(\yh) \right)$, and the ``final radius'' in the IR, $\rh \left(y=0; \wh(\yh), \lh(\yh) \right)$, on the initial data $\left( \wh(\yh), \lh(\yh) \right)$ of the trajectory. These functions are shown in Fig. \ref{A.acht}. 
\begin{figure}[t]
\renewcommand{\baselinestretch}{1}
\epsfxsize=0.49\textwidth
\begin{center}
\leavevmode
\epsffile{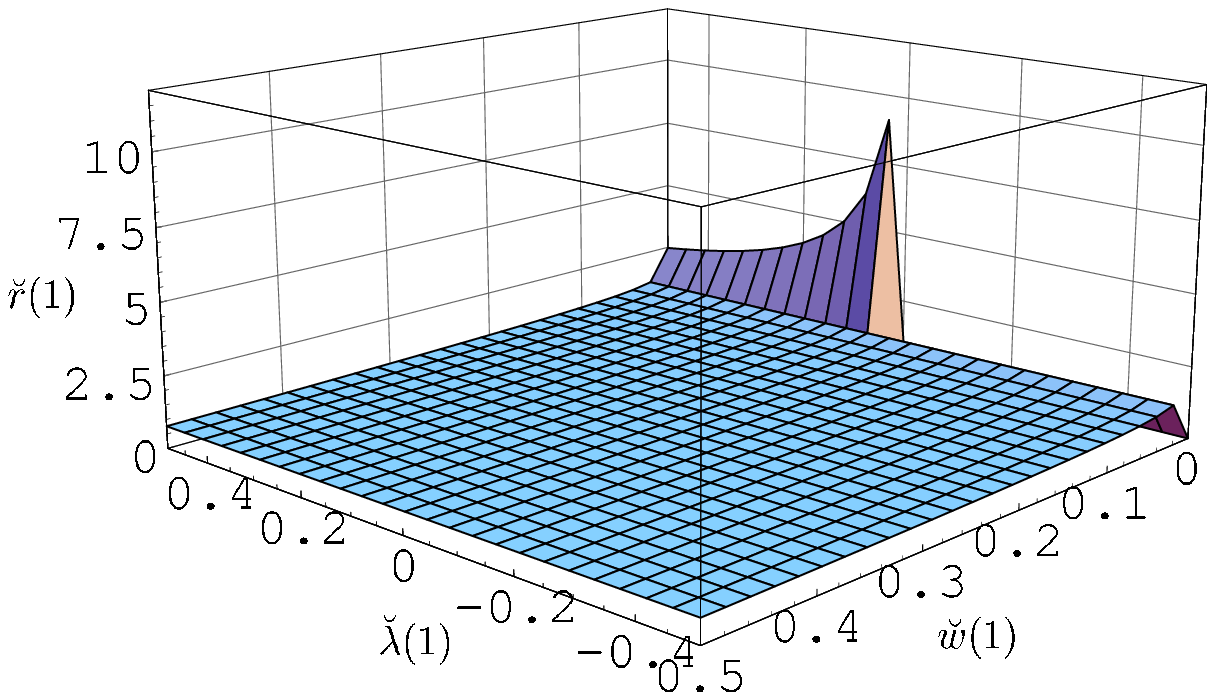}
\epsfxsize=0.48\textwidth
\epsffile{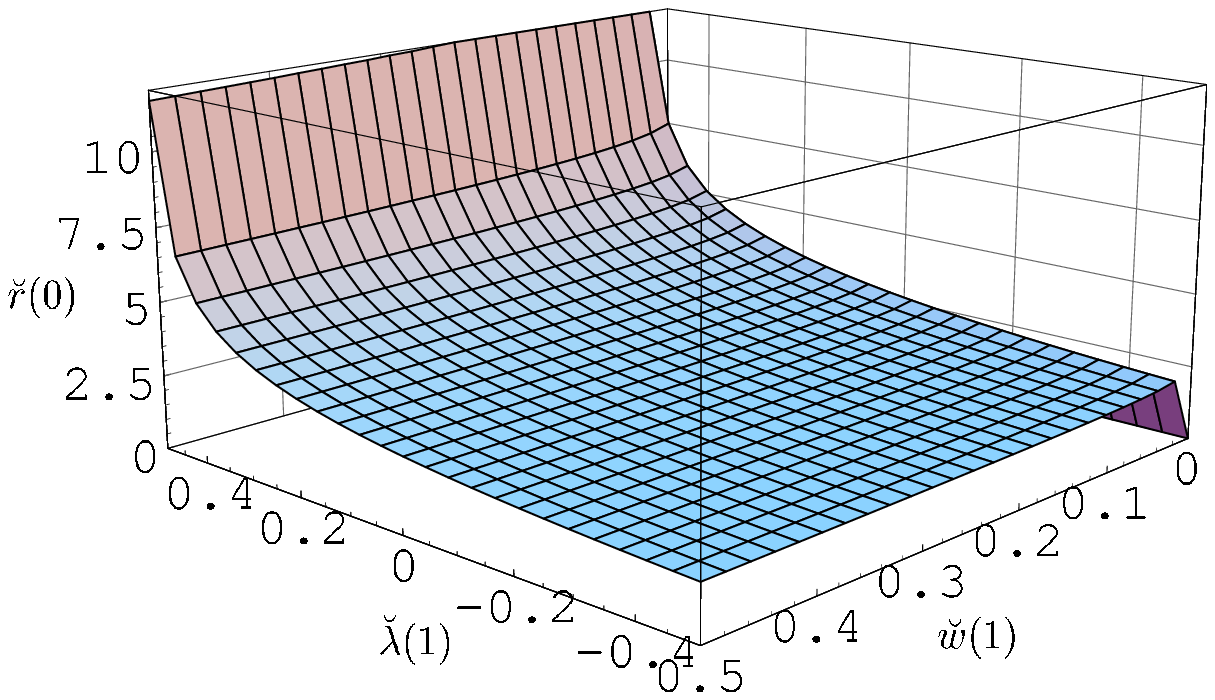}
\end{center}
\parbox[c]{\textwidth}{\caption{\label{A.acht}{\footnotesize Dependence of the radius $\rh$ of the spherical solutions on the initial values $\wh(\yh), \lh(\yh)$ of the trajectories along which $\rh$ is parameterized. The initial values are given at $\yh = 1$. The left and right diagram show the dependence of $\rh(y=1)$ and $\rh(y=0)$ on $\wh(\yh = 1)$ and $\lh(\yh=1)$, respectively. Including the effect of the running coupling constants generally leads to larger values $\rh(0) > \rh(\yh)$.}}}
\end{figure}

In the first diagram we find that, for $\wh(\yh) > 0$, the initial radius $\rh(\yh)$ is generically of order unity. The only exception occurs along the line $\wh(\yh) = 0$. Here positive values $\lh(\yh) > 0$ lead to a stable minimum of $\Gamma^{\rm Sphere}$ at $\rh(\yh) > 1$ which, for $\lh(\yh) \rightarrow 0^+$, is driven to $\rh(\yh) \rightarrow \infty$ while for values $\lh(\yh) < 0$ no stable minima occur.

Regarding the IR value $\rh(0)$ shown in the second diagram of Fig. \ref{A.acht} we find that including the running of the coupling constants generically leads to a moderate increase of the radius: $\rh(0) > \rh(\yh)$. For the trajectories starting close to the boundary singularity, $\lh(\yh) \lesssim \yh/2$, we find comparably large values $\rh(0)$. This is caused by the rapid decrease of $\wh(y)$ at $y \approx \yh$ which drives the minimum of $\Gamma^{\rm Sphere}(\rh)$ towards larger values $\rh(0)$. Along the line $\wh(\yh)=0$ all trajectories lead to negative IR values $\lh(0) < 0$, i.e. to trajectories of Type Ia. No stable minima occur along this line.

Let us now study the $y$-dependence of $\rh(y)$ along the typical trajectory of Type Ia starting with $\lh(\yh) = 0.3$ and $\wh(\yh) = 0.01$. The trajectory as well as $\rh(y)$ along this trajectory are shown in Fig. \ref{A.neun}.
\begin{figure}[t]
\renewcommand{\baselinestretch}{1}
\epsfxsize=0.49\textwidth
\begin{center}
\leavevmode
\epsffile{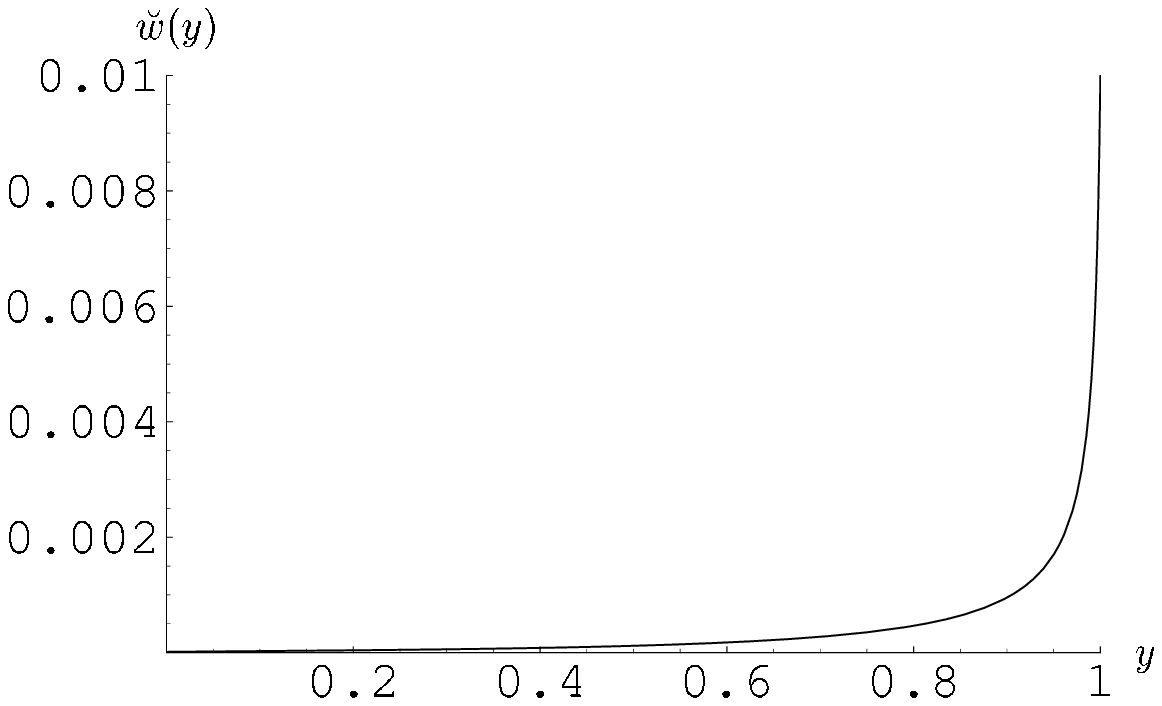}
\epsfxsize=0.48\textwidth
\epsffile{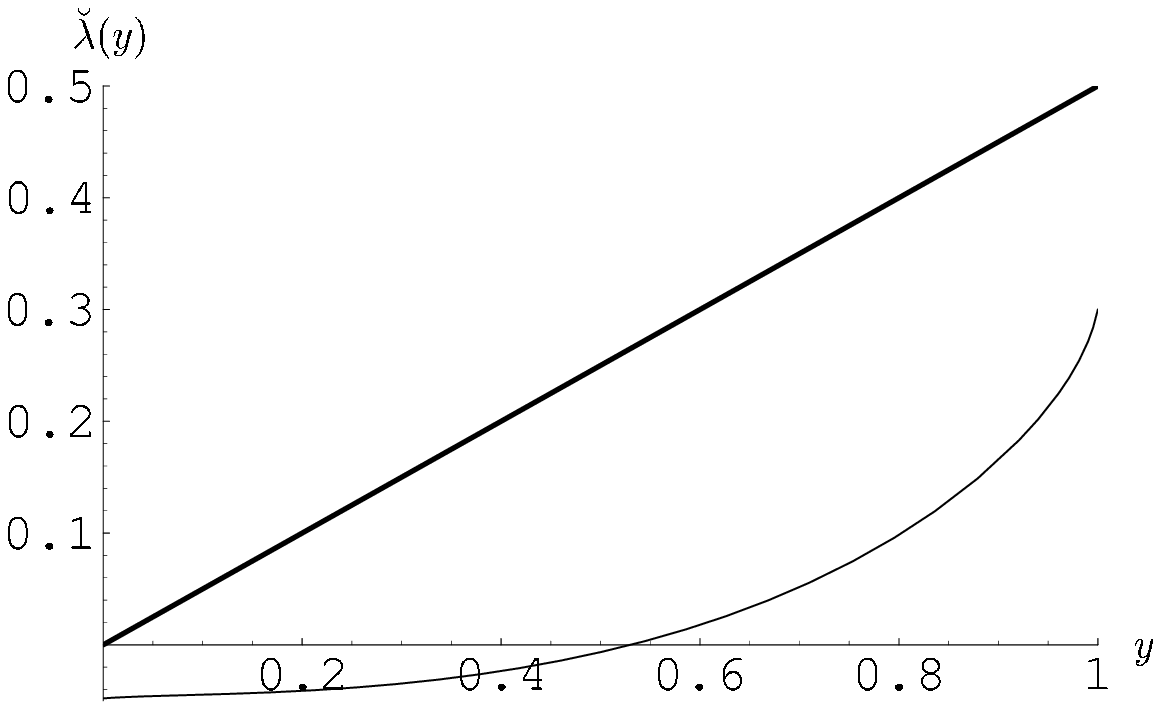}
\end{center}
\vspace{2mm}
\epsfxsize=0.49\textwidth
\begin{center}
\leavevmode
\epsffile{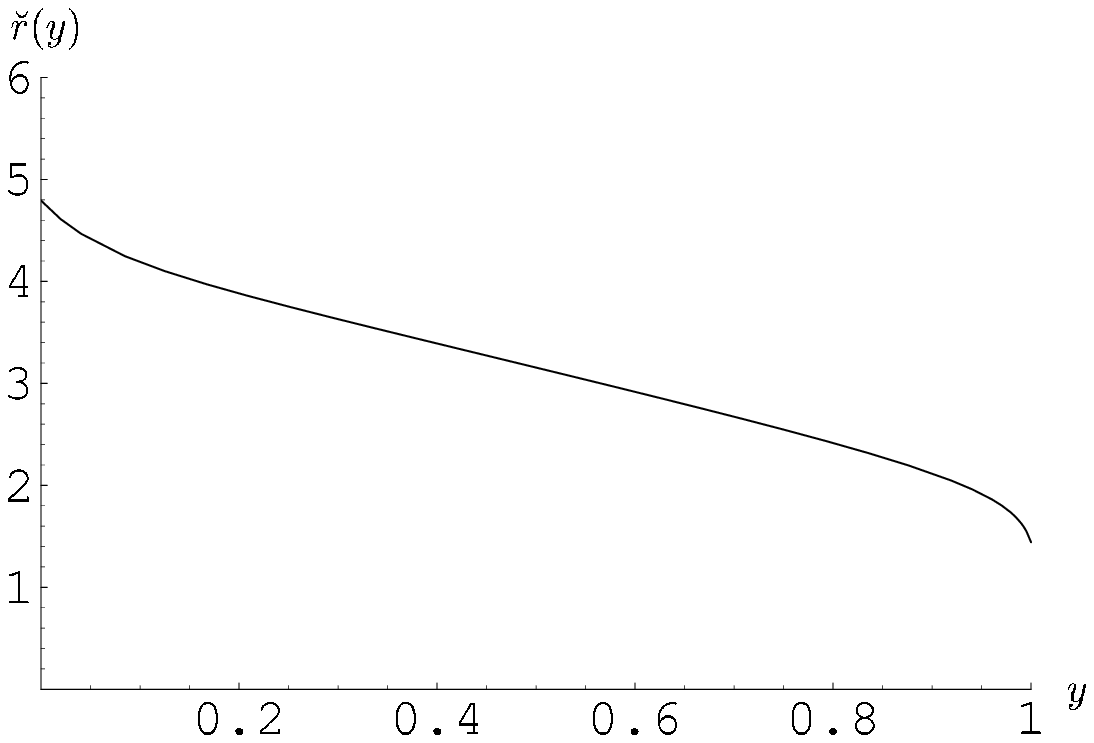}
\epsfxsize=0.48\textwidth
\epsffile{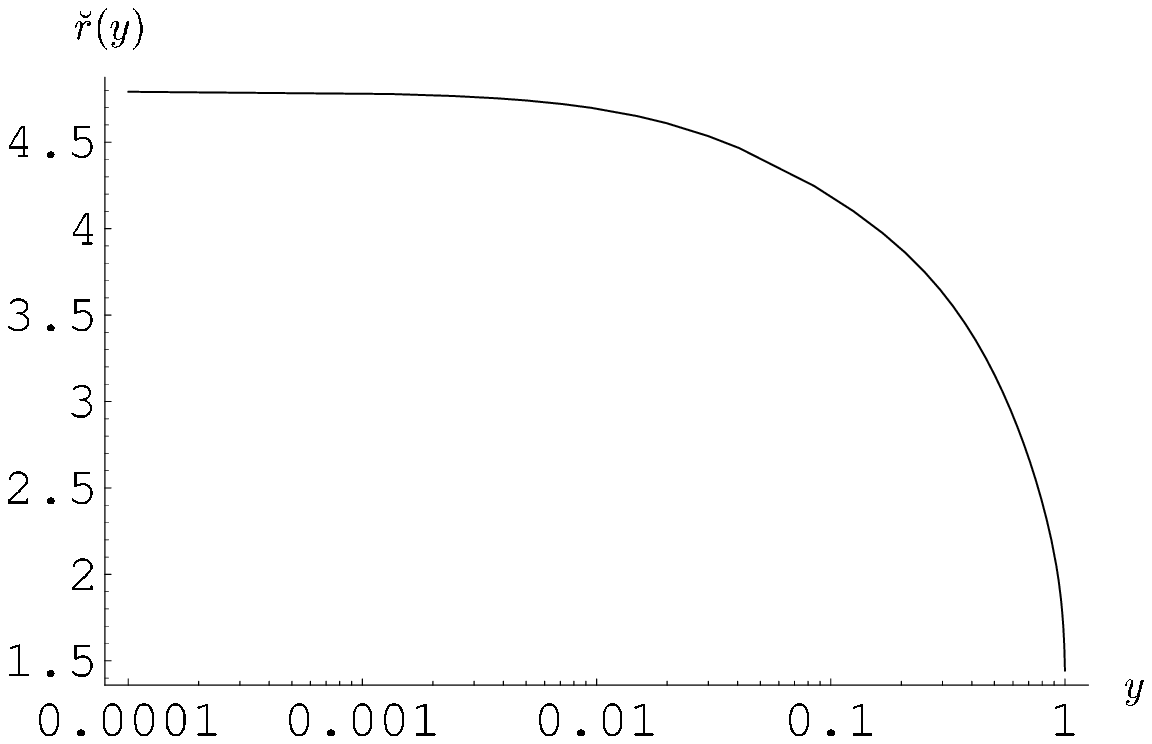}
\end{center}
\parbox[c]{\textwidth}{\caption{\label{A.neun}{\footnotesize Radius $\rh(y)$ along the typical Type Ia-trajectory starting at $\wh(\yh) = 0.01$ and $\lh(\yh) = 0.3$. Outside the IR region ($y > 0.01$), $\rh(y)$ increases by a factor of three due to the running of $\wh(y)$ and $\lh(y)$. For $y < 0.01$, $\rh(y)$ is approximately constant.}}}
\end{figure}

We see that there is a substantial increase of $\rh(y)$ only in the region $0.01 \lesssim y \lesssim 1$. In the IR-region ($y \lesssim 0.01$) we find that $\rh(y)$ is approximately constant. Hence the increase of $\rh(y)$ is {\it not} based on a typical IR effect. Most importantly, we see that by switching on the RG running of the couplings we can increase the radius only by about a factor of 3, for typical initial conditions. This magnification factor has to be compared to the many orders of magnitude we gain by including the RG running in the $V \ln V$--case, see Section IV.B.

Thus we conclude that in order to dynamically generate large or even flat universes from natural initial data fixed at the Planck scale the nonlocal invariant $V \ln V$ is by far more efficient than $V^2$.
\end{subsection}
\end{appendix}

\end{document}